\documentclass[a4paper,11pt]{article}
\pdfoutput=1 
\usepackage{jheppub}
\usepackage{amsmath}
\usepackage{amsfonts}
\usepackage{amssymb}
\usepackage{slashed}
\usepackage{bm}
\usepackage{graphicx}%
\usepackage[T1]{fontenc} % if needed
\newcommand{\dhd}{{\textstyle d}
\lower.03ex\hbox{\kern-0.38em$^{\scriptstyle-}$}\kern-0.05em{}}
\newcommand{\dbar}{{\textstyle \delta}
\lower.03ex\hbox{\kern-0.38em$^{\scriptstyle-}$}\kern-0.05em{}}
\newcommand{\half}{{1\over 2}}
\newcommand{\bu}{{\bullet}}

\newcommand{\barc}{{\bar c}}
\newcommand{\bard}{{\bar d}}
\newcommand{\bare}{{\bar e}}
\newcommand{\barf}{{\bar f}}
\newcommand{\barg}{{\bar g}}
\newcommand{\barh}{{\bar h}}

\newcommand{\bars}{{\bar s}}

\newcommand{\baru}{{\bar u}}

\newcommand{\bsi}{{\bar \psi}}
\newcommand{\Bsi}{{\bar \Psi}}

\newcommand{\bhi}{{\bar \chi}}

\newcommand{\Bxi}{{\bar \Xi}}

\newcommand{\cald}{{\cal D}}  
\newcommand{\calf}{{\cal F}}
 
\newcommand{\calj}{{\cal J}} 
\newcommand{\kal}{{\cal L}}

\newcommand{\calo}{{\cal O}}    
  
\newcommand{\calp}{{\cal P}}

\newcommand{\hatp}{{\hat p}}

\newcommand{\hatA}{{\hat A}}
\newcommand{\hatB}{{\hat B}}
\newcommand{\hatF}{{\hat F}}

\newcommand{\hatD}{{\hat D}}

\newcommand{\hsi}{{\hat \psi}} 
\newcommand{\hbsi}{\hat {\bar\psi}} 
\newcommand{\hamma}{{\breve \gamma}}

\newcommand{\tilh}{{\tilde h}}
\newcommand{\tilj}{{\tilde j}}

\newcommand{\tilA}{{\tilde A}}
\newcommand{\tilB}{{\tilde B}}
\newcommand{\tilC}{{\tilde C}}

\newcommand{\tilF}{{\tilde F}}

\newcommand{\tilJ}{{\tilde J}}

\newcommand{\tilS}{{\tilde S}}

\newcommand{\tigma}{\tilde {\sigma}} 
\newcommand{\tsi}{\tilde {\psi}} 
\newcommand{\tipsi}{\tilde {\psi}}

\pdfoutput=1
\abstract{A typical factorization formula for  production of a particle with a small transverse momentum in hadron-hadron collisions is given
by a convolution of two TMD parton densities with cross section of production of the final particle by the two partons.  
For practical applications at a given transverse momentum, though, one should  estimate at what momenta 
the power corrections to the TMD factorization formula  become essential. In this paper  we calculate 
the first power corrections to TMD factorization formula for Z-boson production and Drell-Yan process in high-energy hadron-hadron collisions. 
At the leading order in $N_c$ power corrections are expressed in terms of leading power TMDs by QCD equations of motion.}
\arxivnumber{}
\affiliation{$^a$ Physics Department, Old Dominion University, Norfolk, VA 23529, USA and Thomas Jefferson National Accelerator Facility, Newport News, VA 23606, USA}
\affiliation{$^b$ Physics Department, Brookhaven National Laboratory, Upton, NY 11973, USA}
\emailAdd{balitsky@jlab.org}
\emailAdd{atarasov@bnl.gov}
\begin{document}

\title{\boldmath Power corrections to TMD factorization for Z-boson production}
\author{I. Balitsky$^a$ and A. Tarasov$^b$}
\preprint{BNL-114875-2017-JA, JLAB-THY-17-2609}
\maketitle

\flushbottom

\section{Introduction\label{aba:sec1}}

\bigskip

A typical analysis of differential cross section of particle production in hadron-hadron collisions at small momentum transfer
of the produced particle is performed with the help of TMD factorization \cite{Collins:2011zzd, Collins:1981uw, Collins:1984kg, Ji:2004wu, GarciaEchevarria:2011rb,Bauer:2000yr,Bauer:2001yt,Bauer:2002nz,Becher:2010tm,Chiu:2012ir}. However, the question of how small should be the momentum transfer in order for  leading power TMD analysis to be successful cannot be resolved at the  leading-power level. The sketch of the 
factorization formula for the differential cross section is \cite{Collins:2011zzd, Collins:2014jpa}
\begin{eqnarray}
&&\hspace{-2mm}
{d\sigma\over  d\eta d^2q_\perp}~=~\sum_f\!\int\! d^2b_\perp e^{i(q,b)_\perp}
\cald_{f/A}(x_A,b_\perp,\eta)\cald_{f/B}(x_B,b_\perp,\eta)\sigma(ff\rightarrow H)
\nonumber\\
&&\hspace{-2mm}
+~{\rm power ~corrections}~+~{\rm Y-terms},
\label{TMDf}
\end{eqnarray}
where $\eta$ is the rapidity, $q$ is the momentum of the produced particle in the hadron frame (see ref. \cite{Collins:2011zzd}), $\cald_{f/A}(x,z_\perp,\eta)$ is the TMD density of  a parton $f$  in hadron $A$, and $\sigma(ff\rightarrow H)$ is the cross section of production of particle $H$ in the scattering of two partons.  The common wisdom is that when we increase transverse momentum $q^2_\perp$ of the produced hadron, at first the leading power TMD analysis
with (nonperturbative) TMDs applies, then at some point  power corrections kick in, and finally at $q_\perp^2\sim Q^2$, where $Q^2 = q^2$, they are transformed into so-called Y-term making
smooth transition to collinear factorization formulas. In this paper we try to answer the question about the first transition, 
namely at what $q_\perp^2$ power corrections become significant.

In our recent paper  \cite{Balitsky:2017flc} we calculated power corrections $\sim {q_\perp^2\over Q^2}$ for Higgs boson production by gluon-gluon fusion. 
The result was a TMD factorization formula with matrix elements of three-gluon operators 
 divided by an extra power of $m_H^2$.
In this paper we calculate power corrections $\sim {q_\perp^2\over Q^2}$ for Z-boson production which are determined by quark-quark-gluon operators. In the leading order Z-boson production was studied in \cite{Landry:1999an,Qiu:2000hf,Landry:2002ix,Mantry:2010bi,Su:2014wpa,DAlesio:2014mrz,Catani:2015vma,Bacchetta:2017gcc,Scimemi:2017etj}.
The interesting (and unexpected) result of our paper is that at the leading-$N_c$ level matrix elements of the relevant quark-quark-gluon operators can be expressed in terms
of leading power quark-antiquark TMDs by QCD equations of motion (see ref. \cite{Mulders:1995dh}). 
The method of calculation is very similar to that of ref. \cite{Balitsky:2017flc} so we will streamline the discussion of the general approach and pay attention to details specific to quark operators.

The paper is organized as follows.  In section  \ref{sec:funt} we derive the TMD factorization
from the double functional integral for the cross section of particle production. 
In section \ref{sect:power}, which is central to our approach, we explain the method of
calculation of power corrections based on a solution of classical Yang-Mills equations.
In section \ref{sec:lhtc} we find the leading power correction to
particle production in the region $s\gg Q^2\gg q_\perp^2$. In section \ref{sec:results}  we perform the order-of-magnitude estimate of power corrections
and in section \ref{sec:DY}  present our result for power corrections to the Drell-Yan cross section.
The necessary technical details  and discussion of subleading power corrections can be found in appendix.

\section{TMD factorization from functional integral \label{sec:funt}}
\begin{figure}[htb]
\begin{center}
\includegraphics[width=131mm]{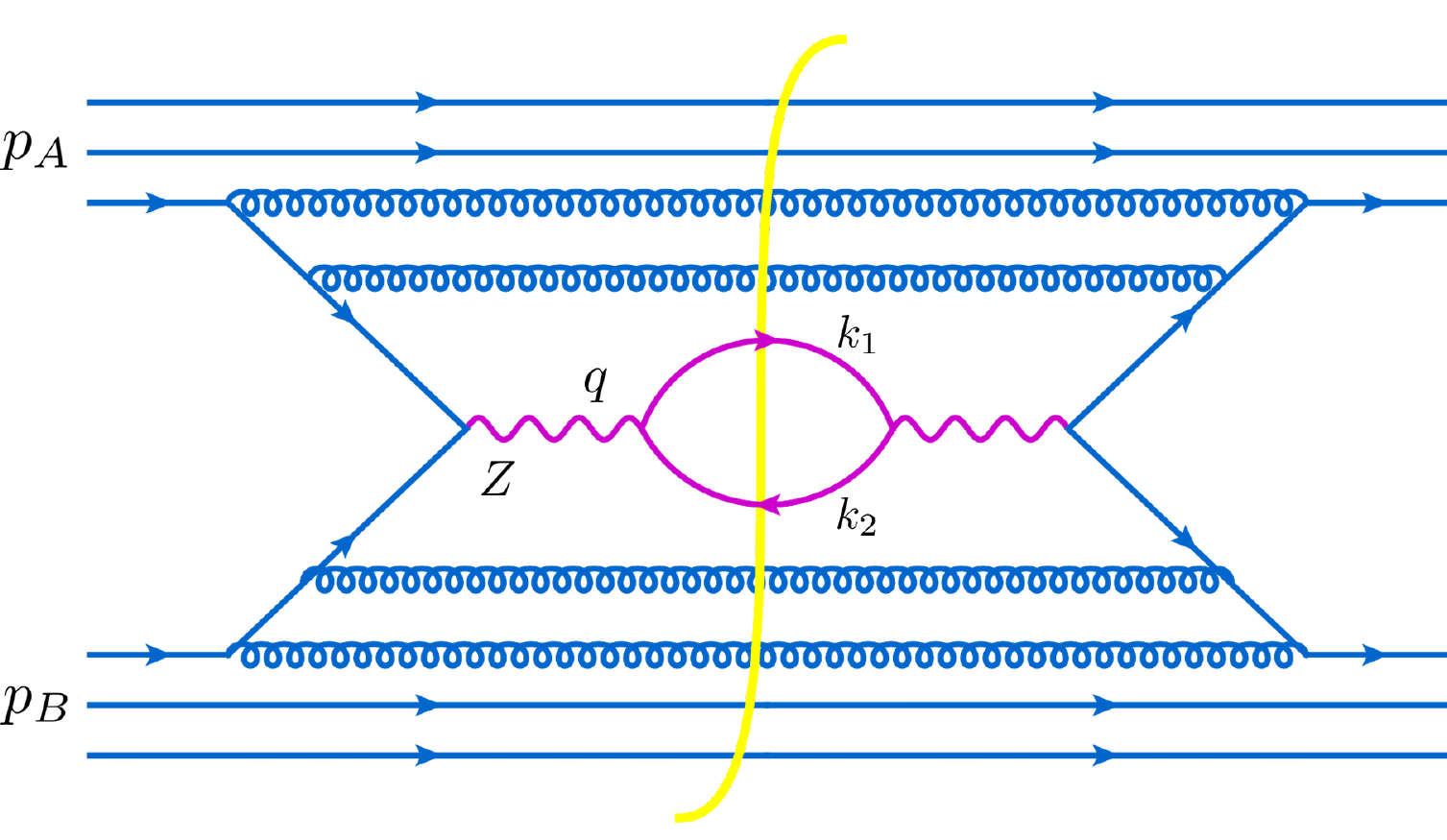}
\end{center}
\caption{Z-boson production in hadron-hadron collisions\label{fig:1}}
\end{figure}

We consider Z-boson production in the Drell-Yan reaction illustrated in figure \ref{fig:1}:
\begin{eqnarray}
h_A(p_A) + h_B(p_A) \to Z(q) + X \to l_1(k_1) + l_2(k_2) + X,
\end{eqnarray}
where $h_{A,B}$ denote the colliding hadrons, and $l_{1, 2}$ the outgoing lepton pair with total momentum $q = k_1 + k_2$.

The relevant term of the Lagrangian for the fermion fields $\psi_i$ describing coupling between fermions  and Z-boson is ($s_W\equiv\sin\theta_W$, $c_W\equiv\cos\theta_W$)
\begin{equation}
\kal_Z~=~ \int\! d^4x~J_\mu Z^\mu(x),~~~~~~~~~J_\mu~=~-{e\over 2s_Wc_W}\sum_i \bsi_i\gamma_\mu(g_i^V-g_i^A\gamma_5)\psi_i,
\end{equation}
where sum goes over different types of fermions, and coupling constants $g_i^V=(t_3^L)_i-2q_is^2_W$ and $g_i^A=(t_3^L)_i$ are defined by week isospin $(t_3^L)_i$ of the fermion $i$, see ref. \cite{Patrignani:2016xqp}. In this paper we take into account only $u,d,s,c$ quarks and $e, \mu$ leptons. We consider all fermions to be massless.
 
%%%%%%%%%%%%%FIGA%%%%%%%%%%%%%%%%%%%%%%%
%%%%%%%%%%%%%%%%%%%%%%%%%%%%%%%%%%%

The differential cross section of Z-boson production with subsequent decay into $e^+e^-$ (or $\mu^+\mu^-$) pair is
\begin{eqnarray}
&&\hspace{-2mm}
{d\sigma\over dQ^2 dy dq_\perp^2}
~
=~{e^2Q^2\over 192s s_W^2c_W^2}
{1-4s_W^2+8s_W^4\over (m_Z^2-Q^2)^2+\Gamma_Z^2m_Z^2}[-W(p_A,p_B,q)],
\end{eqnarray}
where we defined the ``hadronic tensor''  $W(p_A,p_B,q)$ as 
\begin{eqnarray}
\hspace{-1mm}
W(p_A,p_B,q)~&\stackrel{\rm def}{=}&~{1\over (2\pi)^4}\sum_X\!\int\! d^4x~e^{-iqx}
\langle p_A,p_B|J_\mu(x)|X\rangle\langle X|J^\mu(0) |p_A,p_B\rangle
\nonumber\\
~&=&~{1\over (2\pi)^4}\!\int\! d^4x~e^{-iqx}
\langle p_A,p_B|J_\mu(x)J^\mu(0) |p_A,p_B\rangle  
\label{W}.
\end{eqnarray}
As usual,  $\sum_X$ denotes the sum over full set of ``out''  states. It should be mentioned that there is a power correction coming from the leptonic tensor term $\sim q^\mu q^\nu$. However, if we consider quarks to be massless, the only effect of the $q^\mu q^\nu$ term comes from the (square of) axial anomaly which has an extra factor $\alpha_s^2$, and such two-loop factor is beyond our tree approximation.

The sum over full set of ``out''  states in Eq. (\ref{W})  can be represented by a double functional integral
\begin{eqnarray}
&&\hspace{-2mm}
(2\pi)^4W(p_A,p_B,q)~=~\sum_X\!\int\! d^4x~e^{-iqx}
\langle p_A,p_B|J^\mu(x)|X\rangle\langle X|J_\mu(0)|p_A,p_B\rangle
\label{dablfun}\\
&&\hspace{-2mm}
=~\lim_{t_i\rightarrow -\infty}^{t_f\rightarrow\infty}\!\!\int \! d^4x~ e^{-iqx}
\!\int^{\tilA(t_f)=A(t_f)}\!\!  D\tilA_\mu DA_\mu \!\int^{\tsi(t_f)=\psi(t_f)}\! D\tilde{\bar\psi}D\tilde{\psi} D\bsi D\psi 
~\Psi^\ast_{p_A}(\vec{\tilA}(t_i),\tipsi(t_i))
\nonumber\\
&&\hspace{-2mm}
\times~\Psi^\ast_{p_B}(\vec{\tilA}(t_i),\tipsi(t_i))e^{-iS_{\rm QCD}(\tilA,\tipsi)}e^{iS_{\rm QCD}(A,\psi)}
\tilde{J}_\mu(x)J^\mu(0)\Psi_{p_A}(\vec{A}(t_i),\psi(t_i))\Psi_{p_B}(\vec{A}(t_i),\psi(t_i)).
\nonumber
\end{eqnarray}
In this double functional integral the amplitude $\langle X|J_\mu(0)|p_A,p_B\rangle$ is given by the integral over $\psi,A$ fields 
whereas the complex conjugate amplitude $\langle p_A,p_B|J^\mu(x)|X\rangle$ is  represented by the integral over $\tsi,\tilA$ fields. 
Also, $\Psi_p(\vec{A}(t_i),\psi(t_i))$ denotes the proton wave function at the initial time $t_i$ and the boundary conditions
$\tilA(t_f)=A(t_f)$ and $\tsi(t_f)=\psi(t_f)$ reflect the sum over all states $X$, cf. refs. \cite{Balitsky:1988fi,Balitsky:1990ck,Balitsky:1991yz}.

We use 
Sudakov variables $p=\alpha p_1+\beta p_2+p_\perp$, where $p_1$ and $p_2$ are light-like vectors close to $p_A$ and $p_B$, and the notations $x_\bu\equiv x_\mu p_1^\mu$ and $x_\ast\equiv x_\mu p_2^\mu$ 
for the dimensionless light-cone coordinates ($x_\ast=\sqrt{s\over 2}x_+$ and $x_\bu=\sqrt{s\over 2}x_-$). Our metric is $g^{\mu\nu}~=~(1,-1,-1,-1)$ so 
that $p\cdot q~=~(\alpha_p\beta_q+\alpha_q\beta_p){s\over 2}-(p,q)_\perp$ where $(p,q)_\perp\equiv -p_iq^i$. Throughout the paper, the sum over the Latin indices $i$, $j$, ... runs over two transverse components while the sum over Greek indices $\mu$, $\nu$, ... runs over four components as usual.

\begin{figure}[htb]
\begin{center}
\includegraphics[width=151mm]{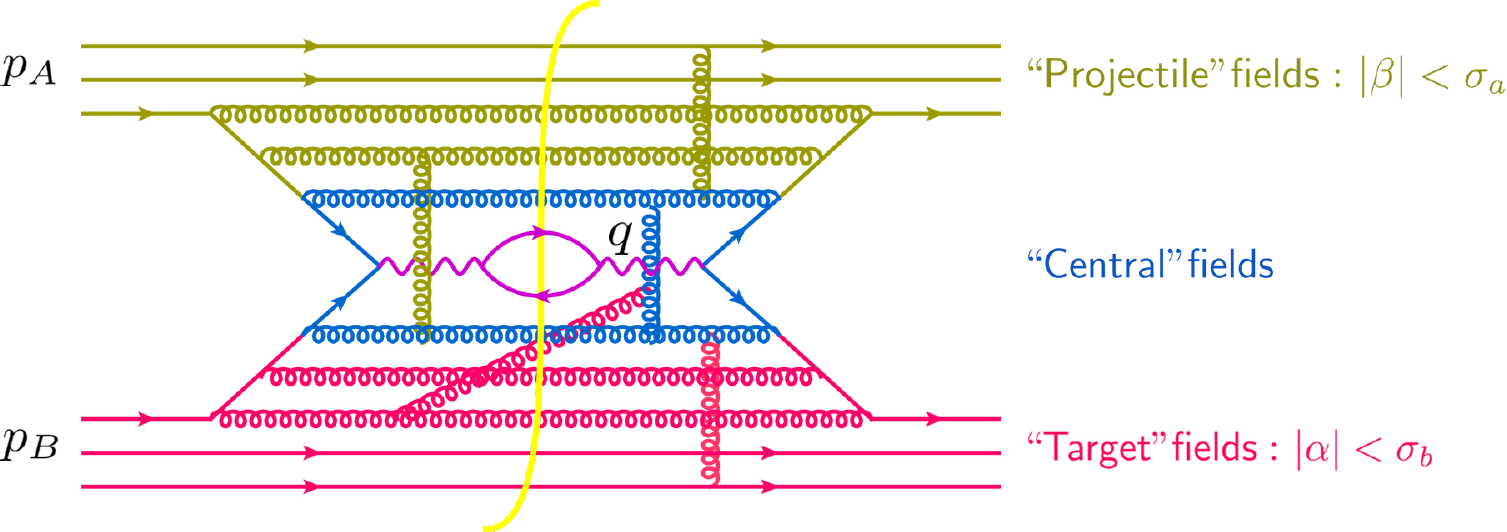}
\end{center}
\caption{Rapidity factorization for particle production \label{fig:2}}
\end{figure}

Following ref. \cite{Balitsky:2017flc}  we separate quark and gluon fields in the functional integral (\ref{dablfun}) into three sectors (see figure \ref{fig:2}): 
``projectile'' fields $A_\mu, \psi_A$ 
with $|\beta|<\sigma_a$, 
``target'' fields $B_\mu, \psi_B$ with $|\alpha|<\sigma_b$ and ``central rapidity'' fields $C_\mu,\psi_C$ with $|\alpha|>\sigma_b$ and $|\beta|>\sigma_a$
and get
%%%%%%%%%%%%%FIGA%%%%%%%%%%%%%%%%%%%%%%%
%%%%%%%%%%%%%%%%%%%%%%%%%%%%%%%%%%%
%
\begin{eqnarray}
&&\hspace{-1mm}
W(p_A,p_B,q)~=~{1\over(2\pi)^4}\!\int \! d^4x e^{-iqx}
\int^{\tilA(t_f)=A(t_f)}\! D\tilA_\mu DA_\mu
\int^{\tipsi_A(t_f)=\psi_A(t_f)}D\bsi_A D\psi_A 
\nonumber\\
&&\hspace{-1mm}
\times 
~D\tilde{\bar\psi}_A D\tilde{\psi}_A 
e^{-iS_{\rm QCD}(\tilA,\tipsi_A)}e^{iS_{\rm QCD}(A,\psi_A)}\Psi^\ast_{p_A}(\vec{\tilA}(t_i),\tipsi_A(t_i))
\Psi_{p_A}(\vec{A}(t_i),\psi_A(t_i))
\nonumber\\
&&\hspace{-1mm}
\times
\int^{\tilB(t_f)=B(t_f)}\! D\tilB_\mu DB_\mu 
\int^{\tipsi_B(t_f)=\psi_B(t_f)}\!\!D\bsi_B D\psi_B D\tilde{\bar\psi}_B D\tilde{\psi}_B 
\nonumber\\
&&\hspace{-1mm}
\times~e^{-iS_{\rm QCD}(\tilB,\tipsi_B)}e^{iS_{\rm QCD}(B,\psi_B)}
\Psi^\ast_{p_B}(\vec{\tilB}(t_i),\tipsi_B(t_i))\Psi_{p_B}(\vec{B}(t_i),\psi_B(t_i))
\label{W2}\\
&&\hspace{-1mm}
\times~
\int\!DC_\mu\! \int^{\tilC(t_f)=C(t_f)} \! D\tilC_\mu \!\int\!D\bsi_C D\psi_C \int^{\tsi_C(t_f)=\psi_C(t_f)}\! D\tilde{\bar\psi}_C D\tsi_C~
\tilJ_\mu(x)J^\mu(0)~e^{-i\tilS_C+iS_C},
\nonumber
\end{eqnarray}
where $S_C=S_{\rm QCD}(C + A+B, \psi_C + \psi_A + \psi_B)-S_{\rm QCD}(A, \psi_A)-S_{\rm QCD}(B, \psi_B)$ and similarly for $\tilS_C$.\footnote{This procedure is obviously gauge-dependent. We have in mind factorization in covariant-type gauge, e.g. Feynman gauge.}

Our goal is to integrate over central fields and get
the amplitude in the factorized form, i.e. as a product of functional integrals over $A$ fields representing projectile matrix elements (TMDs of the projectile) 
and functional integrals over $B$ fields representing target matrix elements (TMDs of the target).

In the spirit of background-field method, we ``freeze'' projectile and target fields and get a sum of diagrams in these external fields. 
Since  $|\beta|<\sigma_a$ in the projectile fields and $|\alpha|<\sigma_b$  in the target fields, at the  tree-level 
one can set with power accuracy $\beta=0$ for the  projectile fields and $\alpha=0$ for the target fields - the corrections will
be $O\big({m^2_N\over\sigma_a s}\big)$ and  $O\big({m^2_N\over\sigma_b s}\big)$, where $m_N$ is the hadron's mass.
Beyond the tree level, one should expect that the integration over $C$ fields will produce
the logarithms of the cutoffs $\sigma_a$ and $\sigma_b$ which will cancel with the corresponding
logs in gluon TMDs of the projectile and the target.
The result of integration over C-fields has the schematic form
\begin{eqnarray}
&&\hspace{-1mm}
\int\!DC_\mu\! \int^{\tilC(t_f)=C(t_f)} D\tilC_\mu \!\int\!D\bsi_C D\psi_C \int^{\tsi_C(t_f)=\psi_C(t_f)}\! 
D\tilde{\bar\psi}_C D\tsi_C~\tilJ_\mu(x)J^\mu(0)~e^{-i\tilS_C+iS_C}
\nonumber\\
&&\hspace{-1mm}
=~e^{S_{\rm eff}(A,\psi_A,\tilde{A},\tipsi_A; B, \psi_B, \tilde{B},\tipsi_B)} \calo(q,x; A,\psi_A,\tilde{A},\tipsi_A; B, \psi_B, \tilde{B},\tipsi_B),
\label{intc}
\end{eqnarray}
where $\calo(q,x; A,\psi_A,\tilde{A},\tipsi_A; B, \psi_B, \tilde{B},\tipsi_B)$
is a sum of diagrams connected to $\tilJ_\mu(x)J^\mu(0)$ and
 $e^{S_{\rm eff}}$ represents
a sum of disconnected diagrams  (``vacuum bubbles'') in external fields.
As usual, since the rapidities of central $C$ fields and of $A$, $B$ fields are very different,  the result of
integration over $C$ fields is expressed in terms of Wilson-line operators made form $A$ and $B$ fields. 

After integration over $C$ fields the amplitude (\ref{dablfun}) can be rewritten as
\begin{eqnarray}
&&\hspace{-1mm}
W(p_A,p_B,q)~=~{1\over (2\pi)^4}\!\int \! d^4x e^{-iqx}\!
\int^{\tilA(t_f)=A(t_f)}\! D\tilA_\mu DA_\mu
\int^{\tipsi_A(t_f)=\psi_A(t_f)}D\bsi_A D\psi_A D\tilde{\bar\psi}_A D\tilde{\psi}_A 
\nonumber\\
&&\hspace{-1mm}
\times 
~e^{-iS_{\rm QCD}(\tilA,\tipsi_A)}e^{iS_{\rm QCD}(A,\psi_A)}\Psi^\ast_{p_A}(\vec{\tilA}(t_i),\tipsi_A(t_i))
\Psi_{p_A}(\vec{A}(t_i),\psi_A(t_i)) \int^{\tilB(t_f)=B(t_f)}\! D\tilB_\mu DB_\mu
\nonumber\\
&&\hspace{-1mm}
\times \int^{\tipsi_B(t_f)=\psi_B(t_f)}\!\!D\bsi_B D\psi_B D\tilde{\bar\psi}_B D\tilde{\psi}_B e^{-iS_{\rm QCD}(\tilB,\tipsi_B)}e^{iS_{\rm QCD}(B,\psi_B)}\Psi^\ast_{p_B}(\vec{\tilB}(t_i),\tipsi_B(t_i))
\nonumber\\
&&\hspace{-1mm}
\times~\Psi_{p_B}(\vec{B}(t_i),\psi_B(t_i)) e^{S_{\rm eff}(A,\psi_A,\tilde{A},\tipsi_A; B, \psi_B, \tilde{B},\tipsi_B)} \calo(q,x; A,\psi_A,\tilde{A},\tipsi_A; B, \psi_B, \tilde{B},\tipsi_B).
\label{W3}
\end{eqnarray}
From integrals over projectile and target fields in the above equation we see that 
the functional integral over $C$ fields  should
be done in the background of $A$ and $B$ fields satisfying 
\begin{equation}
\tilA(t_f)~=~A(t_f),~~~\tipsi_A(t_f)~=~\psi_A(t_f)~~{\rm and} ~~~\tilB(t_f)~=~B(t_f),~~~\tipsi_B(t_f)~=~\psi_B(t_f).
\label{baukon}
\end{equation}
Combining this with our approximation
that at the tree level $\beta=0$ for $A$, $\tilA$ fields and $\alpha=0$ for $B$, $\tilB$ fields, 
which corresponds to  $A=A(x_\bu,x_\perp),~\tilA=\tilA(x_\bu,x_\perp)$ and $B=B(x_\ast,x_\perp),~\tilB=\tilB(x_\ast,x_\perp)$, 
we see that for the purpose of calculation of the functional integral over central fields (\ref{intc}) we can set 
\begin{eqnarray}
&&
A(x_\bu,x_\perp)=\tilA(x_\bu,x_\perp),~~~~\psi_A(x_\bu,x_\perp)=\tipsi_A(x_\bu,x_\perp)
\nonumber\\
&&
{\rm and}
\nonumber\\
&&
B(x_\ast,x_\perp)=\tilB(x_\ast,x_\perp),~~~~\psi_B(x_\ast,x_\perp)=\tipsi_B(x_\ast,x_\perp).
\label{baukond}
\end{eqnarray}
In other words, since $A$, $\psi$ and $\tilA$, $\tipsi$ do not depend on $x_\ast$, if they coincide at $x_\ast=\infty$ they should coincide everywhere.
Similarly, since $B$, $\psi_B$ and $\tilB$, $\tipsi_B$ do not depend on $x_\bu$, if they coincide at $x_\bu=\infty$ they should be equal.
  
Next, in ref. \cite{Balitsky:2017flc} it was demonstrated that due to eqs. (\ref{baukond}) the effective action  
$S_{\rm eff}(A,\psi_A,\tilde{A},\tipsi_A; B, \psi_B, \tilde{B},\tipsi_B)$ vanishes 
for background fields satisfying conditions (\ref{baukon}).
\footnote{It corresponds to cancellation of so-called ``Glauber gluons'', see discussion in ref. \cite{Collins:2011zzd}.}

Summarizing, we see that at the tree level in our approximation
\begin{eqnarray}
&&\hspace{-1mm}
\int\!DC_\mu\! \int^{\tilC(t_f)=C(t_f)} D\tilC_\mu \!\int\!D\bsi_C D\psi_C \int^{\tsi_C(t_f)=\psi_C(t_f)}~D\tilde{\bar\psi}_C D\tsi_C ~ \tilJ_\mu (x)J^\mu(0)~e^{-i\tilS_C
+iS_C}
\nonumber\\
&&\hspace{-1mm}
=~
\calo(q,x;A,\psi_A;B,\psi_B),
\label{funtc}
\end{eqnarray}
where now $S_C~=~S_{\rm QCD}(C + A+B, \psi_C + \psi_A + \psi_B)-S_{\rm QCD}(A, \psi_A)-S_{\rm QCD}(B, \psi_B)$ and 
$\tilS_C~=~S_{\rm QCD}(\tilC+A+B, \tilde{\psi}_C + \psi_A + \psi_B)-S_{\rm QCD}(A, \psi_A)-S_{\rm QCD}(B, \psi_B)$. It is well known that 
in the tree approximation the double functional integral (\ref{funtc}) is given by a set of 
retarded Green functions in the background fields \cite{Gelis:2003vh,Gelis:2006yv,Gelis:2007kn} (see also appendix A of ref. \cite{Balitsky:2017flc} for the proof).
Since the double functional integral (\ref{funtc}) is given by a set of retarded Green functions 
(in the background fields $A$ and $B$), calculation of the tree-level contribution to $\bsi\gamma_\mu\psi$  
in the r.h.s. of Eq. (\ref{funtc}), is equivalent to solving YM equation for 
$\psi(x)$ (and $A_\mu(x)$) with boundary conditions such that the solution has the same asymptotics at $t\rightarrow -\infty$ 
as the superposition of incoming projectile and  target background fields. 

The hadronic tensor  (\ref{W3}) can now be represented as
\footnote{
As discussed in ref. \cite{Balitsky:2017flc}, 
there is a subtle point in the promotion of background fields to operators. 
 When we calculate 
$\calo$ as the r.h.s. of eq. (\ref{funtc}) the fields $\Phi_A$ and $\Phi_B$ are c-numbers;
on the other hand, after functional integration in eq. (\ref{dablfun}) they become operators which must be time-ordered
in the right sector and anti-time-ordered in the left sector. Fortunately, as we shall see below, all these operators are separated 
either by space-like distances or light-cone distances so all of them (anti) commute and thus can be treated as $c$-numbers.}
\begin{eqnarray}
&&\hspace{-1mm}  
W(p_A,p_B,q)~=~{1\over (2\pi)^4}\!\int \! d^4x ~e^{-iqx}
 \langle p_A|\langle p_B| \hat\calo(q,x;\hatA,\hsi_A;\hatB,\hsi_B)|p_A\rangle |p_B\rangle,   
\label{W4}
\end{eqnarray}
where $\hat \calo(q,x;\hatA,\hsi_A;\hatB,\hsi_B)$ should be expanded in a series in $\hatA$, $\hsi_A$, $\hatB$, $\hsi_B$ operators
and evaluated between the corresponding (projectile or target) states: if
\begin{equation}
\hat\calo(q,x;\hatA,\hsi_A;\hatB,\hsi_B)
~=~\sum_{m,n}\! \int\! dz_mdz'_n c_{m,n}(q,x)\hat\Phi_A(z_m)\hat\Phi_B(z'_n)
\end{equation}
(where $c_{m,n}$ are coefficients and $\Phi$ can be any of $A_\mu$, $\psi$ or $\bsi$) then
\begin{equation}
\hspace{-1mm}  
W~=~\frac{1}{(2\pi)^4}\!\int \! d^4x  e^{-iqx}
\sum_{m,n}\! \int\! dz_m c_{m,n}(q,x)
\langle p_A|\hat\Phi_A(z_m)|p_A\rangle\!\int\! dz'_n\langle p_B| \hat\Phi_B(z'_n)|p_B\rangle.   
\label{W5}
\end{equation}
As we will demonstrate below, the relevant operators are  quark and gluon fields with Wilson-line type gauge links 
collinear to either $p_2$ for $A$ fields or  $p_1$ for $B$ fields.

\section{Power corrections and solution of classical YM equations \label{sect:power}}
\subsection{Power counting for background fields}
As we discussed in previous section, to get the hadronic tensor in the form  (\ref{W4}) we need to calculate 
the functional integral (\ref{funtc}) in the background of the fields (\ref{baukond}). Since we integrate over fields 
(\ref{baukond}) afterwards, we may assume that they satisfy Yang-Mills equations
\footnote{As we mentioned, for the purpose of calculation of integral over $C$ fields the projectile and target fields are
``frozen''.}
\begin{eqnarray}
&&\hspace{-1mm}
i\slashed{D}_A \psi_A~=~0,~~~D_A^\nu A_{\mu\nu}^a~=~g\sum_f\bar\psi^f_A\gamma_\mu t^a\psi^f_A,
\nonumber\\
&&\hspace{-1mm}
i\slashed{D}_B \psi_B~=~0,~~~D_B^\nu B_{\mu\nu}^a~=~g\sum_f\bar\psi^f_B\gamma_\mu t^a\psi^f_B,  
\label{YMs}
\end{eqnarray}
where $A_{\mu\nu}\equiv\partial_\mu A_\nu-\partial_\nu A_\mu-ig[A_\mu, A_\nu]$, 
$D_A^\mu\equiv (\partial^\mu-ig[A^\mu,)$ and similarly for $B$ fields. 

It is convenient to choose a gauge where $A_\ast=0$ for projectile fields and $B_\bu=0$ for target fields.
The rotation from
a general gauge  (Feynman gauge in our case, see footnote on p. 5) to this gauge is performed by the matrix $\Omega(x_\ast,x_\bu,x_\perp)$ satisfying boundary conditions
\begin{equation}
\hspace{-1mm}
\Omega(x_\ast,x_\bu,x_\perp)~\stackrel{x_\ast\rightarrow -\infty}{\rightarrow}~[x_\bu,-\infty_\bu]_x^{A_\ast},~~~~~
\Omega(x_\ast,x_\bu,x_\perp)~\stackrel{x_\bu\rightarrow -\infty}{\rightarrow}~[x_\ast,-\infty_\ast]_x^{B_\bu},
\label{Omega}
\end{equation}
where $A_\ast(x_\bu,x_\perp)$ and  $B_\bu(x_\ast,x_\perp)$ are projectile and target fields in an arbitrary gauge and $[x_\bullet, y_\bullet]^{A_\ast}_z$ denotes a gauge link constructed from $A$ fields ordered along a light-like line:
\begin{eqnarray}
[x_\bullet, y_\bullet]^{A_\ast}_z = Pe^{\frac{2ig}{s}\int^{x_\bullet}_{y_\bullet} dz_\bullet A_\ast (z_\bullet, z_\perp)}
\end{eqnarray}
and similarly for $[x_\ast, y_\ast]^{B_\bullet}_z$.

The existence of matrix $\Omega(x_\ast,x_\bu,x_\perp)$ was proved in appendix B of ref. \cite{Balitsky:2017flc} by explicit construction.
The relative 
strength of Lorentz components of projectile and target fields in this gauge was found in ref. \cite{Balitsky:2017flc}
\begin{eqnarray}
&&\hspace{-1mm}
\slashed{p}_1\psi_A(x_\bu,x_\perp)~\sim~m_\perp^{5/2}, ~~~\gamma_i\psi_A(x_\bu,x_\perp)~\sim~m_\perp^{3/2}, ~~~~~
\slashed{p}_2\psi_A(x_\bu,x_\perp)~\sim~s\sqrt{m_\perp},
\nonumber\\
&&\hspace{-1mm}
\slashed{p}_1\psi_B(x_\ast,x_\perp)~\sim~s\sqrt{m_\perp}, ~~~\gamma_i\psi_B(x_\ast,x_\perp)~\sim~m_\perp^{3/2}, ~~~~~
\slashed{p}_2\psi_B(x_\ast,x_\perp)~\sim~m_\perp^{5/2},
\nonumber\\
&&\hspace{-1mm}
A_\bu(x_\bu,x_\perp)~\sim~B_\ast(x_\ast,x_\perp)~\sim~m_\perp^2,~~~~~~A_i(x_\bu,x_\perp)~\sim~B_i(x_\ast,x_\perp)~\sim m_\perp.
\label{fildz}
\end{eqnarray}
Here $m_\perp$ is a scale of order of $m_N$ or $q_\perp$.
In general, we consider $W(p_A,p_B,q)$ in the region where $s,Q^2\gg q_\perp^2,m^2_N$, 
while the relation between $q_\perp^2$ and $m^2_N$ and between $Q^2$ and $s$ may be arbitrary. 
So, for the purpose of counting of powers of $s$, we will not distinguish between $s$ and $Q^2$ 
(although at the final step we will be able to tell the difference since our final expressions for 
power corrections will have either $s$ or $Q^2$ in denominators). Similarly, for the 
purpose of power counting we will not distinguish between $m_N$ and $q_\perp$ so we
introduce $m_\perp$ which may be of order of $m_N$ or $q_\perp$ depending on matrix element.

Note also that in our gauge
\begin{eqnarray}
&&\hspace{-5mm}
A_i(x_\bu,x_\perp)~=~{2\over s}\!\int_{-\infty}^{x_\bu}\! dx'_\bu ~A_{\ast i}(x'_\bu,x_\perp)
,~~~~
B_i(x_\ast,x_\perp)~=~{2\over s}\!\int_{-\infty}^{x_\ast}\! dx'_\ast ~B_{\bu i}(x'_\ast,x_\perp),
\label{AfromF}
\end{eqnarray}
where $A_{\ast i}\equiv F^{(A)}_{\ast i}$ and $B_{\bu i}\equiv F^{(B)}_{\bu i}$ are field strengths for $A$ and $B$ fields respectively.

Thus, to find TMD factorization formula with power corrections at the tree level we need to calculate the functional integral (\ref{dablfun}) in the background fields of the strength given by eqs. (\ref{fildz}).

\subsection{Approximate solution of classical equations \label{sect:ApprSol}}
\begin{figure}[htb]
\begin{center}
\includegraphics[width=131mm]{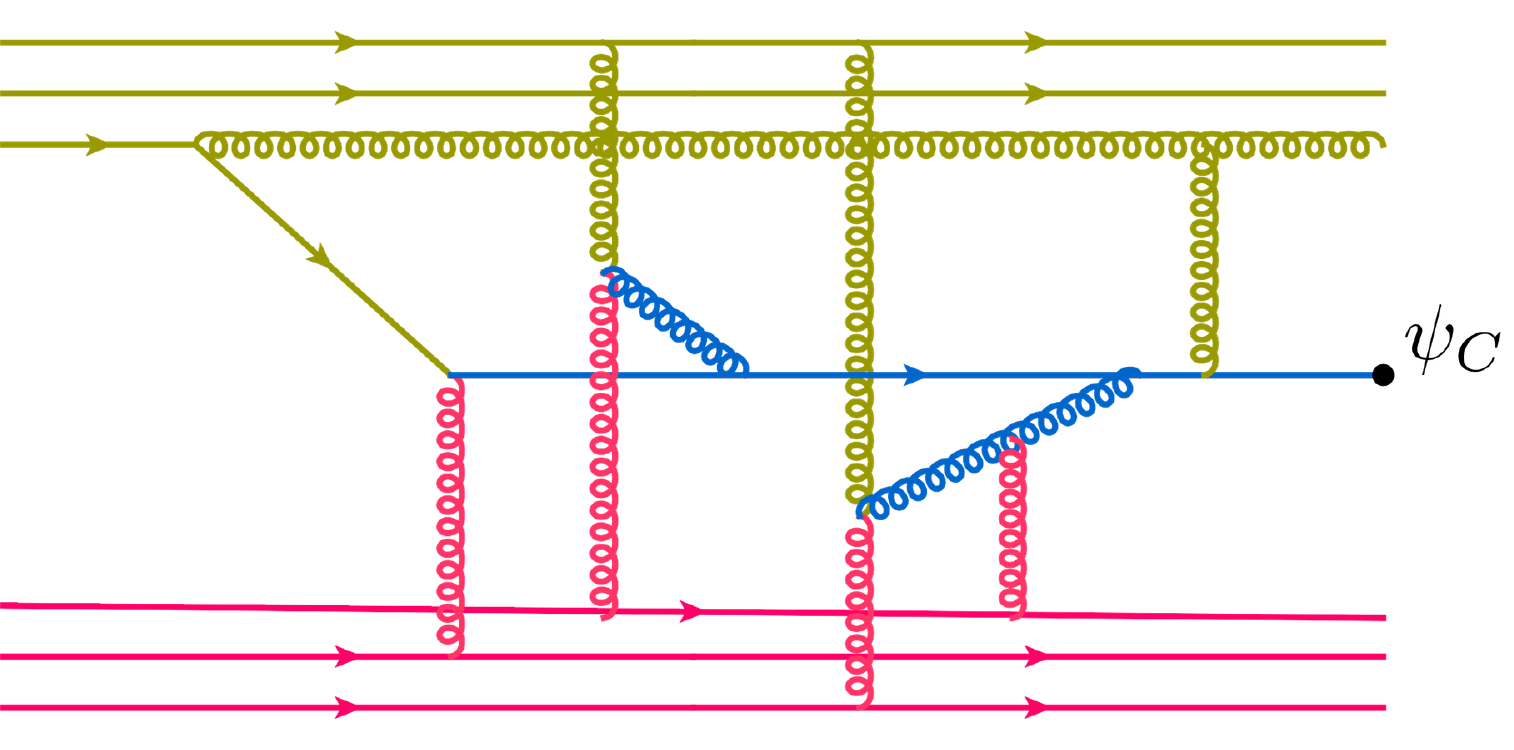}
\end{center}
\caption{Typical diagram for the classical field with projectile/target sources. The Green functions of the central fields are given by retarded propagators.   \label{fig:3}}
\end{figure}

As we discussed in section \ref{sec:funt}, the calculation of the functional integral (\ref{funtc}) over $C$-fields 
in the tree approximation reduces to finding fields $C_\mu$ and $\psi_C$ as solutions of Yang-Mills equations for the action
 $S_C~=~S_{\rm QCD}(C+A+B, \psi_C + \psi_A + \psi_B)-S_{\rm QCD}(A, \psi_A)-S_{\rm QCD}(B, \psi_B)$
\begin{eqnarray}
&&\hspace{-1mm}
(i\slashed{\partial}+g\slashed{A}+g\slashed{B}+g\slashed{C})(\psi^f_A+\psi^f_B+\psi^f_C)~=~0,
\label{yd}\\
&&\hspace{-1mm}
D^\nu F_{\mu\nu}^a(A+B+C)~=~g\sum_f(\bsi^f_A+\bsi^f_B+\bsi^f_C)\gamma_\mu t^a(\psi^f_A+\psi^f_B+\psi^f_C).
\nonumber
\end{eqnarray}
As we discussed above, the solution of eq. (\ref{yd}) which we need corresponds to the sum of set of diagrams
in background field $A + B$ with {\it retarded} Green functions, see figure \ref{fig:3}.
%%%%%%%%%%%%%FIGA%%%%%%%%%%%%%%%%%%%%%%%
%%%%%%%%%%%%%%%%%%%%%%%%%%%%%%%%%%%
 The retarded Green
functions (in the background-Feynman gauge) are defined as 
\begin{eqnarray}
&&\hspace{-1mm}
(x|{1\over \mathcal{P}^2g^{\mu\nu}+2ig\mathcal{F}^{\mu\nu}+i\epsilon p_0}|y)
~\equiv~(x|{1\over p^2+i\epsilon p_0}|y)
-g(x|{1\over p^2+i\epsilon p_0}\calo_{\mu\nu}{1\over p^2+i\epsilon p_0}|y)
\nonumber\\
&&\hspace{-1mm}
+~g^2(x|{1\over p^2+i\epsilon p_0}\calo_{\mu\xi}
{1\over p^2+i\epsilon p_0}\calo^\xi_{~\nu}{1\over p^2+i\epsilon p_0}|y)
+...~,
\end{eqnarray}
where
\begin{eqnarray}
&&\hspace{-1mm}
\mathcal{P}_\mu~\equiv~i\partial_\mu+gA_\mu+gB_\mu, ~~~~
\mathcal{F}_{\mu\nu}~=~\partial_\mu(A + B)_\nu-\mu\leftrightarrow\nu-ig[A_\mu+B_\mu, A_\nu+B_\nu],
\nonumber\\
&&\hspace{-1mm}
\calo_{\mu\nu}~\equiv~\big(\{p^\xi, A_\xi + B_\xi\}+g(A + B)^2\big)g_{\mu\nu}+2i\mathcal{F}_{\mu\nu}
\label{3.4}
\end{eqnarray}
and similarly for quarks.

Hereafter we use Schwinger's notations for propagators in external fields normalized according to
\begin{equation}
 (x|F(p)|y)\equiv \int\!\dhd^4 p~e^{-ip(x-y)}F(p),
\end{equation}
 where we use  space-saving notation $\dhd^np\equiv {d^np\over (2\pi)^n}$. Moreover, when it will not lead to a confusion,
 we will use short-hand notation
 \begin{eqnarray}
 {1\over \calo}\calo'(x)~\equiv~\int\! d^4z~(x|{1\over\calo}|z)\calo'(z).
\end{eqnarray}

The solution of eqs. (\ref{yd}) in terms of retarded Green functions gives fields $C_\mu$ and $\psi_C$ that vanish at $t\rightarrow -\infty$. Thus, we are solving the usual classical YM equations
\footnote{We take into account only $u,d,s,c$ quarks and consider them massless.}
\begin{equation}
\mathbb{D}^\nu \mathbb{F}^a_{\mu\nu}~=~\sum_fg\bar{\Psi}^f t^a\gamma_\mu \Psi^f,~~~~\slashed{\mathbb{P}}\Psi^f~=~0,
\label{kleqs}
\end{equation}
where
\begin{eqnarray}
&&\mathbb{A}_\mu = C_\mu + A_\mu + B_\mu,\ \ \ \Psi^f = \psi^f_C + \psi^f_A + \psi^f_B,
\nonumber\\
&&\mathbb{P}_\mu~\equiv~i\partial_\mu+gC_\mu+gA_\mu+gB_\mu, ~~~~
\mathbb{F}_{\mu\nu}~=~\partial_\mu \mathbb{A}_\nu-\mu\leftrightarrow\nu-ig[\mathbb{A}_\mu, \mathbb{A}_\nu],
\end{eqnarray}
with boundary conditions
\footnote{It is convenient to fix redundant gauge transformations by requirements $A_i(-\infty_\bu,x_\perp)=0$ for the projectile and $B_i(-\infty_\ast,x_\perp)=0$ for the target, 
see the discussion in ref. \cite{Belitsky:2002sm}.
}
\begin{eqnarray}
&&\hspace{-11mm}
\mathbb{A}_\mu(x)\stackrel{x_\ast\rightarrow -\infty}{=}A_\mu(x_\bu,x_\perp),~~~~
\Psi(x)\stackrel{x_\ast\rightarrow -\infty}{=}\psi_A(x_\bu,x_\perp),
\nonumber\\
&&\hspace{-11mm}
\mathbb{A}_\mu(x)\stackrel{x_\bu\rightarrow -\infty}{=}B_\mu(x_\ast,x_\perp),~~~~
\Psi(x)\stackrel{x_\bu\rightarrow -\infty}{=}\psi_B(x_\ast,x_\perp)
\label{inicondi}
\end{eqnarray}
following from $C_\mu,\psi_C\stackrel{t\rightarrow -\infty}{\rightarrow} 0$.
These boundary conditions reflect the fact that at $t\rightarrow -\infty$ we have only incoming hadrons with $A$ and $B$ fields.

As discussed in ref. \cite{Balitsky:2017flc}, for our case of particle production with ${q_\perp\over Q}\ll 1$ it is possible to find 
the approximate solution of  (\ref{kleqs}) as a series in this small parameter. 
We will solve eqs. (\ref{kleqs}) iteratively,  order by order in perturbation theory,  starting from the 
zero-order approximation in the form of the sum of projectile and target fields
\begin{eqnarray}
&&\hspace{-1mm}
\mathbb{A}_\mu^{[0]}(x)~=~A_\mu(x_\bu,x_\perp)+B_\mu(x_\ast,x_\perp),
\nonumber\\
&&\hspace{-1mm}
\Psi^{[0]}(x)~=~\psi_A(x_\bu,x_\perp)+\psi_B(x_\ast,x_\perp)
\label{trials}
\end{eqnarray}
and improving it by calculation of Feynman diagrams with retarded propagators in the background fields (\ref{trials}).

The first step is the calculation of the linear term for the trial configuration (\ref{trials}). 
The quark part of the linear term has the form
\begin{eqnarray}
&&\hspace{-1mm}
L_\psi~\equiv~\slashed{\mathcal{P}} \Psi^{[0]}~=~L_\psi^{(0)}+L_\psi^{(1)},~~~~
\nonumber\\
&&\hspace{-1mm}
L_\psi^{(0)}~=~g\gamma^iA_i\psi_B + g\gamma^iB_i\psi_A
,~~~~
L_\psi^{(1)}~=~{2g\over s}\slashed{p}_2 A_\bu\psi_B + {2g\over s}\slashed{p}_1 B_\ast\psi_A,
\label{kvlinterm}
\end{eqnarray}
where 
\begin{equation}
\mathcal{P}_\bu~=~i\partial_\bu+gA_\bu,~~~~\mathcal{P}_\ast~=~i\partial_\ast+gB_\ast,~~~
\mathcal{P}_i~=~i\partial_i+gA_i+gB_i
\label{3.12}
\end{equation}
are operators in external zero-order fields (\ref{trials}).
Here we denote the order of expansion in the parameter ${m_\perp^2\over s}$ by $(...)^{(n)}$,
and the order of perturbative expansion is labeled by $(...)^{[n]}$ as usual. 
The power-counting estimates for linear term in eq. (\ref{kvlinterm}) comes from eq. (\ref{fildz})
in the form
\begin{equation}
\hspace{-1mm}
L_\psi^{(0)}~\sim~m_\perp^{5/2},~~~~~~~L_\psi^{(1)}~\sim~{m^{9/2}_\perp\over s}.
\label{kvlintermvalues}
\end{equation}
The gluon linear term is
\begin{eqnarray}
&&\hspace{-1mm}
L^a_\mu~\equiv~\mathcal{D}^\xi \calf_{\xi\mu}^a+g\Bsi^{[0]} \gamma_\mu t^a\Psi^{[0]}~=~L^{(-1)a}_\mu+L^{(0)a}_\mu+L^{(1)a}_\mu,
\label{glinterm}\\
&&\hspace{-1mm}
L_\mu^{(-1)a}~~=~{2g\over s}p_{1\mu}f^{abc}A_{\ast j}^b B^{cj}+{2g\over s}p_{2\mu}f^{abc}B_{\bu j}^b A^{cj}~\sim~sm_\perp^2.
\nonumber
\end{eqnarray}
The explicit form of gluon linear terms $L^{(0)a}_\mu$ and $L^{(1)a}_\mu$  is presented in eq. (3.26) from our paper  \cite{Balitsky:2017flc}. 
For our purposes we need only the leading term $L_\mu^{(-1)a}$.

With the linear terms (\ref{kvlinterm}) and (\ref{glinterm}), a couple of first terms in our perturbative series are
\begin{eqnarray}
&&\hspace{-1mm}
\Psi^{[1]}(x)~=~-\!\int\! d^4z~(x|{1\over \slashed{\mathcal{P}}}|z)L_\psi(z),~~~~
\Psi^{[2]}(x)~=~-g\!\int\! d^4z~(x|{1\over \slashed{\mathcal{P}}}|z)\slashed{\mathbb{A}}^{[1]}(z)\Psi^{[0]}(z)
\nonumber\\
\label{kvarkfildz}
\end{eqnarray}
for quark fields and
\begin{eqnarray}
&&\hspace{-1mm}
\mathbb{A}_\mu^{[1]a}(x)~=~\int\!d^4z~(x|{1\over \calp^2 g^{\mu\nu}+2ig\calf^{\mu\nu}}|z)^{ab}L^{b\nu}(z),
\label{fields}\\
&&\hspace{-1mm}
\mathbb{A}_\mu^{[2]a}(x)~=~g\int\!d^4z~\Big[
-~i(x|{1\over \calp^2 g^{\mu\eta}+2ig\calf^{\mu\eta}}\calp^{\xi}|z)^{aa'}f^{a'bc}\mathbb{A}^{[1]b}_\xi \mathbb{A}^{[1]c\eta}
\nonumber\\
&&\hspace{15mm}
+~(x|{1\over \calp^2 g^{\mu\eta}+2ig\calf^{\mu\eta}}|z)^{aa'}f^{a'bc}\mathbb{A}^{[1]b\xi} 
(\cald_\xi \mathbb{A}^{[1]c\eta}-\cald^\eta \mathbb{A}^{[1]c}_\xi)\Big]
\nonumber
\end{eqnarray}
for gluon fields (in the background-Feynman gauge). Next iterations, like $\Psi^{[3]}(x)$ and $\mathbb{A}_\mu^{[3]a}(x)$, will give us a set of tree-level Feynman diagrams 
in the background fields $A_\mu+B_\mu$ and $\psi_A + \psi_B$.

Let us consider the fields in the first order in perturbative expansion:
\begin{eqnarray}
&&\hspace{-1mm}
\Psi^{[1]}~=~-{1\over \slashed{\calp}}L_\psi~=~-{1
\over  \alpha \slashed{p}_1+\beta \slashed{p}_2+{2g\over s}(B_\ast \slashed{p}_1+A_\bu \slashed{p}_2)+ \slashed{\calp}_\perp+i\epsilon p_0}L_\psi,
\nonumber\\
&&\hspace{-1mm}
\mathbb{A}_\mu^{[1]}~=~{1\over \calp^2 g^{\mu\nu}+2ig\calf^{\mu\nu}}L^\nu
\label{fields1}\\
&&\hspace{13mm}
=~
{1\over [\{\alpha+{2g\over s}B_\ast,\beta+{2g\over s}A_\bu\}{s\over 2}
-\calp_\perp^2]g^{\mu\nu}+2ig\calf^{\mu\nu}+i\epsilon p_0}L^\nu.
\nonumber
\end{eqnarray}
Here $\alpha$, $\beta$, and $p_\perp$ are understood as differential operators
$\alpha=i{\partial\over\partial x_\bu}$, $\beta=i{\partial\over\partial x_\ast}$  and $p_i=i{\partial\over\partial x^i}$.

Now comes the central point of our approach. Let us expand quark and gluon propagators in powers of background fields, then 
we get a set of diagrams shown in figure \ref{fig:3}.
 The typical bare gluon propagator in figure \ref{fig:3} is
\begin{equation}
{1\over p^2+i\epsilon p_0}~=~{1\over\alpha\beta s-p_\perp^2+i\epsilon(\alpha+\beta)}.
\label{gluonpropagator}
\end{equation}
 Since we do not consider loops of $C$-fields in this paper, the  transverse momenta 
in tree diagrams  are determined by further integration over projectile (``A'') and target (``B'') fields 
in eq. (\ref{W3}) which converge on  either $q_\perp$ or $m_N$. On the other hand, the integrals over 
 $\alpha$ converge on either $\alpha_q$ or $\alpha\sim 1$ and similarly the characteristic $\beta$'s
 are either $\beta_q$ or $\beta \sim 1$.
Since $\alpha_q\beta_qs=Q_\parallel^2\gg q_\perp^2$, one can expand gluon and quark propagators 
in powers of ${p_\perp^2\over \alpha\beta s}$ 
\begin{eqnarray}
&&
{1\over p^2+i\epsilon p_0}~=~{1\over s(\alpha+i\epsilon)(\beta+i\epsilon)}
\Big(1
+{p_\perp^2/s\over(\alpha+i\epsilon)(\beta+i\epsilon)}+...\Big),
\label{propexpan}\\
&&
{\slashed{p}\over p^2+i\epsilon p_0}~=~{1\over s}\Big({\slashed{p}_1\over \beta+i\epsilon}
+{\slashed{p}_2\over \alpha+i\epsilon}+{\slashed{p}_\perp\over (\alpha+i\epsilon)(\beta+i\epsilon)}\Big)
\Big(1
+{p_\perp^2/s\over(\alpha+i\epsilon)(\beta+i\epsilon)}+...\Big).
\nonumber
\end{eqnarray}
The explicit form of operators ${1\over \alpha+i\epsilon}$, ${1\over \beta+i\epsilon}$, and 
${1\over (\alpha+i\epsilon)(\beta+i\epsilon)}$ is
\begin{eqnarray}
(x|{1\over \alpha+i\epsilon}|y)
&=&~{s\over 2}\!\int\! \dhd^2p_\perp\!\int\! {\dhd\alpha\over\alpha+i\epsilon}\dhd\beta
~e^{-i\alpha(x-y)_\bu-i\beta(x-y)_\ast+i(p,x-y)_\perp}
\nonumber\\
&&=~-i{s\over 2}(2\pi)^2\delta^{(2)}(x_\perp-y_\perp)
\theta(x_\bu-y_\bu)\delta(x_\ast-y_\ast),
\nonumber\\
(x|{1\over \beta+i\epsilon}|y)
&=&~{s\over 2}\!\int\! \dhd^2p_\perp\!\int\! \dhd\alpha{\dhd\beta\over\beta+i\epsilon}
~e^{-i\alpha(x-y)_\bu-i\beta(x-y)_\ast+i(p,x-y)_\perp}
\nonumber\\
&&=~-i{s\over 2}(2\pi)^2\delta^{(2)}(x_\perp-y_\perp)
\theta(x_\ast-y_\ast)\delta(x_\bu-y_\bu),
\nonumber\\
(x|{1\over (\alpha+i\epsilon)(\beta+i\epsilon)}|y)
&=&~{s\over 2}\!\int\! \dhd^2p_\perp\!\int\! {\dhd\alpha\over\alpha+i\epsilon}{\dhd\beta\over\beta+i\epsilon} 
~e^{-i\alpha(x-y)_\bu-i\beta(x-y)_\ast+i(p,x-y)_\perp}
\nonumber\\
&&=~-{s\over 2}(2\pi)^2\delta^{(2)}(x_\perp-y_\perp)
\theta(x_\ast-y_\ast)\theta(x_\bu-y_\bu).
\label{propexplicit}
\end{eqnarray}
After the expansion (\ref{propexpan}), the dynamics in the transverse space effectively becomes trivial: 
all background fields stand either at $x$ or at $0$. 
The formula (\ref{fields1}) turns into expansion
\begin{eqnarray}
&&\hspace{-2mm}
\Psi_f^{[1]}~=~-\big({\slashed{p}_1\over\beta s}+{\slashed{p}_2\over\alpha s}\big)L_\psi
+{2g\over s^2}\big(B_\ast{\slashed{p}_2\over \alpha^2}+A_\bu{\slashed{p}_1\over\beta^2}\big)L_\psi
+{1\over s^2}\big({\slashed{p}_1\over\beta}\slashed{\calp}_\perp{\slashed{p}_2\over\alpha}
+{\slashed{p}_2\over\alpha}\slashed{\calp}_\perp{\slashed{p}_1\over\beta}\big)L_\psi+\dots,
\nonumber\\
&&\hspace{-2mm}
\mathbb{A}_\mu^{[1]}~=~{1\over \alpha\beta s}L_\mu
+{1\over \alpha\beta s}\big(\big[\calp_\perp^2 -g\{\alpha,A_\bu\}-g\{\beta,B_\ast\}\big]g_{\mu\nu}-2ig\calf_{\mu\nu}
\big){1\over \alpha\beta s}L^\nu~+~\dots,
\label{fields1a}
\end{eqnarray}
where ${1\over\alpha}$ and ${1\over\beta}$ are understood as  ${1\over\alpha+i\epsilon}$ and ${1\over\beta+i\epsilon}$ respectively.

One may question why we do not cut the integrals in eq. (\ref{propexplicit}) to $|\alpha|>\sigma_b$ and $|\beta|>\sigma_a$
according to the definition of $C$ fields in section \ref{sec:funt}. 
\footnote{
Such cutoffs for integrals over $C$ fields are introduced explicitly 
in the framework of soft-collinear effective theory (SCET), see review \cite{Rothstein:2016bsq}.}
The reason is that in the diagrams like figure \ref{fig:3}
with retarded propagators (\ref{propexplicit}) one can 
shift the contour of integration over $\alpha$ and/or $\beta$  to the complex plane away to avoid the region of 
small $\alpha$ or $\beta$.  It should be mentioned, however, that such shift may not be possible if there is pinching of 
poles in the integrals over $\alpha$ or $\beta$. For example, if after the expansion (\ref{propexpan}) we encounter 
${1\over (\alpha+i\epsilon)(\alpha-i\epsilon)}$, the expansion was not justified since actual $\alpha$'s in the integral are $\sim {p_\perp^2\over s}$ 
and hence the field was misidentified: we have a propagator of $B$-field rather than of $C$-field. Fortunately, at the tree level all
propagators are retarded and  the pinching of poles never occurs. In the higher orders in perturbation theory Feynman propagators in the loops 
cannot be replaced by retarded propagators so after the expansion (\ref{propexpan}) we can get ${1\over (\alpha+i\epsilon\beta)(\alpha+i\epsilon\beta')}$.
In such case the pinching may occur so one needs to formulate a subtraction program to get rid of pinched poles and avoid double counting of the 
fields.

Note that the background fields are also smaller than typical $p_\parallel^2\sim s$. Indeed, from eq. (\ref{fildz}) 
we see that $p_\bu={s\over 2}\beta\gg A_\bu\sim m^2_\perp$ ( because $\alpha\geq\alpha_q\gg {m^2_\perp \over s}$) and similarly
$p_\ast\gg B_\ast$. Also $(p_i+A_i+B_i)^2~\sim~q_\perp^2\ll p_\parallel^2$.  
\footnote{The only exception is the fields
$B_{\bu i}$ or $A_{\ast i}$ which are of order of $sm_\perp$ but we saw in ref.  \cite{Balitsky:2017flc} that effectively the expansion in powers of
these fields is cut at the second term.}

\subsection{Power expansion of classical quark fields}

Now we expand the classical  quark fields  in powers of  ${p_\perp^2\over p_\parallel^2}\sim{m_\perp^2\over s}$ 
(the expansion of classical gluon fields is presented in eqs. (3.35)-(3.38) in ref. \cite{Balitsky:2017flc}). 
From the previous section it is clear that  the leading power correction comes only from 
the first term displayed in eq. (\ref{kvarkfildz}).
Expanding it in powers of $p^2_\perp/p_\parallel^2$ as explained in the previous section, we obtain
\begin{eqnarray}
&&\hspace{-1mm}
\Psi(x)~=~\Psi^{[0]}(x)+\Psi^{[1]}(x)+\Psi^{[2]}(x)+\dots~=~\Psi_A^{(0)}+\Psi_B^{(0)}
+\Psi_A^{(1)}+\Psi_B^{(1)}+\dots,
\end{eqnarray}
where
\begin{eqnarray}
&&\hspace{-5mm}
\Psi_A^{(0)}~=~\psi_A+\Xi_{2A},~~~~
\Xi_{2A}~=~-{g\slashed{p}_2\over s}\gamma^iB_i{1\over \alpha+i\epsilon}\psi_A,
\nonumber\\
&&\hspace{-5mm}
\Bsi_A^{(0)}~=~\bar\psi_A+\Bxi_{2A},~~~~
\Bxi_{2A}~=~-\big(\bar\psi_A{1\over\alpha-i\epsilon}\big)\gamma^iB_i{g\slashed{p}_2\over s},
\nonumber\\
&&\hspace{-5mm}
\Psi_B^{(0)}~=~\psi_B+\Xi_{1B},~~~~
\Xi_{1B}~=~-{g\slashed{p}_1\over s}\gamma^iA_i{1\over\beta+i\epsilon}\psi_B,
\nonumber\\
&&\hspace{-5mm}
\Bsi_B^{(0)}~=~\bar\psi_B+\Bxi_{1B},~~~~
\Bxi_{1B}~=~-\big(\bar\psi_B{1\over \beta-i\epsilon}\big)\gamma^iA_i{g\slashed{p}_1\over s}.
\label{fildz0}
\end{eqnarray}
In this formula
\begin{eqnarray}
&&\hspace{-1mm}
{1\over \alpha+i\epsilon}\psi_A(x_\bu,x_\perp)~\equiv~-i\!\int_{-\infty}^{x_\bu}\! dx'_\bu~\psi_A(x'_\bu,x_\perp),
\nonumber\\
&&\hspace{-1mm}
\Big(\bsi_A{1\over \alpha-i\epsilon}\Big)(x_\bu,x_\perp)~\equiv~i\!\int_{-\infty}^{x_\bu}\! dx'_\bu~\bsi_A(x'_\bu,x_\perp)
\label{3.25}
\end{eqnarray}
and similarly for ${1\over\beta\pm i\epsilon}$. 
From now on we will denote $\big(\bar\psi_A{1\over\alpha}\big)(x)\equiv \big(\bsi_A{1\over \alpha-i\epsilon}\big)(x)$ and
$\big(\bar\psi_B{1\over\beta}\big)(x)\equiv \big(\bsi_B{1\over \beta-i\epsilon}\big)(x)$  while in all other places 
$\big({1\over\alpha}\calo\big)\equiv\big({1\over\alpha+i\epsilon}\calo\big)$
and  $\big({1\over\beta}\calo\big)\equiv\big({1\over\beta+i\epsilon}\calo\big)$.

It is easy to see that  power counting of these quark fields has the form
\begin{eqnarray}
&&\Psi_A^{(0)}\sim\Psi_B^{(0)}~\sim~m_\perp^{3/2}.
\label{psi0}
\end{eqnarray}
As to quark fields $\Psi^{(1)}$, we  present their explicit form in  appendix \ref{sec:nloquarks} and prove in appendix \ref{sec:sublead} that their contribution is
small in the kinematic region $s\gg Q^2$. 

\section{Leading power corrections at  $s\gg Q^2\gg q_\perp^2$\label{sec:lhtc}}

As we mentioned in the introduction, our method is relevant to calculation of power corrections
at any $s,Q^2\gg q_\perp^2,m^2_N$. However, the expressions are greatly simplified 
in the physically interesting case $s\gg Q^2\gg q_\perp^2$ which we consider in this paper.
\footnote
{We also assume that Z-boson is emitted in the central region of rapidity  so
$\alpha_qs\sim\beta_qs\gg Q^2$. \label{foot}}
 
 As we noted above, we take into account only $u,d,s,c$ quarks and consider them massless. 
The hadronic tensor takes the form
\begin{eqnarray}
&&\hspace{-1mm}
W(p_A, p_B, q)~=~\int\!d^2x_\perp ~e^{i(q,x)_\perp} W(\alpha_q,\beta_q, x_\perp),
\label{defW}\\
&&\hspace{-1mm}
W(\alpha_q,\beta_q,x_\perp)~\equiv~
{1\over(2\pi)^4}{2\over s}\!\int\! dx_\bu dx_\ast ~e^{-i\alpha_qx_\bu-i\beta_q x_\ast}\langle p_A,p_B|J_\mu(x_\bu,x_\ast,x_\perp)J^\mu(0)|p_A,p_B\rangle,
\nonumber
\end{eqnarray}
where ($c_W\equiv\cos\theta_W$, $s_W\equiv\sin\theta_W$)
\footnote{We denote the weak coupling constant by $e/s_W$ and reserve the notation ``$g$'' for QCD coupling constant.}
\begin{eqnarray}
&&\hspace{-1mm}
J_\mu~=~{e\over 4s_Wc_W}\big[-\baru\gamma_\mu(1-{8\over 3}s_W^2-\gamma_5)u-\barc\gamma_\mu(1-{8\over 3}s_W^2-\gamma_5)c
\nonumber\\
&&\hspace{11mm}
+~\bard\gamma_\mu(1-{4\over 3}s_W^2-\gamma_5)d+\bars\gamma_\mu(1-{4\over 3}s_W^2-\gamma_5)s\big].
\end{eqnarray}
After integration over central fields in the tree approximation we obtain
\begin{equation}
\hspace{-1mm}
W(\alpha_q,\beta_q,x_\perp)~\equiv~
{2\over (2\pi)^4s}\!\int\! dx_\bu dx_\ast ~e^{-i\alpha_qx_\bu-i\beta_q x_\ast}
\langle p_A|\langle p_B|\calj_\mu(x_\bu,x_\ast,x_\perp)\calj^\mu(0)|p_A\rangle|p_B\rangle,
\label{4.3}
\end{equation}
where
\begin{eqnarray}
&&\hspace{-1mm}
\calj^\mu~=~\calj^\mu_A+\calj^\mu_B+\calj^\mu_{AB}+\calj^\mu_{BA},
\nonumber\\
&&\hspace{-1mm}
\calj^\mu_A~=~{e\over 4s_Wc_W}\big[-\bar\Psi_{Au}\hamma^\mu\Psi_{Au}-\bar\Psi_{Ac}\hamma^\mu\Psi_{Ac}
+\bar\Psi_{Ad}\hamma^\mu\Psi_{Ad}+\bar\Psi_{As}\hamma^\mu\Psi_{As}\big],
\nonumber\\
&&\hspace{-1mm}
\calj^\mu_{AB}~=~{e\over 4s_Wc_W}\big[-\bar\Psi_{Au}\hamma^\mu\Psi_{Bu}-\bar\Psi_{Ac}\hamma^\mu\Psi_{Bc}
+\bar\Psi_{Ad}\hamma^\mu\Psi_{Bd}+\bar\Psi_{As}\hamma^\mu\Psi_{Bs}\big],
\label{kaljs}
\end{eqnarray}
and similarly for $\calj^\mu_B$ and $\calj^\mu_{BA}$. 
Hereafter we use notation $\hamma_\mu\equiv\gamma_\mu(a-\gamma_5)$ where $a$ is one of $a_{u,c}=(1-{8\over 3}s_W^2)$ or $a_{d,s}=(1-{4\over 3}s_W^2)$ depending on quark's flavor.

The quark fields are given by a series in the parameter
${m_\perp^2\over s}$, see eqs. (\ref{fildz0}) and (\ref{fildz1}), where $\Psi$ can be any of $u,d,s$ or $c$ quarks.
\footnote{
As we mentioned, we will need only first two terms of the expansion given by eqs. (\ref{fildz0})
and (\ref{fildz1}).} Accordingly, the currents (\ref{kaljs}) can be expressed as a series in this parameter, e.g.
\begin{eqnarray}
&&\hspace{-1mm}
\calj^{(0)\mu}_{AB}~=~{e\over 4s_Wc_W}\big[-\bar\Psi^{(0)}_{Au}\hamma^\mu\Psi^{(0)}_{Bu}-\bar\Psi^{(0)}_{Ac}\hamma^\mu\Psi^{(0)}_{Bc}
+\bar\Psi^{(0)}_{Ad}\hamma^\mu\Psi^{(0)}_{Bd}+\bar\Psi^{(0)}_{As}\hamma^\mu\Psi^{(0)}_{Bs}\big],
\nonumber\\
&&\hspace{-1mm}
\calj^{(1)\mu}_{AB}~=~{e\over 4s_Wc_W}\big[-\bar\Psi^{(1)}_{Au}\hamma^\mu\Psi^{(0)}_{Bu}-\bar\Psi^{c(0)}_{Au}\hamma^\mu\Psi^{c(1)}_{Bu}
-\bar\Psi^{(1)}_{Ac}\hamma^\mu\Psi^{(0)}_{Bc}-\bar\Psi^{(0)}_{Ac}\hamma^\mu\Psi^{(1)}_{Bc}
\nonumber\\
&&\hspace{10mm}
+~\bar\Psi^{(1)}_{Ad}\hamma^\mu\Psi^{d(0)}_{Bd}+~\bar\Psi^{(0)}_{Ad}\hamma^\mu\Psi^{(1)}_{Bd}
+\bar\Psi^{(1)}_{As}\hamma^\mu\Psi^{(0)}_{Bs}+\bar\Psi^{(0)}_{As}\hamma^\mu\Psi^{(1)}_{Bs}\big].
\label{kaljeis}
\end{eqnarray}

The leading power contribution comes only from product $\calj_{AB}^\mu(x)\calj_{BA\mu}(0)$ (or $\calj_{BA}^\mu(x)\calj_{AB\mu}(0)$), 
while power corrections may come from other terms like $\calj_A^\mu(x)\calj_{B\mu}(0)$. We will consider all terms in turn.

\subsection{Leading contribution and power corrections from $\calj_{AB}^\mu(x)\calj_{BA\mu}(0)$ terms}
Power expansion of $\calj_{AB}^\mu(x)\calj_{BA\mu}(0)$ reads
\begin{eqnarray}
&&\hspace{-1mm}
\Bsi_A(x)\hamma^\mu\Psi_B(x)\Bsi_B(0)\hamma_\mu\Psi_A(0)
~=~\Bsi_A^{(0)}(x)\hamma^\mu\Psi_B^{(0)}(x)\Bsi_B^{(0)}(0)\hamma_\mu\Psi_A^{(0)}(0)
\label{4.6}\\
&&\hspace{-1mm}
+~\Bsi_A^{(1)}(x)\hamma^\mu\Psi_B^{(0)}(x)\Bsi_B^{(0)}(0)\hamma_\mu\Psi_A^{(0)}(0)
+\Bsi_A^{(0)}(x)\hamma^\mu\Psi_B^{(1)}(x)\Bsi_B^{(0)}(0)\hamma_\mu\Psi_A^{(0)}(0)
\nonumber\\
&&\hspace{-1mm}
+~\Bsi_A^{(0)}(x)\hamma^\mu\Psi_B^{(0)}(x)\Bsi_B^{(1)}(0)\hamma_\mu\Psi_A^{(0)}(0)
+\Bsi_A^{(0)}(x)\hamma^\mu\Psi_B^{(0)}(x)\Bsi_B^{(0)}(0)\hamma_\mu\Psi_A^{(1)}(0)~+~...
\nonumber
\end{eqnarray}
In appendix \ref{psi1sup} we demonstrate that terms $\sim \Psi^{(1)}$  lead to power corrections $\sim {q_\perp^2\over \alpha_qs}$ or
$\sim {q_\perp^2\over \beta_qs}$ which are much smaller than ${q_\perp^2\over \alpha_q\beta_qs}={q_\perp^2\over Q^2_\|}\sim{q_\perp^2\over Q^2}$ if Z-boson 
is emitted in the central region of rapidity. Note that since we want to calculate the leading power corrections, hereafter we substitute $Q^2_\|$ with $Q^2$. In the limit $s\gg Q^2\gg q_\perp^2$ this change of variables can only lead to errors of the order of subleading power terms.

As to terms $\sim \Bsi_A^{(0)}(x)\gamma_\mu\Psi_B^{(0)}(x)\Bsi_B^{(0)}(0)\gamma^\mu\Psi_A^{(0)}(0)$, they can be decomposed
using eq.  (\ref{fildz0}) as follows:
\begin{eqnarray}
&&\hspace{-1mm}
\big[\big(\bar\psi_A +\Bxi_{2A}\big)(x)\hamma_\mu\big(\psi_B+\Xi_{1B}\big)(x)\big]
[\big(\bar\psi_B+\Bxi_{1B}\big)(0)\hamma^\mu\big(\psi_A+\Xi_{2A}\big)(0)\big]~+~x\leftrightarrow 0
\nonumber\\
&&\hspace{-1mm}
=~[\bar\psi_A(x)\hamma_\mu\psi_B(x)\big]\big[\bar\psi_B(0)\hamma^\mu\psi_A(0)\big]
\label{7lines}\\
&&\hspace{-1mm}
+~[\Bxi_{2A}(x)\hamma_\mu\psi_B(x)\big]\big[\bar\psi_B(0)\hamma^\mu\psi_A(0)\big]
+[\bar\psi_A(x)\hamma_\mu\Xi_{1B}(x)\big]\big[\bar\psi_B(0)\hamma^\mu\psi_A(0)\big]
\nonumber\\
&&\hspace{-1mm}
+~
[\bar\psi_A(x)\hamma_\mu\psi_B(x)\big]\big[\Bxi_{1B}(0)\hamma^\mu\psi_A(0)\big]
+[\bar\psi_A(x)\hamma_\mu\psi_B(x)\big]\big[\bar\psi_B(0)\hamma^\mu\Xi_{2A}(0)\big]
\nonumber\\
&&\hspace{-1mm}
+~[\Bxi_{2A}(x)\hamma_\mu\psi_B(x)\big]\big[\bar\psi_B(0)\hamma^\mu\Xi_{2A}(0)\big]
+[\bar\psi_A(x)\hamma_\mu\Xi_{1B}(x)\big]\big[\Bxi_{1B}(0)\hamma^\mu\psi_A(0)\big]
\nonumber\\
&&\hspace{-1mm}
+~[\Bxi_{2A}(x)\hamma_\mu\psi_B(x)\big]\big[\Bxi_{1B}(0)\hamma^\mu\psi_A(0)\big]
+[\bar\psi_A(x)\hamma_\mu\Xi_{1B}(x)\big]\big[\bar\psi_B(0)\hamma^\mu\Xi_{2A}(0)\big]
\nonumber\\
&&\hspace{-1mm}
+~[\Bxi_{2A}(x)\hamma_\mu\Xi_{1B}(x)\big]\big[\bar\psi_B(0)\hamma^\mu\psi_A(0)\big]
+[\bar\psi_A(x)\hamma_\mu\psi_{B}(x)\big]\big[\Bxi_{1B}(0)\hamma^\mu\Xi_{2A}(0)\big]
~+~x\leftrightarrow 0.
\nonumber
\end{eqnarray}
First, let us consider the leading power term coming from the first term in the r.h.s. of this equation.

\subsection{Leading power contribution}
As we mentioned, the leading-power term comes from $\calj_{AB}^{(0)\mu}(x)\calj^{(0)}_{BA\mu}(0)$ and \\
$\calj_{BA}^{(0)\mu}(x)\calj^{(0)}_{AB\mu}(0)$. Using Fierz transformation 
\begin{eqnarray}
&&\hspace{-1mm}
(\bsi_A\gamma^\mu[a-\gamma_5]\chi_A)(\bhi_B\gamma_\mu[a-\gamma_5]\psi_B)
\label{fierz}\\
&&\hspace{11mm}
=~
{1+a^2\over 2}[(\bsi_A\gamma^\alpha\psi_B)(\bhi_B\gamma_\alpha\chi_A)
+(\bsi_A\gamma^\alpha\gamma_5\psi_B)(\bhi_B\gamma_\alpha\gamma_5\chi_A)]
\nonumber\\
&&\hspace{21mm}
-~a[(\bsi_A\gamma^\alpha\psi_B)(\bhi_B\gamma_\alpha\gamma_5\chi_A)
+(\bsi_A\gamma^\alpha\gamma_5\psi_B)(\bhi_B\gamma_\alpha\chi_A)]
\nonumber\\
&&\hspace{31mm}
+~(1-a^2)[(\bsi_A\psi_B)(\bhi_B\chi_A)-(\bsi_A\gamma_5\psi_B)(\bhi_B\gamma_5\chi_A)]
\nonumber
\end{eqnarray}
with $a_{u,c}=(1-{8\over 3}s_W^2)$ and  $a_{d,s}=(1-{4\over 3}s_W^2)$ one obtains
\begin{eqnarray}
&&\hspace{-1mm}
N_c{16s_W^2c_W^2\over e^2}\langle p_A|\langle p_B|\calj_{AB\mu}^{(0)}(x)\calj^{(0)\mu}_{BA}(0)+(x\leftrightarrow 0)|p_A\rangle|p_B\rangle~=~
\nonumber\\
&&\hspace{-1mm}
=~\Big[\Big\{{1+a_u^2\over 2}
\big(\langle \bsi_{Au}(x)\gamma_\mu \psi_{Au}(0)\rangle\langle \bsi_{Bu}(0)\gamma^\mu \psi_{Bu}(x)\rangle+
\gamma_\mu\otimes\gamma^\mu\leftrightarrow \gamma_\mu\gamma_5\otimes\gamma^\mu\gamma_5\big)
\nonumber\\
&&\hspace{-1mm}
-~a_u\big(\langle \bsi_{Au}(x)\gamma_\mu \psi_{Au}(0)\rangle\langle \bsi_{Bu}(0)\gamma^\mu\gamma_5 \psi_{Bu}(x)\rangle+
\gamma_\mu\otimes\gamma^\mu\gamma_5\leftrightarrow \gamma_\mu\gamma_5\otimes\gamma^\mu\big)
\nonumber\\
&&\hspace{-1mm}
+~(1-a_u^2)\big(\langle \bsi_{Au}(x)\psi_{Au}(0)\rangle\langle \bsi_{Bu}(0)\psi_{Bu}(x)\rangle
-\langle \bsi_{Au}(x)\gamma_5 \psi_{Au}(0)\rangle\langle \bsi_{Bu}(0)\gamma_5 \psi_{Bu}(x)\rangle\big)\Big\}
\nonumber\\
&&\hspace{-1mm}
+~\Big\{u\leftrightarrow c\Big\}+\Big\{u\leftrightarrow d\Big\}+\Big\{u\leftrightarrow s\Big\}\Big]~+~\big[x\leftrightarrow 0\big],
\end{eqnarray}
where
\begin{eqnarray}
&&\hspace{-2mm}
\langle \bsi_{Au}(x)\gamma_\mu \psi_{Au}(0)\rangle~\equiv~\langle A|\hat{\bar\psi}_u(x)\gamma_\mu \hat\psi_u(0)|A\rangle,~~~
\langle\bsi_{Bu}(x)\gamma_\mu \psi_{Bu}(0)\rangle~\equiv~\langle B|\hat{\bar\psi}_u(x)\gamma_\mu \hat\psi_u(0)|B\rangle
\nonumber\\
\label{fla4.7}
\end{eqnarray}
and similarly for other matrix elements (summation over color and Lorentz indices is implied).

As usual, after integration over background fields $A$ and $B$ we promote $A$, $\psi_A$ and $B$, $\psi_B$ to operators $\hatA$, $\hat\psi$.
A subtle point is that our operators are not under T-product ordering so one should be careful while changing the order of operators in
 formulas like Fierz transformation. Fortunately, all our operators are separated either by space-like intervals or light-like intervals so they commute
 with each other.

In a general gauge for
projectile and target fields these expressions read (see eq. (\ref{Omega}))
\begin{eqnarray}
&&\hspace{-2mm}
\langle A|\hat{\bar\psi}_{f}(x)\gamma_\mu \hsi_{f}(0)|A\rangle
~=~\langle A|\hat{\bar\psi}_f(x_\bu,x_\perp)\gamma_\mu[x_\bu,-\infty_\bu]_x[x_\perp,0_\perp]_{-\infty_\bu}[-\infty_\bu,0_\bu]_0\hsi_f(0)|A\rangle,
\nonumber\\
&&\hspace{-2mm}
\langle B|\hat{\bar\psi}_{f}(x)\gamma_\mu \hsi_{f}(0)|B\rangle
~=~\langle B|\hat{\bar\psi}_f(x_\ast,x_\perp)\gamma_\mu[x_\ast,-\infty_\ast]_x[x_\perp,0_\perp]_{-\infty_\ast}[-\infty_\ast,0_\ast]_0\hsi_f(0)|B\rangle
\nonumber\\
\label{gaugelinks}
\end{eqnarray}
and similarly for $\langle A|\hat{\bar\psi}_{f}(0)\gamma_\mu \hsi_{f}(x)|A\rangle$ and $\langle B|\hat{\bar\psi}_{f}(0)\gamma_\mu \hsi_{f}(x)|B\rangle$.

From parametrization of two-quark operators in section \ref{paramlt},
it is clear that the leading power contribution to $W(q)$ of eq. (\ref{defW}) comes from the product of two $f_1's$ in eq. (\ref{Amael}) and (\ref{Bmael}).
It has the form \cite{Tangerman:1994eh}
\begin{eqnarray}
&&\hspace{-1mm}W^{\rm lt}(\alpha_q,\beta_q,q_\perp)
\nonumber\\
&&\hspace{-1mm}
=~-{e^2\over 8 s_W^2c_W^2N_c}\!\int\! d^2k_\perp\Big(\big\{(1+a_u^2)
\big[f_1^u(\alpha_q,k_\perp)\barf_1^u(\beta_q,q_\perp-k_\perp) 
\nonumber\\
&&\hspace{-1mm}
+\barf_1^u(\alpha_q,k_\perp)f_1^u(\beta_q,q_\perp-k_\perp)\big]
\big\}
+\{u\leftrightarrow c\}+\{u\leftrightarrow d\}+\{u\leftrightarrow s\}\Big).
\label{WLT}
\end{eqnarray}
All other terms in the product of eqs. (\ref{Amael}) and (\ref{Bmael}) give higher power contributions  
$\sim {q_\perp^2\over s}W^{\rm lt}(q)$ 
(but \underline{not} $\sim{q_\perp^2\over Q^2}W^{\rm lt}(q)$)\footnote{The trivial but important point is that any $f(x,k_\perp)$ may have only logarithmic dependence on Bjorken $x$ but
not the power dependence $\sim{1\over x}$. Indeed, at small $x$ the cutoff of corresponding longitudinal integrals comes from 
the rapidity cutoff $\sigma_a$, see the discussion in section \ref{sec:funt}. Thus, at small $x$ one can safely put $x=0$ and the corresponding 
logarithmic contributions would be proportional to powers of $\alpha_s\ln\sigma_a$ (or, in some cases, $\alpha_s\ln^2\sigma_a$, see e.g. 
ref. \cite{Kovchegov:2015zha}).  Also, a more technical version of this argument was presented on page 
12.}
 so they can be neglected at $Q^2\ll s$. Similarly, the contribution of two matrix elements in eq. (\ref{mael5}) 
 is  $\sim{m_\perp^2\over s}$ in comparison to $W^{\rm lt}(q)$
so it can be neglected as well.
 
 \subsubsection{Parametrization of leading matrix elements \label{paramlt}}
 Let us  first consider matrix elements of operators without $\gamma_5$. The standard parametrization of quark TMDs reads
\begin{eqnarray}
&&\hspace{-1mm}
{1\over 16\pi^3}\!\int\!dx_\bu d^2x_\perp~e^{-i\alpha x_\bu+i(k,x)_\perp}
~\langle A|\hbsi_f(x_\bu,x_\perp)\gamma^\mu\hsi_f(0)|A\rangle
\label{Amael}\\
&&\hspace{27mm}
=~p_1^\mu f_1^f(\alpha,k_\perp^2)
+k_\perp^\mu f_\perp^f(\alpha,k_\perp^2)+p_2^\mu{2m^2_N\over s}f_3^f(\alpha,k_\perp^2),
\nonumber\\
&&\hspace{-1mm}
{1\over 16\pi^3}\!\int\!dx_\bu d^2x_\perp~e^{-i\alpha x_\bu+i(k,x)_\perp}
~\langle A|\hbsi_f(x_\bu,x_\perp)\hsi_f(0)|A\rangle
~=~m_Ne^f(\alpha,k_\perp^2)
\nonumber
\end{eqnarray}
for quark distributions in the projectile and 
\begin{eqnarray}
&&\hspace{-1mm}
{1\over 16\pi^3}\!\int\!dx_\bu d^2x_\perp~e^{i\alpha x_\bu-i(k,x)_\perp}
~\langle A|\hbsi_f(x_\bu,x_\perp)\gamma^\mu\hsi_f(0)|A\rangle
\label{baramael}\\
&&\hspace{27mm}
=~-p_1^\mu \barf_1^f(\alpha,k_\perp^2)
-k_\perp^\mu\barf _\perp^f(\alpha,k_\perp^2)-p_2^\mu{2m^2_N\over s}\barf_3^f(\alpha,k_\perp^2),
\nonumber\\
&&\hspace{-1mm}
{1\over 16\pi^3}\!\int\!dx_\bu d^2x_\perp~e^{i\alpha x_\bu-i(k,x)_\perp}
~\langle A|\hbsi_f(x_\bu,x_\perp)\hsi_f(0)|A\rangle
~=~
m_N\bare^f(\alpha,k_\perp^2)
\nonumber
\end{eqnarray}
for the antiquark distributions. 
\footnote{In an arbitrary gauge, there are gauge links to $-\infty$ as displayed in  eq. (\ref{gaugelinks}).}

The corresponding matrix elements for the target are obtained by trivial replacements $p_1\leftrightarrow p_2$, $x_\bu\leftrightarrow x_\ast$
and $\alpha\leftrightarrow\beta$:
\begin{eqnarray}
&&\hspace{-1mm}
{1\over 16\pi^3}\!\int\!dx_\ast d^2x_\perp~e^{-i\beta x_\ast+i(k,x)_\perp}
~\langle B|\hbsi_f(x_\ast,x_\perp)\gamma^\mu\hsi_f(0)|B\rangle
\label{Bmael}\\
&&\hspace{27mm}
=~p_2^\mu f_1^f(\beta,k_\perp^2)+k_\perp^\mu f_\perp^f(\beta,k_\perp^2)
+p_1^\mu{2m^2_N\over s}f_3^f(\beta,k_\perp^2),
\nonumber\\
&&\hspace{-1mm}
{1\over 16\pi^3}\!\int\!dx_\ast d^2x_\perp~e^{-i\beta x_\ast+i(k,x)_\perp}
~\langle B|\hbsi_f(x_\ast,x_\perp)\hsi_f(0)|B\rangle
~=~m_Ne^f(\beta,k_\perp^2),
\nonumber
\end{eqnarray}
and
\begin{eqnarray}
&&\hspace{-1mm}
{1\over 16\pi^3}\!\int\!dx_\ast d^2x_\perp~e^{i\beta x_\ast-i(k,x)_\perp}
~\langle B|\hbsi_f(x_\ast,x_\perp)\gamma^\mu\hsi_f(0)|B\rangle
\label{barbmael}\\
&&\hspace{27mm}
=~-p_2^\mu \barf_1^f(\beta,k_\perp^2)
-k_\perp^\mu\barf _\perp^f(\beta,k_\perp^2)-p_1^\mu{2m^2_N\over s}\barf_3^f(\beta,k_\perp^2),
\nonumber\\
&&\hspace{-1mm}
{1\over 16\pi^3}\!\int\!dx_\ast d^2x_\perp~e^{i\beta x_\ast-i(k,x)_\perp}
~\langle B|\hbsi_f(x_\ast,x_\perp)\hsi_f(0)|B\rangle
~=~
m_N\bare^f(\beta,k_\perp^2).
\nonumber
\end{eqnarray}

Matrix elements of operators with $\gamma_5$ are parametrized as follows: 
\begin{eqnarray}
&&\hspace{-1mm}
{1\over 16\pi^3}\!\int\!dx_\bu d^2x_\perp~e^{-i\alpha x_\bu+i(k,x)_\perp}
~\langle A|\hbsi_f(x_\bu,x_\perp)\gamma^\mu\gamma_5\hsi_f(0)|A\rangle
~=~i\epsilon_{\mu \nu\lambda\rho}{2\over s}p_1^\nu p_2^\lambda k^\rho 
g^\perp_f(\alpha,k_\perp^2),
\nonumber\\
&&\hspace{-1mm}
{1\over 16\pi^3}\!\int\!dx_\bu d^2x_\perp~e^{i\alpha x_\bu-i(k,x)_\perp}
~\langle A|\hbsi_f(x_\bu,x_\perp)\gamma^\mu\gamma_5\hsi_f(0)|A\rangle
~=~i\epsilon_{\mu \nu\lambda\rho}{2\over s}p_1^\nu p_2^\lambda k^\rho 
\barg^\perp_f(\alpha,k_\perp^2).
\nonumber\\
\label{mael5}
\end{eqnarray}
The corresponding matrix elements for the target are obtained by trivial replacements $p_1\leftrightarrow p_2$, $x_\bu\leftrightarrow x_\ast$
and $\alpha\leftrightarrow\beta$ similarly to eq. (\ref{barbmael}).

Finally, for future use we present the parametrization of time-odd TMDs
\begin{eqnarray}
&&\hspace{-1mm}
{1\over 16\pi^3}\!\int\!dx_\bu d^2x_\perp~e^{-i\alpha x_\bu+i(k,x)_\perp}
~\langle A|\hbsi_f(x_\bu,x_\perp)\sigma^{\mu \nu}\hsi_f(0)|A\rangle
\nonumber\\
&&\hspace{11mm}
=~{1\over m_N}(k_\perp^\mu p_1^\nu -\mu\leftrightarrow\nu)h_{1f}^\perp(\alpha,k_\perp^2)
+{2m_N\over s}(p_1^\mu p_2^\nu-\mu\leftrightarrow\nu)h_{f}(\alpha,k_\perp^2)
\nonumber\\
&&\hspace{33mm}
+~{2m_N\over s}(k_\perp^\mu p_2^\nu -\mu\leftrightarrow\nu)h_{3f}^\perp(\alpha,k_\perp^2),
\nonumber\\
&&\hspace{-1mm}
{1\over 16\pi^3}\!\int\!dx_\bu d^2x_\perp~e^{i\alpha x_\bu-i(k,x)_\perp}
~\langle A|\hbsi_f(x_\bu,x_\perp)\sigma^{\mu \nu}\hsi_f(0)|A\rangle
\nonumber\\
&&\hspace{11mm}
=~-{1\over m_N}(k_\perp^\mu p_1^\nu -\mu\leftrightarrow\nu)\barh_{1f}^\perp(\alpha,k_\perp^2)
-{2m_N\over s}(p_1^\mu p_2^\nu-\mu\leftrightarrow\nu)\barh_{f}(\alpha,k_\perp^2)
\nonumber\\
&&\hspace{33mm}
-~{2m_N\over s}(k_\perp^\mu p_2^\nu -\mu\leftrightarrow\nu)\barh_{3f}^\perp(\alpha,k_\perp^2)
\label{hmael}
\end{eqnarray}
and similarly for the target with usual replacements   $p_1\leftrightarrow p_2$, $x_\bu\leftrightarrow x_\ast$
and $\alpha\leftrightarrow\beta$.

Note that
the coefficients in front of $f_3$,  $g^\perp_f$, $h$ and $h_3^\perp$ in eqs. (\ref{Amael}),  (\ref{Bmael}), (\ref{mael5}),  and  (\ref{hmael}) 
 contain an extra ${1\over s}$ since $p_2^\mu$ enters only through the direction
of gauge link so the result should not depend on rescaling $p_2\rightarrow\lambda p_2$. For this reason,  these functions do not contribute to $W(q)$ in our approximation.

\subsection{Power corrections from $\calj_{AB}^\mu(x)\calj_{BA\mu}(0)$ terms \label{tw3first}}

The  terms in eq. (\ref{7lines}) proportional to  $\Xi$ fields are
\begin{eqnarray}
&&\hspace{-1mm}
\big[\big(\bar\psi_A (x)+\Bxi_{2A}(x)\big)\hamma_\mu\big(\psi_B(x)+\Xi_{1B}(x)\big)\big]
\label{6lines}\\
&&\hspace{44mm}
\times~[\big(\bar\psi_B(0)+\Bxi_{1B}(0)\big)\hamma^\mu\big(\psi_A(x)+\Xi_{2A}(0)\big)\big]~+~x\leftrightarrow 0
\nonumber\\
&&\hspace{-1mm}
\stackrel{\rm tw3}{=}~[\Bxi_{2A}(x)\hamma_\mu\psi_B(x)\big]\big[\bar\psi_B(0)\hamma^\mu\psi_A(0)\big]
+[\bar\psi_A(x)\hamma_\mu\Xi_{1B}(x)\big]\big[\bar\psi_B(0)\hamma^\mu\psi_A(0)\big]
\nonumber\\
&&\hspace{-1mm}
+~
[\bar\psi_A(x)\hamma_\mu\psi_B(x)\big]\big[\Bxi_{1B}(0)\hamma^\mu\psi_A(0)\big]
+[\bar\psi_A(x)\hamma_\mu\psi_B(x)\big]\big[\bar\psi_B(0)\hamma^\mu\Xi_{2A}(0)\big]
\nonumber\\
&&\hspace{-1mm}
+~[\Bxi_{2A}(x)\hamma_\mu\psi_B(x)\big]\big[\bar\psi_B(0)\hamma^\mu\Xi_{2A}(0)\big]
+[\bar\psi_A(x)\hamma_\mu\Xi_{1B}(x)\big]\big[\Bxi_{1B}(0)\hamma^\mu\psi_A(0)\big]
\nonumber\\
&&\hspace{-1mm}
+~[\Bxi_{2A}(x)\hamma_\mu\psi_B(x)\big]\big[\Bxi_{1B}(0)\hamma^\mu\psi_A(0)\big]
+[\bar\psi_A(x)\hamma_\mu\Xi_{1B}(x)\big]\big[\bar\psi_B(0)\hamma^\mu\Xi_{2A}(0)\big]
\nonumber\\
&&\hspace{-1mm}
+~[\Bxi_{2A}(x)\hamma_\mu\Xi_{1B}(x)\big]\big[\bar\psi_B(0)\hamma^\mu\psi_A(0)\big]
+[\bar\psi_A(x)\hamma_\mu\psi_B(x)\big]\big[\Bxi_{1B}(0)\hamma^\mu\Xi_{2A}(0)\big]
~+~x\leftrightarrow 0.
\nonumber
\end{eqnarray}
First,  as we demonstrate in appendix \ref{234lines}, 
the terms in the second, third, and fourth lines lead to negligible power corrections $\sim {q_\perp^2\over \alpha_qs}$ or
$\sim {q_\perp^2\over \beta_qs}$, so we are left with contribution of the fifth and sixth lines.

\subsubsection{Fifth line in eq. (\ref{6lines}): the leading term in ${1\over N_c}$ \label{fifthline}}
Let us start with the term 
$\big[\bar\psi_A(x)\hamma_\mu\Xi_{1B}(x)\big]\big[\bar\psi_B(0)\hamma^\mu\Xi_{2A}(0)\big]$. 
Performing Fierz transformation (\ref{fierz}) we obtain
\begin{eqnarray}
&&\hspace{-1mm}
\big[\bar\psi_A^{m}(x)\gamma_\mu(a-\gamma_5)\Xi^{m}_{1B}(x)\big]\big[\bar\psi_B^{n}(0)\gamma^\mu(a-\gamma_5)\Xi^{n}_{2A}(0)\big]
\\
&&\hspace{-1mm}
=~{1+a^2\over 2}\big\{\big[\bar\psi_A^{m}(x)\gamma_\alpha\Xi^{n}_{2A}(0)\big]\big[\bar\psi_B^{n}(0)\gamma^\alpha\Xi^{m}_{1B}(x)\big]
~+~(\gamma_\alpha\otimes\gamma^\alpha\leftrightarrow\gamma_\alpha\gamma_5\otimes\gamma^\alpha\gamma_5)\big\}
\nonumber\\
&&\hspace{-1mm}
-~a\big(\big[\bar\psi_A^{m}(x)\gamma_\alpha\Xi^{n}_{2A}(0)\big]\big[\bar\psi_B^{n}(0)\gamma^\alpha\gamma_5\Xi^{m}_{1B}(x)\big]
+(\gamma_\alpha\otimes\gamma^\alpha\gamma_5\leftrightarrow\gamma_\alpha\gamma_5\otimes\gamma^\alpha)\big\}
\nonumber\\
&&\hspace{-1mm}
+~(1-a^2)\big(\big[\bar\psi_A^{m}(x)\Xi^{n}_{2A}(0)\big]\big[\bar\psi_B^{n}(0)\Xi^{m}_{1B}(x)\big]
-(1\otimes1\leftrightarrow\gamma_5\otimes\gamma_5)\big\}
\nonumber\\
&&\hspace{-1mm}
=~{1+a^2\over 2}\big\{\big[\bar\psi_A^{m}(x)\gamma_i\Xi^{n}_{2A}(0)\big]\big[\bar\psi_B^{n}(0)\gamma^i\Xi^{m}_{1B}(x)\big]
~+~(\gamma_i\otimes\gamma^i\leftrightarrow\gamma_i\gamma_5\otimes\gamma^i\gamma_5)\big\}
\nonumber\\
&&\hspace{-1mm}
+~(1-a^2)\big(\big[\bar\psi_A^{m}(x)\Xi^{n}_{2A}(0)\big]\big[\bar\psi_B^{n}(0)\Xi^{m}_{1B}(x)\big]
-(1\otimes1\leftrightarrow\gamma_5\otimes\gamma_5)\big\}
\nonumber\\
&&\hspace{-1mm}
-~a\big(\big[\bar\psi_A^{m}(x)\gamma_i\Xi^{n}_{2A}(0)\big]\big[\bar\psi_B^{n}(0)\gamma^i\gamma_5\Xi^{m}_{1B}(x)\big]
+(\gamma_i\otimes\gamma^i\gamma_5\leftrightarrow\gamma_i\gamma_5\otimes\gamma^i)\big\}
~+~O\big({m_\perp^8\over s}\big).
\nonumber
\end{eqnarray}

Next, separating color-singlet contributions 
\begin{eqnarray}
&&\hspace{-1mm}
\langle A,B|(\bsi_A^m (B_j)^{nk}\psi_A^k)(\bsi_B^n  (A_i)^{ml}\psi_B^l)|A,B\rangle
~=~\langle A,B|(\bsi_A^m (A_i)^{ml}\psi_A^k)(\bsi_B^n (B_j)^{nk} \psi_B^l)|A,B\rangle
\nonumber\\
&&\hspace{-1mm}
=~{1\over N_c}\langle A|(\bsi_A^m A_i^{ml}\psi_A^l)|A\rangle\langle B|(\bsi_B^n B_j^{nk}\psi_B^k)|B\rangle
\label{4.23}
\end{eqnarray}
we get
\begin{eqnarray}
&&\hspace{-1mm}
s^2N_cg^{-2}\big[\bar\psi_A(x)\hamma_\mu\Xi_{1B}(x)\big]\big[\bar\psi_B(0)\hamma^\mu\Xi_{2A}(0)\big]
\label{4.24}\\
&&\hspace{-1mm}
=~{1+a^2\over 2}
\big\{\big[\bar\psi_A(x)A_k(x)\gamma_i\slashed{p}_2\gamma^j{1\over\alpha}\psi_A(0)\big]
\big[\bar\psi_B(0)B_j(0)\gamma^i\slashed{p}_1\gamma^k{1\over\beta}\psi_B(x)\big]
\nonumber\\
&&\hspace{77mm}
+~~(\gamma_i\otimes\gamma^i\leftrightarrow\gamma_i\gamma_5\otimes\gamma^i\gamma_5)\big\}
\nonumber\\
&&\hspace{-1mm}
+~(1-a^2)\big\{\big[\bar\psi_A(x)A_k(x)\slashed{p}_2\gamma^j{1\over\alpha}\psi_A(0)\big]
\big[\bar\psi_B(0)B_j(0)\slashed{p}_1\gamma^k{1\over\beta}\psi_B(x)\big]
\nonumber\\
&&\hspace{77mm}
-~~(\gamma^j\otimes\gamma^k\leftrightarrow\gamma^j\gamma_5\otimes\gamma^k\gamma_5)\big\}
\nonumber\\
&&\hspace{-1mm}
-~a\big\{\big[\bar\psi_A(x)A_k(x)\gamma_i\gamma_5\slashed{p}_2\gamma^j{1\over\alpha}\psi_A(0)\big]
\big[\bar\psi_B(0)B_j(0)\gamma^i\slashed{p}_1\gamma^k{1\over\beta}\psi_B(x)\big]
\nonumber\\
&&\hspace{77mm}
+~~(\gamma_i\gamma_5\otimes\gamma^i\leftrightarrow\gamma_i\otimes\gamma^i\gamma_5)\big\}.
\nonumber
\end{eqnarray}
Using equations (\ref{9.9}), (\ref{formula}), and (\ref{9.13}) from appendix \ref{diracs} we can rewrite eq. (\ref{4.24}) as
\begin{eqnarray}
&&\hspace{-1mm}
g^{-2}N_c\big\{\bar\psi_A(x)\gamma_\mu(a-\gamma_5)\Xi_{1B}(x)\big\}\big\{\bar\psi_B(0)\gamma^\mu(a-\gamma_5)\Xi_{2A}(0)\big\}
\label{5.14}\\
&&\hspace{-1mm}
=~{1+a^2\over s^2}
\big\{\bar\psi_A(x)\slashed{p}_2[A_i(x)-i\gamma_5\tilde A_i(x)]{1\over\alpha}\psi_A(0)\big\}
\big\{\bar\psi_B(0)\slashed{p}_1[B^i(0)-i\gamma_5\tilde B^i(0)]{1\over\beta}\psi_B(x)\big\}
\nonumber\\
&&\hspace{-1mm}
+~{1-a^2\over s^2}\big\{\bar\psi_A(x)A_k(x)\slashed{p}_2\gamma_j{1\over\alpha}\psi_A(0)\big\}
\nonumber\\
&&\hspace{44mm}
\times~\big\{\bar\psi_B(0)[B^j(0)\slashed{p}_1\gamma^k-j\leftrightarrow k+g^{jk}B^i(0)\slashed{p}_1\gamma_i]{1\over\beta}\psi_B(x)\big\}
\nonumber\\
&&\hspace{-1mm}   
-~{2a\over s^2}\big\{\bar\psi_A(x)\slashed{p}_2[\gamma_5A_i(x)-i\tilde A_i(x)]{1\over\alpha}\psi_A(0)\big\}
\nonumber\\
&&\hspace{44mm}
\times~
\big\{\bar\psi_B(0)\slashed{p}_1[B^i(0)-i\gamma_5\tilde B^i(0)]{1\over\beta}\psi_B(x)\big\}
~+~O\big({m_\perp^8\over s}\big).
\nonumber
\end{eqnarray}
For forward matrix elements we get
\begin{eqnarray}
&&\hspace{-1mm}
\int\! dx_\bu~e^{-i\alpha_qx_\bu}
\langle A|\hat{\bar\psi}(x_\bu,x_\perp)\slashed{p}_2[\hatA_i(x_\bu,x_\perp)-i\gamma_5\hat\tilA_i(x_\bu,x_\perp)]{1\over\alpha}\hat\psi(0)|A\rangle
\nonumber\\
&&\hspace{-1mm}
=~{1\over\alpha_q}\!\int\! dx_\bu~e^{-i\alpha_q x_\bu}
\langle A|\hat{\bar\psi}(x_\bu,x_\perp)\slashed{p}_2[\hatA_i(x_\bu,x_\perp)-i\gamma_5\hat\tilA_i(x_\bu,x_\perp)]\hat\psi(0)|A\rangle,
\nonumber\\
&&\hspace{-1mm}
\int\! dx_\ast~e^{-i\beta_qx_\ast}\langle B|\hat{\bar\psi}(0)\slashed{p}_1[\hatA_j(0)-i\gamma_5\hat\tilA_j(0)]{1\over\beta}\hsi(x_\ast,x_\perp)|B\rangle
\nonumber\\
&&\hspace{-1mm}
=~-{1\over\beta_q}\!\int\! dx_\ast~e^{-i\beta_qx_\ast}
\langle B|\hat{\bar\psi}(0)\slashed{p}_1[\hatA_j(0)-i\gamma_5\hat\tilA_j(0)]\hsi(x_\ast,x_\perp)|B\rangle
\label{4.26}
\end{eqnarray}
and similarly for other Lorentz structures in eq. (\ref{5.14}). The corresponding contribution of the r.h.s of eq. (\ref{5.14}) to 
$W(\alpha_q,\beta_q,x_\perp)$ takes the form
\footnote{After specifying the projectile and target matrix elements the  $``A"$ and $``B"$ labels of the fields become redundant.}
\begin{eqnarray}
&&\hspace{-1mm}
-{e^2g^2(2\pi)^{-4}\over 8s_W^2c_W^2N_cQ^2}\!\int\! dx_\bu dx_\ast~e^{-i\alpha_qx_\bu-i\beta_q x_\ast}
\Big\{{1+a^2\over s^2}\langle A|\hat{\bar\psi}(x_\bu,x_\perp)\slashed{p}_2(\hatA_i-i\gamma_5\hat\tilA_i)(x_\bu,x_\perp)\hsi(0)|A\rangle
\nonumber\\
&&\hspace{-1mm}
\times~\langle B|\hat{\bar\psi}(0)\slashed{p}_1(\hatA^i-i\gamma_5\hat\tilA^i)(0)\hsi(x_\ast,x_\perp)|B\rangle
~+~{1-a^2\over s^2}\langle A|\hat{\bar\psi}(x_\bu,x_\perp)\hatA^j(x_\bu,x_\perp)\slashed{p}_2\gamma_j\hsi(0)|A\rangle
\nonumber\\
&&\hspace{-1mm}
\times~\langle B|\hat{\bar\psi}(0)\hatA^k(0)\slashed{p}_1\gamma_k\hsi(x_\ast,x_\perp)|B\rangle\Big\}
\Big(1+O\big({m_\perp^2\over s}\big)\Big).
\label{4.27}
\end{eqnarray}
Note that for unpolarized hadrons $\langle B|\hat{\bar\psi}(0)(\hatA^j(0)\slashed{p}_1\gamma^k-j\leftrightarrow k)\hsi(x_\ast,x_\perp)|B\rangle~=~0$.
Also, it is easy to see that the last line of eq. (\ref{5.14}) 
\begin{equation}
\hspace{-1mm}
-~{2a\over s^2}\langle A|\hat{\bar\psi}(x)\slashed{p}_2[\hatA_i(x)-i\gamma_5\hat{\tilde{A}}_i(x)]\hsi(0)|A\rangle
\langle B|\hat{\bar\psi}(0)\slashed{p}_1(\gamma_5\hat{A}^i(0)-i\hat{\tilde{A}}^i(0)]\hsi(x)|B\rangle
\label{5.17}
\end{equation}
gives zero contribution. Indeed, let us consider the first term in the r.h.s. of this equation. Since
\begin{eqnarray}
&&\hspace{-1mm}
\langle A|\hat{\bar\psi}(x)\slashed{p}_2[\hatA_i(x)-i\gamma_5\hat{\tilde{A}}_i(x)]\hsi(0)|A\rangle~\sim~x_i,
\nonumber\\
&&\hspace{-1mm}
\langle B|\hat{\bar\psi}(0)\slashed{p}_1(\gamma_5\hat{A}^i(0)-i\hat{\tilde{A}}^i(0)]\hsi(x)|B\rangle~\sim~\epsilon^{ij}x_j,
\label{4.29}
\end{eqnarray}
this term  vanishes (and similarly all other terms in the r.h.s. of eq. (\ref{5.17}) do vanish too).

Repeating the same steps for the second term in the fifth line in eq. (\ref{6lines}) we get
\begin{eqnarray}
&&\hspace{-1mm}
N_cg^{-2}\big\{\Bxi_{2A}(x)\gamma_\mu(a-\gamma_5)\psi_B(x)\big\}\big\{\Bxi_{1B}(0)\gamma^\mu(a-\gamma_5)\psi_A(0)\big\}
\label{5.18}\\
&&\hspace{-1mm}
=~{1+a^2\over s^2}
\big\{\big(\bar\psi_A{1\over\alpha}\big)(x)\slashed{p}_2[A_i(0)+i\gamma_5\tilde A_i(0)]\psi_A(0)\big\}
\big\{\big(\bar\psi_B{1\over\beta}\big)(0)\slashed{p}_1[B^i(x)+i\gamma_5\tilde B^i(x)]\psi_B(x)\big\}
\nonumber\\
&&\hspace{-1mm}
+~{1-a^2\over s^2}\big\{\big(\bar\psi_A{1\over\alpha}\big)(x)A_k(0)\slashed{p}_2\gamma_j\psi_A(0)\big\}
\nonumber\\
&&\hspace{44mm}
\times~
\big\{\big(\bar\psi_B{1\over\beta}\big)(0)[B^j(x)\slashed{p}_1\gamma^k-j\leftrightarrow k+g^{jk}B^i(x)\slashed{p}_1\gamma_i]\psi_B(x)\big\}
\nonumber\\
&&\hspace{-1mm}
-~{2a\over s^2}\big\{\big(\bar\psi_A{1\over\alpha}\big)(x)\slashed{p}_2[\gamma_5A_i(0)+i\tilde A_i(0)]\psi_A(0)\big\}
\nonumber\\
&&\hspace{44mm}
\times~
\big\{\big(\bar\psi_B{1\over\beta}\big)(0)\slashed{p}_1[B^i(x)+i\gamma_5\tilde B^i(x)]\psi_B(x)\big\}
~+~O\big({m_\perp^8\over s}\big).
\nonumber
\end{eqnarray}
Hereafter $\big(\bar\psi_A{1\over\alpha}\big)(x)\equiv \big(\bsi_A{1\over \alpha-i\epsilon}\big)(x)$ and 
$\big(\bar\psi_B{1\over\beta}\big)(x)\equiv \big(\bsi_B{1\over \beta-i\epsilon}\big)(x)$ (see eq. (\ref{3.25})) while in all other places 
$\big({1\over\alpha}\calo\big)\equiv\big({1\over\alpha+i\epsilon}\calo\big)$ and  $\big({1\over\beta}\calo\big)\equiv\big({1\over\beta+i\epsilon}\calo\big)$.

For forward matrix elements this gives
\begin{eqnarray}
&&\hspace{-1mm}
\int\! dx_\bu~e^{-i\alpha_qx_\bu}\langle A|\big(\hat{\bar\psi}{1\over\alpha}\big)(x_\bu,x_\perp)\slashed{p}_2[\hatA_i(0)+i\gamma_5\hat{\tilde{A}}_i(0)]\psi(0)|A\rangle
\nonumber\\
&&\hspace{-1mm}
=~{1\over\alpha_q}\!\int\! dx_\bu~e^{-i\alpha_q x_\bu}
\langle A|\hat{\bar\psi}(x_\bu,x_\perp)\slashed{p}_2[\hatA_i(0)+i\gamma_5\hat{\tilde{A}}_i(0)]\psi(0)|A\rangle,
\nonumber\\
&&\hspace{-1mm}
\int\! dx_\ast~e^{-i\beta_qx_\ast}
\langle B|\big(\hat{\bar\psi}{1\over\beta}\big)(0)\slashed{p}_1[\hatA_j(x_\ast,x_\perp)+i\gamma_5\hat{\tilde{A}}_j(x_\ast,x_\perp)]
\hsi(x_\ast,x_\perp)|B\rangle
\nonumber\\
&&\hspace{-1mm}
=~-{1\over\beta_q}\!\int\! dx_\ast~e^{-i\beta_qx_\ast}
\langle B|\hat{\bar\psi}(0)\slashed{p}_1[\hatA_j(x_\ast,x_\perp)+i\gamma_5\hat{\tilde{A}}_j(x_\ast,x_\perp)]
\hsi(x_\ast,x_\perp)|B\rangle,
\label{4.31}
\end{eqnarray}
and similarly for other Lorentz structures in eq. (\ref{5.18}). Similarly to eq. (\ref{4.27}), 
we get the contribution to $W(\alpha_q,\beta_q,x_\perp)$ 
in the form
\begin{eqnarray}
&&\hspace{-1mm}
-{e^2g^2\over 8(2\pi)^4s_W^2c_W^2N_cQ^2}\!\int\! dx_\bu dx_\ast~e^{-i\alpha_qx_\bu-i\beta_q x_\ast}
\label{4.32}\\
&&\hspace{-1mm}
\times~\Big\{{1+a^2\over s^2}
\langle A|\hat{\bar\psi}(x_\bu,x_\perp)\slashed{p}_2[\hatA_i(0)+i\gamma_5\hat{\tilde{A}}_i(0)]\hsi(0)|A\rangle
\nonumber\\
&&\hspace{44mm}
\times~
\langle B|\hat{\bar\psi}(0)\slashed{p}_1[\hatA^i(x_\ast,x_\perp)+i\gamma_5\hat{\tilde{A}}^i(x_\ast,x_\perp)]\hsi(x_\ast,x_\perp)|B\rangle
\nonumber\\
&&\hspace{-1mm}
+~{1-a^2\over s^2}\langle A|\hat{\bar\psi}(x_\bu,x_\perp)\hatA_j(0)\slashed{p}_2\gamma^j\hsi(0)|A\rangle
\nonumber\\
&&\hspace{44mm}
\times~\langle B|\hat{\bar\psi}(0)\hatA^k(x_\ast,x_\perp)\slashed{p}_1\gamma_k\hsi(x_\ast,x_\perp)|B\rangle\Big\}\Big(1 
+ O\big({m_\perp^2\over s}\big)\Big).
\nonumber
\end{eqnarray}

In section \ref{paramfifth}, we demonstrated that the matrix elements of quark-antiquark-gluon operators in eqs. (\ref{4.27}) and (\ref{4.32}) 
reduce to the leading-power TMDs from section \ref{paramlt}.
Using  parametrizations from section \ref{paramfifth} we obtain the contribution of the 5th line in eq. (\ref{6lines})  to  $W(q)$ in the form:
\begin{eqnarray}
&&\hspace{-1mm}
W^{\rm 5th}(\alpha_q,\beta_q,q_\perp)
\nonumber\\
&&\hspace{-1mm}
=~{e^2\over 4s_W^2c_W^2N_cQ^2}\!\int\! d^2k_\perp ~
\Big[\Big\{(1+a_u^2)(k,q-k)_\perp
f_{1u}(\alpha_q,k_\perp)\barf_{1u}(\beta_q,q_\perp-k_\perp) 
\nonumber\\
&&\hspace{22mm}
+~{1\over m^2_N}(1-a_u^2)k_\perp^2(q-k)_\perp^2h^\perp_{1u}(\alpha_q,k_\perp)\barh^\perp_{1u}(\beta_q,q_\perp-k_\perp) 
+~(\alpha_q\leftrightarrow\beta_q)\Big\}
\nonumber\\
&&\hspace{33mm}
~+~\Big\{u\leftrightarrow c\Big\}+\Big\{u\leftrightarrow d\Big\}+\Big\{u\leftrightarrow s\Big\}\Big]\Big(1 
+ O\big({m_\perp^2\over s}\big)\Big),
\label{firstPC}
\end{eqnarray}
where quark$\leftrightarrow$antiquark ($\alpha_q\leftrightarrow\beta_q$) term comes from $x\leftrightarrow 0$  contribution in eq. (\ref{6lines}).
As we will demonstrate later, the power corrections which reduce to the leading-power TMDs come with the leading power of ${1\over N_c}$
in the large-$N_c$ approximation - all other power corrections are $\sim {1\over N_c^2}$ or ${1\over N_c^3}$.

\subsubsection{Parametrization of  matrix elements from section \ref{fifthline} \label{paramfifth}}

 In this section we will demonstrate that matrix elements of quark-antiquark-gluon operators 
 from section \ref{fifthline} can be expressed in terms of leading-power
 matrix elements from section \ref{paramlt}. Let us start with matrix element  (\ref{4.26}) which can be rewritten as (see ref. \cite{Mulders:1995dh})
\begin{eqnarray}
&&\hspace{-1mm}
g\!\int\! dx_\bu dx_\perp~e^{-i\alpha_qx_\bu+i(k,x)_\perp}\langle A|\hat{\bar\psi}(x_\bu,x_\perp)\slashed{p}_2[\hatA_i(x_\bu,x_\perp)-i\gamma_5\hat\tilA_i(x_\bu,x_\perp)]\hat\psi(0)|A\rangle
\label{tw3mael1}\\
&&\hspace{-1mm}
=~\int\! dx_\bu dx_\perp~e^{-i\alpha_qx_\bu+i(k,x)_\perp}
\nonumber\\
&&\hspace{31mm}
\times~\big[k^j\langle A|\hbsi(x_\bu,x_\perp)\gamma_j\slashed{p}_2\gamma_i\hsi(0)|A\rangle
+i
\langle A|\hbsi(x_\bu,x_\perp)\!\stackrel{\leftarrow}{\hatD^j}\!\gamma_j\slashed{p}_2\gamma_i\hsi(0)|A\rangle\big].
\nonumber
\end{eqnarray}
Using QCD equations of motion (\ref{YMs}) we can rewrite the r.h.s. of eq. (\ref{tw3mael1}) as
\begin{eqnarray}
&&\hspace{-2mm}
\int\! dx_\bu dx_\perp~e^{-i\alpha_qx_\bu+i(k,x)_\perp}\big[k^j\langle A|\hbsi(x_\bu,x_\perp)\gamma_j\slashed{p}_2\gamma_i\hsi(0)|A\rangle
+\alpha_q\langle A|\hbsi(x_\bu,x_\perp)\slashed{p}_1\slashed{p}_2\gamma_i\hsi(0)|A\rangle\big]
\nonumber\\
&&\hspace{-1mm}
=~\int\! dx_\bu dx_\perp~e^{-i\alpha_qx_\bu+i(k,x)_\perp}
\Big[
-k_i\langle A|\hbsi(x_\bu,x_\perp)\slashed{p}_2\hsi(0)|A\rangle
+~\alpha_q{s\over 2} \langle A|\hbsi(x_\bu,x_\perp)\gamma_i\hsi(0)|A\rangle
\nonumber\\
&&\hspace{6mm}
+~{2i\over s}\epsilon_{\bu\ast ij}k^j\langle A|\hbsi(x_\bu,x_\perp)\slashed{p}_2\gamma_5\hsi(0)|A\rangle
-i\alpha\epsilon_{\bu\ast ij}
\langle A|\hbsi(x_\bu,x_\perp)\gamma^j\gamma_5\hsi(0)|A\rangle\Big]
\nonumber\\
&&\hspace{14mm}
=~-k_i8\pi^3sf_1(\alpha_q,k_\perp^2)+8\pi^3s\alpha_q k_i\big[ f_\perp(\alpha_q,k_\perp^2)+g^\perp(\alpha_q,k_\perp^2)\big],
\label{tw3mael2}
\end{eqnarray}
where we used parametrizations (\ref{Amael}) and (\ref{mael5}) for the leading power matrix elements. 

Now, the second term in eq. (\ref{tw3mael2}) contains extra $\alpha_q$ with respect to the first term, so
 it should be neglected in our kinematical region $s\gg Q^2\gg q_\perp^2$  and we get 
\begin{eqnarray}
&&\hspace{-1mm}
{g\over 8\pi^3s}\!\int\! dx_\bu dx_\perp~e^{-i\alpha_qx_\bu+i(k,x)_\perp}\langle A|\hat{\bar\psi}^f(x_\bu,x_\perp)\slashed{p}_2[\hatA_i(x_\bu,x_\perp)-i\gamma_5\hat\tilA_i(x_\bu,x_\perp)]\hat\psi^f(0)|A\rangle
\nonumber\\
&&\hspace{11mm}
=~-k_if_1^f(\alpha_q,k_\perp^2)~+~O(\alpha_q).
\label{9.22}
\end{eqnarray}
By complex conjugation
\begin{eqnarray}
&&\hspace{-1mm}
{g\over 8\pi^3s}\!\int\! dx_\perp dx_\bu~e^{-i\alpha_q x_\bu+i(k,x)_\perp}
\langle A|\hat{\bsi}_f(x_\bu,x_\perp)\slashed{p}_2[\hatA_i(0)+i\gamma_5\hat{\tilde A}_i(0)]\hsi_f(0)|A\rangle
\nonumber\\
&&\hspace{11mm}
=~-k_i
f_{1f}(\alpha_q,k_\perp^2).
\label{9.23}
\end{eqnarray}

For the  corresponding antiquark distributions we get 
\begin{eqnarray}
&&\hspace{0mm}
{g\over 8\pi^3s}\!\int\! dx_\perp dx_\bu~e^{-i\alpha x_\bu+i(k,x)_\perp}
\langle A|\hat{\bsi}_f(0)\slashed{p}_2[\hatA_i(x_\bu,x_\perp)+i\gamma_5\hat{\tilde A}_i(x_\bu,x_\perp)]\hsi_f(x_\bu,x_\perp)|A\rangle
\nonumber\\
&&\hspace{11mm}
=~{1\over 8\pi^3s}\!\int\! dx_\bu dx_\perp e^{-i\alpha_qx_\bu+i(k,x)_\perp}\Big[-k_j\langle A|\hat{\bar\psi}(0)\gamma_i\slashed{p}_2\gamma^j\hat{\psi}(x_\bu,x_\perp)|A\rangle
\nonumber\\
&&\hspace{22mm}
-~i\langle A|\hbsi(0)\gamma_i\slashed{p}_2\gamma^j\hatD_j\hsi(x_\bu,x_\perp)|A\rangle\Big]
~=~-k_i\barf_{1f}(\alpha_q,k_\perp^2)
\label{9.26}
\end{eqnarray}
and
\begin{eqnarray}
&&\hspace{0mm}
{g\over 8\pi^3s}\!\int\! dx_\perp dx_\bu~e^{-i\alpha x_\bu+i(k,x)_\perp}
\langle A|\hat{\bsi}_f(0)\slashed{p}_2[\hatA_i(0)-i\gamma_5\hat{\tilde A}_i(0)]\hsi_f(x_\bu,x_\perp)|A\rangle
\nonumber\\
&&\hspace{11mm}
=~
-k_i\barf_{1f}(\alpha,k_\perp^2).
\label{9.25}
\end{eqnarray}

The corresponding target matrix elements are obtained by trivial replacements 
$x_\ast\leftrightarrow x_\bu$, $\alpha_q\leftrightarrow\beta_q$ and
$\slashed{p}_2\leftrightarrow\slashed{p}_1$.

Next, let us consider
\begin{eqnarray}
&&\hspace{-1mm}
{g\over 8\pi^3s}\!\int\! dx_\bu dx_\perp~e^{-i\alpha_qx_\bu+i(k,x)_\perp}
\langle A|\hsi(x_\bu,x_\perp)\slashed{p}_2\gamma^i \hatA_i(x_\bu,x_\perp)\hsi(0)|A\rangle
\\
&&\hspace{11mm}
=~{1\over 8\pi^3s}\!\int\! dx_\bu dx_\perp~e^{-i\alpha_qx_\bu+i(k,x)_\perp}
\nonumber\\
&&\hspace{22mm}
\times~\Big[k_i\langle A|\hsi(x_\bu,x_\perp)\gamma^i\slashed{p}_2\hsi(0)|A\rangle
+i\langle A|\hsi(x_\bu,x_\perp)\stackrel{\leftarrow}{\hatD_i}\gamma^i\slashed{p}_2\hsi(0)|A\rangle\Big].
\nonumber
\end{eqnarray}
Using QCD equation of motion and parametrization (\ref{hmael}), one can rewrite the r.h.s. of this equation as
\begin{eqnarray}
&&\hspace{-3mm}
{1\over 8\pi^3s}\!\int\! dx_\bu dx_\perp~e^{-i\alpha_qx_\bu+i(k,x)_\perp}
\Big[k_i\langle A|\hbsi(x_\bu,x_\perp)\gamma^i\slashed{p}_2\hsi(0)|A\rangle
+\alpha_q\langle A|\hbsi(x_\bu,x_\perp)\slashed{p}_1\slashed{p}_2\hsi(0)|A\rangle\Big]
\nonumber\\
&&\hspace{-1mm}
=~i{k_\perp^2\over m_N}h_{1}^\perp(\alpha_q,k_\perp^2)+\alpha_q m_N\big[e(\alpha,k_\perp^2)+i h(\alpha,k_\perp^2)\big].
\end{eqnarray}
Again, only the first term contributes in our kinematical region so we finally get
\begin{eqnarray}
&&\hspace{-2mm}
{g\over 8\pi^3s}\!\int\! dx_\bu dx_\perp~e^{-i\alpha_qx_\bu+i(k,x)_\perp}
\langle A|\hbsi^f(x_\bu,x_\perp)\slashed{p}_2\gamma^i \hatA_i(x_\bu,x_\perp)\hsi^f(0)|A\rangle
~=~i{k_\perp^2\over m_N}h_{1f}^\perp(\alpha_q,k_\perp^2).
\nonumber\\
\label{9.28}
\end{eqnarray}
By complex conjugation we obtain
\begin{eqnarray}
&&\hspace{-1mm}
{g\over 8\pi^3s}\!\int\! dx_\bu dx_\perp~e^{-i\alpha_qx_\bu+i(k,x)_\perp}
\langle A|\hbsi^f(x_\bu,x_\perp)\slashed{p}_2\gamma^i \hatA_i(0)\hsi^f(0)|A\rangle
~=~i{k_\perp^2\over m_N}h_{1f}^\perp(\alpha_q,k_\perp^2).
\nonumber\\
\label{9.29}
\end{eqnarray}

For corresponding antiquark distributions one gets in a similar way
\begin{eqnarray}
&&\hspace{-2mm}
{g\over 8\pi^3s}\!\int\! dx_\bu dx_\perp~e^{-i\alpha_qx_\bu+i(k,x)_\perp}
\langle A|\hbsi^f(0)\slashed{p}_2\gamma^i\hatA_i(x_\bu,x_\perp)\hsi^f(x_\bu,x_\perp)|A\rangle
=~
i{k_\perp^2\over m_N}\barh_{1f}^\perp(\alpha_q,k_\perp^2),
\nonumber\\
&&\hspace{-2mm}
{g\over 8\pi^3s}\!\int\! dx_\bu dx_\perp~e^{-i\alpha_qx_\bu+i(k,x)_\perp}
\langle A|\hbsi^f(0)\slashed{p}_2\gamma^i\hatA_i(0)\hsi^f(x_\bu,x_\perp)|A\rangle
=~
i{k_\perp^2\over m_N}\barh_{1f}^\perp(\alpha_q,k_\perp^2).
\nonumber\\
\label{9.30}
\end{eqnarray}
The target matrix elements are obtained by usual replacements 
$x_\ast\leftrightarrow x_\bu$, $\alpha_q\leftrightarrow\beta_q$ and
$\slashed{p}_2\leftrightarrow\slashed{p}_1$.

\subsubsection{Sixth line in eq. (\ref{6lines}) \label{sixline}}
In this section we consider $[\bar\psi_A(x)\hamma_\mu\psi_{B}(x)\big]\big[\Bxi_{1B}(0)\hamma^\mu\Xi_{2A}(0)\big]$
which turns to 
\begin{eqnarray}
&&\hspace{-1mm}
[\bar\psi_A^m(x)\gamma_\mu(a-\gamma_5)\psi^m_{B}(x)\big]\big[\Bxi_{1B}^{n}(0)\gamma^\mu(a-\gamma_5)\Xi^{n}_{2A}(0)\big]
\\
&&\hspace{-1mm}
=~{1+a^2\over 2}\big\{\big[\bar\psi_A^{m}(x)\gamma_i\Xi^{n}_{2A}(0)\big]\big[\Bxi_{1B}^{n}(0)\gamma^i\psi^m_{B}(x)\big]
~+~(\gamma_i\otimes\gamma^i\leftrightarrow\gamma_i\gamma_5\otimes\gamma^i\gamma_5)\big\}
\nonumber\\
&&\hspace{-1mm}
+~(1-a^2)\big\{\big[\bar\psi_A^{m}(x)\Xi^{n}_{2A}(0)\big]\big[\Bxi_{1B}^{n}(0)\psi^m_{B}(x)\big]
-(1\otimes1\leftrightarrow\gamma_5\otimes\gamma_5)\big\}
\nonumber\\
&&\hspace{-1mm}
-~a\big\{\big[\bar\psi_A^{m}(x)\gamma_i\Xi^{n}_{2A}(0)\big]\big[\Bxi_{1B}^{n}(0)\gamma^i\gamma_5\psi^m_{B}(x)\big]
+(\gamma_i\otimes\gamma^i\gamma_5\leftrightarrow\gamma_i\gamma_5\otimes\gamma^i)\big\}
+~O\big({m_\perp^8\over s}\big)
\nonumber
\end{eqnarray}
after Fierz transformation (cf. eq. (\ref{fierz})). After separation of color singlet contributions
\begin{eqnarray}
&&\hspace{-1mm}
\langle A,B|(\Bsi_A^m (A_i)^{ln}\psi_A^k)(\Bsi_B^l (B_j)^{nk} \psi_B^m)|A,B\rangle~=~\langle A,B|(\Bsi_A^m (A_i)^{nk}\psi_A^k)(\Bsi_B^l  (B_j)^{ln}\psi_B^m)|A,B\rangle
\nonumber\\
&&\hspace{-1mm}
+~if^{abc}\langle A,B|(\Bsi_A^m (t^c)^{lk}A_i^a\psi_A^k)(\Bsi_B^l  B_j^b\psi_B^{m})|A,B\rangle
\nonumber\\
&&\hspace{-1mm}
=~{1\over N_c}\langle A|\Bsi_A A_i\psi_A|A\rangle\langle B|\Bsi_B B_j\psi_B|B\rangle
~+~2if^{abc}\langle A|\Bsi_A t^dt^cA^a_i\psi_A|A\rangle\langle B|\Bsi_B t^dB^b_j\psi_B|B\rangle
\nonumber\\
&&\hspace{-1mm}
=~-{1\over N_c(N_c^2-1)}\langle A|\Bsi_A A_i\psi_A|A\rangle\langle B|\Bsi_B B_j\psi_B|B\rangle
\label{4.35}
\end{eqnarray}
we obtain
\begin{eqnarray}
&&\hspace{-1mm}
-g^{-2}N_c(N_c^2-1)[\bar\psi_A(x)\hamma_\mu\psi_{B}(x)\big]\big[\Bxi_{1B}(0)\hamma^\mu\Xi_{2A}(0)\big]
\label{4.36}\\
&&\hspace{-1mm}
=~{1+a^2\over 2s^2}
\big\{\big[\bar\psi_A(x)\gamma_i\slashed{p}_2\gamma^j A_k(0){1\over\alpha}\psi_A(0)\big]
\big[\big(\bar\psi_B{1\over\beta}\big)(0)\gamma^k\slashed{p}_1\gamma^i B_j(0)\psi_B(x)\big]
\nonumber\\
&&\hspace{55mm}
+~(\gamma_i\otimes\gamma^i\leftrightarrow\gamma_i\gamma_5\otimes\gamma^i\gamma_5)\big\}
\nonumber\\
&&\hspace{-1mm}
+~{a^2-1\over s^2}\big\{\big[\bar\psi_A(x)\slashed{p}_2\gamma^jA_k(0){1\over\alpha}\psi_A(0)\big]
\big[\big(\bar\psi_B{1\over\beta}\big)(0)\slashed{p}_1\gamma^kB_j(0)\psi_B(x)\big]
\nonumber\\
&&\hspace{55mm}
-~(\gamma^j\otimes\gamma^k\leftrightarrow\gamma^j\gamma_5\otimes\gamma^k\gamma_5)\big\}\nonumber\\
&&\hspace{-1mm}
-~{a\over s^2}\big\{\big[\bar\psi_A(x)\gamma_i\slashed{p}_2\gamma^jA_k(0){1\over\alpha}\psi_A(0)\big]
\big[\big(\bar\psi_B{1\over\beta}\big)(0)\gamma^k\slashed{p}_1\gamma^i\gamma_5B_j(0)\psi_B(x)\big]
\nonumber\\
&&\hspace{55mm}
+~(\gamma_i\otimes\gamma^i\gamma_5\leftrightarrow\gamma_i\gamma_5\otimes\gamma^i)\big\}
~+~O\big({m_\perp^8\over s}\big),
\nonumber
\end{eqnarray}
which can be rewritten as
\begin{eqnarray}
&&\hspace{-1mm}
-g^{-2}N_c(N_c^2-1)[\bar\psi_A(x)\hamma_\mu\psi_{B}(x)\big]\big[\Bxi_{1B}(0)\hamma^\mu\Xi_{2A}(0)\big]
=~{a^2-1\over s^2}\big[\bar\psi_A(x)A_k(0)\slashed{p}_2\gamma_j{1\over\alpha}\psi_A(0)\big]
\nonumber\\
&&\hspace{-1mm}
\big[\big(\bar\psi_B{1\over\beta}\big)(0)(B^j(0)\slashed{p}_1\gamma^k-j\leftrightarrow k+g^{jk}B^i(0)\slashed{p}_1\gamma_i)\psi_B(x)\big]
~+~O\big({m_\perp^8\over s}\big),
\label{4.37}
\end{eqnarray}
where we again used formulas (\ref{9.9}), (\ref{9.11}), and (\ref{9.13}) from appendix \ref{diracs}.

Next, it is easy to see that ${1\over\alpha}$ and ${1\over\beta}$ in eq. (\ref{4.36}) give ${1\over\alpha_q}$ and $-{1\over\beta_q}$:
\begin{eqnarray}
&&\hspace{-1mm}
\int\! dx_\bu~e^{-i\alpha_qx_\bu}\langle A|\hat{\bar\psi}(x_\bu,x_\perp)\Gamma\hatA_i(0){1\over\alpha}\hsi(0)|A\rangle
\label{4.38}\\
&&\hspace{-1mm}
=~{1\over\alpha_q}\!\int\! dx_\bu~e^{-i\alpha_qx_\bu}\langle A|\hat{\bar\psi}(x_\bu,x_\perp)\Gamma
\Big[\hatA_i(0)\hsi(0)+{2\over s}\hat{F}_{\ast i}(0)\!\int_{-\infty}^0\!\! dx'_\bu~\hsi(x'_\bu,0_\perp)\Big]|A\rangle,
\nonumber\\
&&\hspace{-1mm}
\int\! dx_\ast~e^{-i\beta_qx_\ast}\langle B|\big(\hat{\bar\psi}{1\over\beta}\big)(0)\hatA_i(0)\Gamma\hsi(x_\ast,x_\perp)|B\rangle
\nonumber\\
&&\hspace{-1mm}
=~-{1\over\beta_q}\!\int\! dx_\ast~e^{-i\beta_qx_\ast}\langle B|
\Big[\hat{\bar\psi}(0)\hatA_i(0)+\!\int_{-\infty}^0\!\!dx'_\ast
~\hat{\bar\psi}(x'_\ast,0_\perp){2\over s}\hat{F}_{\bu i}(0)\Big]\Gamma\hsi(x_\ast,x_\perp)|B\rangle,
\nonumber
\end{eqnarray}
where $\Gamma$ is any of the Dirac matrices in eq. (\ref{4.36}).

The corresponding contribution to 
$W(\alpha_q,\beta_q,x_\perp)$ takes the form
\begin{eqnarray}
&&\hspace{-1mm}
{e^2g^2\over 8(2\pi)^4s^2_Wc_W^2N_c(N_c^2-1)Q^2}\!\int\! dx_\bu dx_\ast~e^{-i\alpha_qx_\bu-i\beta_q x_\ast}
\label{5.33}\\
&&\hspace{-1mm}
\times~{a^2-1\over s^2}\langle A|\hat{\bar\psi}(x_\bu,x_\perp)\slashed{p}_2\gamma^j
\Big[\hatA_j(0)\hsi(0)+{2\over s}\hat{F}_{\ast j}(0)\!\int_{-\infty}^0\!\! dx'_\bu~\hsi(x'_\bu,0_\perp)\Big]|A\rangle
\nonumber\\
&&\hspace{-1mm}
\times~\langle B|\Big[\hat{\bar\psi}(0)\hatA_k(0)+\!\int_{-\infty}^0\!\!dx'_\ast
~\hat{\bar\psi}(x'_\ast,0_\perp){2\over s}\hat{F}_{\bu k}(0)\Big]\slashed{p}_1\gamma^k\hsi(x)|B\rangle\Big(1 
+ O\big({m_\perp^2\over s}\big)\Big),
\nonumber
\end{eqnarray}
where we have used the fact that 
\begin{equation}
\hspace{-1mm}   
\langle B|\big\{\big[\hat{\bar\psi}(0)\hatA_j(0)
+\!\int_{-\infty}^0\!\!dx'_\ast
~\hat{\bar\psi}(x'_\ast,0_\perp){2\over s}\hat{F}_{\bu j}(0)\big]\slashed{p}_1\gamma_k-j\leftrightarrow k\big\}
\hsi(x_\ast,x_\perp)|B\rangle~=~0     
\end{equation}
for the unpolarized hadron.

Similarly,
\begin{eqnarray}
&&\hspace{-3mm}
-g^{-2}N_c(N_c^2-1)[\Bxi_{2A}(x)\hamma_\mu\Xi_{1B}(x)\big]\big[\bar\psi_B(0)\hamma^\mu\psi_A(0)\big]
~=~
{a^2-1\over s^2}\big[\big(\bar\psi_A{1\over\alpha}\big)(x)A_k(x)\slashed{p}_2\gamma_j\psi_A(0)\big]
\nonumber\\
&&\hspace{-3mm}
\times~\big[\bar\psi_B(0)(B^j(x)\slashed{p}_1\gamma^k-j\leftrightarrow k+g^{jk}B^i(x)\slashed{p}_1\gamma_i){1\over\beta}\psi_B(x)\big]
~+~O\big({m_\perp^8\over s}\big),
\label{5.26}
\end{eqnarray}
so the corresponding contribution to 
$W(\alpha_q,\beta_q,x_\perp)$ is 
\begin{eqnarray}
&&\hspace{-1mm}
{g^2e^2\over 8(2\pi)^4s_W^2c_W^2N_c(N_c^2-1)Q^2}{a^2-1\over s^2}\!\int\! dx_\bu dx_\ast~e^{-i\alpha_qx_\bu-i\beta_q x_\ast}
\nonumber\\
&&\hspace{-1mm}
\times~\langle A|
\Big[\hat{\bar\psi}(x_\bu,x_\perp)\hatA_j(x_\bu,x_\perp)
+\!\int_{-\infty}^{x_\bu}\!\!dx'_\bu
~\hat{\bar{\psi}}(x'_\bu,x_\perp){2\over s}\hat{F}_{\ast j}(x_\bu,x_\perp)\Big]\slashed{p}_2\gamma^j\hsi(0)
|A\rangle
\nonumber\\
&&\hspace{-1mm}
\times~\langle B|\hat{\bar\psi}(0)\slashed{p}_1\gamma^k\Big[\hatA_k(x_\ast,x_\perp)\hsi(x_\ast,x_\perp)
+\!\int_{-\infty}^{x_\ast}\!\!dx'_\ast
~{2\over s}\hat{F}_{\bu k}(x_\ast,x_\perp)\hsi(x'_\ast,x_\perp)\Big]|B\rangle\Big(1 
+ O\big({m_\perp^2\over s}\big)\Big).
\nonumber
\end{eqnarray}

Using parametrizations (\ref{9.33}) and (\ref{paramh34})  from appendix \ref{param7} we obtain the contribution of the 6th line in  eq. (\ref{6lines}) in the form
\begin{eqnarray}
&&\hspace{-1mm}
W^{\rm 6th}(\alpha_q,\beta_q,q_\perp)
~=~-{e^2\over 4s_W^2c_W^2N_c(N_c^2-1)Q^2}\!\int\! d^2k_\perp ~k_\perp^2(q-k)_\perp^2
\label{2ndPC}\\
&&\hspace{-1mm}
\times~\Big[\Big\{{1\over m^2_N}(a_u^2-1)[h_{u}^{\rm tw3}(\alpha_q,k_\perp)
\barh_{u}^{\rm tw3}(\beta_q,q_\perp-k_\perp) 
+\tilh_{u}^{\rm tw3}(\alpha_q,k_\perp)\tilde\barh_{u}^{\rm tw3}(\beta_q,q_\perp-k_\perp)]
\nonumber\\
&&\hspace{-1mm}
+~(\alpha_q\leftrightarrow\beta_q)\Big\}~+~\Big\{u\leftrightarrow c\Big\}+\Big\{u\leftrightarrow d\Big\}+\Big\{u\leftrightarrow s\Big\}\Big]\Big(1 
+ O\big({m_\perp^2\over s}\big)\Big),
\nonumber
\end{eqnarray}
where quark$\leftrightarrow$antiquark ($\alpha_q\leftrightarrow\beta_q$) term comes from $x\leftrightarrow 0$ replacement,
cf. eq. (\ref{firstPC}).

\subsubsection{Parametrization of matrix elements from section \ref{sixline} \label{param7}}

In this section we present parametrization of matrix elements from section \ref{sixline}. Similarly  to eqs. (\ref{9.28})-(\ref{9.30}) we define

\begin{eqnarray}
&&\hspace{-1mm}
{g\over 8\pi^3s}\!\int\! dx_\perp dx_\bu~e^{-i\alpha x_\bu+i(k,x)_\perp}
\langle A|\hat{\bsi}_f(x_\bu,x_\perp)\slashed{p}_2\gamma^i\Big\{\hatA_i(0)\hsi_f(0)
\label{9.33}\\
&&\hspace{11mm}
+~{2\over s}\hat{F}_{\ast i}(0)\!\int_{-\infty}^0\!\! dx'_\bu~\hsi_f(x'_\bu,0_\perp)\Big\}|A\rangle
~=~
i{k_\perp^2\over m_N}
\big[h_{f}^{\rm tw3}(\alpha,k_\perp^2)+i\tilh_{f}^{\rm tw3}(\alpha,k_\perp^2)\big],
\nonumber\\
&&\hspace{-1mm}
{g\over 8\pi^3s}\!\int\! dx_\perp dx_\bu~e^{-i\alpha x_\bu+i(k,x)_\perp}
\langle A|\hat{\bsi}_f(0)\slashed{p}_2\gamma^i\Big\{\hatA_i(x_\bu,x_\perp)\hsi_f(x_\bu,x_\perp)
\nonumber\\
&&\hspace{11mm}
+~{2\over s}\hat{F}_{\ast i}(x_\bu,x_\perp)
\!\int_{-\infty}^{x_\bu}\!\! dx'_\bu~\hsi_f(x'_\bu,x_\perp)\Big\}|A\rangle
~=~i{k_\perp^2\over m_N}
\big[\barh_{f}^{\rm tw3}(\alpha,k_\perp^2)+i\tilde\barh_{f}^{\rm tw3}(\alpha,k_\perp^2)\big]
\nonumber
\end{eqnarray}
and similarly for the target matrix elements.
Note that unlike two-quark matrix elements,
quark-quark-gluon ones may have  imaginary parts which we denote by functions with tildes.

By complex conjugation we get
\begin{eqnarray}
&&\hspace{-2mm}
{g\over 8\pi^3s}\!\int\! dx_\perp dx_\bu~e^{-i\alpha x_\bu+i(k,x)_\perp}
\langle A|\Big\{\hat{\bsi}_f(x_\bu,x_\perp)\hatA_i(x_\bu,x_\perp)
\label{paramh34}\\
&&\hspace{2mm}
+\!\int_{-\infty}^{x_\bu}\!\!dx'_\bu
~\hat{\bsi}_f(x'_\bu,x_\perp){2\over s}\hat{F}_{\ast i}(x_\bu,x_\perp)\Big\}\slashed{p}_2\gamma^i\hsi_f(0)|A\rangle
~
=~
i{k_\perp^2\over m_N}
\big[h_{f}^{\rm tw3}(\alpha,k_\perp^2)-i\tilh_{f}^{\rm tw3}(\alpha,k_\perp^2)\big],
\nonumber\\
&&\hspace{-2mm}
{g\over 8\pi^3s}\!\int\! dx_\perp dx_\bu~e^{-i\alpha x_\bu+i(k,x)_\perp}
\langle A|\Big\{\hat{\bsi}_f(0)\hatA_i(0)
\nonumber\\
&&\hspace{2mm}
+\!\int_{-\infty}^0\!\!dx'_\bu
~\hat{\bsi}_f(x'_\bu,0_\perp){2\over s}\hat{F}_{\ast i}(0)\Big\}\slashed{p}_2\gamma^i\hsi_f(x_\bu,x_\perp)|A\rangle
~
=~i{k_\perp^2\over m_N}
\big[\barh_{f}^{\rm tw3}(\alpha,k_\perp^2)-i\tilde\barh_{f}^{\rm tw3}(\alpha,k_\perp^2)\big]
\nonumber
\end{eqnarray}
and similarly for the target matrix elements.

For completeness, let us present the structure of gauge links in an arbitrary gauge, for example:
\begin{eqnarray}
&&\hspace{-1mm}
\langle A|\Big\{\hat{\bsi}_f(x_\bu,x_\perp)\hatA_j(x_\bu,x_\perp)
+\!\int_{-\infty}^{x_\bu}\!\!dx'_\bu
~\hat{\bsi}_f(x'_\bu,x_\perp){2\over s}\hat{F}_{\ast j}(x_\bu,x_\perp)\Big\}\slashed{p}_2\gamma_i\hsi_f(0)|A\rangle
\label{gaugelinkh3}\\
&&\hspace{-1mm}   
\rightarrow~{2\over s}\!\int_{-\infty}^{x_\bu}\!\!dx'_\bu\langle A|\Big\{\hat{\bsi}_f(x_\bu,x_\perp)[x_\bu,x'_\bu]_x
\hatF_{\ast j}(x'_\bu,x_\perp)[x'_\bu,-\infty]_x
\nonumber\\
&&\hspace{-1mm}   
+~\hat{\bsi}_f(x'_\bu,x_\perp)[x'_\bu,x_\bu]_x\hat{F}_{\ast j}(x_\bu,x_\perp)[x_\bu,-\infty]_x\Big\}
[x_\perp,0_\perp]_{-\infty_\bu}[-\infty_\bu, 0_\bu]_{0_\perp}\slashed{p}_2\gamma_i\hsi_f(0)|A\rangle.
\nonumber
\end{eqnarray}

\subsection{Power corrections from $\calj_{A}^\mu(x)\calj_{B\mu}(0)$ terms \label{2ndtype}}
Power corrections of the second type come from the terms
\begin{eqnarray}
&&\hspace{-1mm}
\Bsi_A(x)\hamma^\mu\Psi_A(x)\Bsi_B(0)\hamma_\mu\Psi_B(0)~+~x\leftrightarrow 0
~=~\Bsi_A^{(0)}(x)\hamma^\mu\Psi_A^{(0)}(x)\Bsi_B^{(0)}(0)\hamma_\mu\Psi_B^{(0)}(0)
\\
&&\hspace{-1mm}
+~\Bsi_A^{(1)}(x)\hamma^\mu\Psi_A^{(0)}(x)\Bsi_B^{(0)}(0)\hamma_\mu\Psi_B^{(0)}(0)
+\Bsi_A^{(0)}(x)\hamma^\mu\Psi_A^{(1)}(x)\Bsi_B^{(0)}(0)\hamma_\mu\Psi_B^{(0)}(0)
\nonumber\\
&&\hspace{-1mm}
+~\Bsi_A^{(0)}(x)\hamma^\mu\Psi_A^{(0)}(x)\Bsi_B^{(1)}(0)\hamma_\mu\Psi_B^{(0)}(0)
+\Bsi_A^{(0)}(x)\hamma^\mu\Psi_A^{(0)}(x)\Bsi_B^{(0)}(0)\hamma_\mu\Psi_B^{(1)}(0)~+~x\leftrightarrow 0~+~...
\nonumber
\end{eqnarray}
In appendix \ref{psi1sup}, we will demonstrate that terms $\sim \Psi^{(1)}$ are small in our kinematical
region $s\gg Q^2\gg q_\perp^2$. 

Terms $\sim  \Psi^{(0)}$ read
\begin{eqnarray}
&&\hspace{-1mm}
\big[\big(\bar\psi_A +\Bxi_{2A}\big)(x)\hamma_\mu\big(\psi_A+\Xi_{2A}\big)(x)\big]
[\big(\bar\psi_B+\Bxi_{1B}\big)(0)\hamma^\mu\big(\psi_B+\Xi_{1B}\big)(0)\big]~+~x\leftrightarrow 0
\nonumber\\
&&\hspace{-1mm}
=~[\bar\psi_A(x)\hamma_\mu\psi_A(x)\big]\big[\bar\psi_B(0)\hamma^\mu\psi_B(0)\big]
\label{7newlines}\\
&&\hspace{-1mm}
+~[\Bxi_{2A}(x)\hamma_\mu\psi_A(x)\big]\big[\bar\psi_B(0)\hamma^\mu\psi_B(0)\big]
+[\bar\psi_A(x)\hamma_\mu\Xi_{2A}(x)\big]\big[\bar\psi_B(0)\hamma^\mu\psi_B(0)\big]
\nonumber\\
&&\hspace{-1mm}
+~
[\bar\psi_A(x)\hamma_\mu\psi_A(x)\big]\big[\Bxi_{1B}(0)\hamma^\mu\psi_B(0)\big]
+[\bar\psi_A(x)\hamma_\mu\psi_A(x)\big]\big[\bar\psi_B(0)\hamma^\mu\Xi_{1B}(0)\big]
\nonumber\\
&&\hspace{-1mm}
+~[\Bxi_{2A}(x)\hamma_\mu\Xi_{2A}(x)\big]\big[\bar\psi_B(0)\hamma^\mu\psi_B(0)\big]
+[\bar\psi_A(x)\hamma_\mu\psi_{A}(x)\big]\big[\Bxi_{1B}(0)\hamma^\mu\Xi_{1B}(0)\big]
\nonumber\\
&&\hspace{-1mm}
+~[\Bxi_{2A}(x)\hamma_\mu\psi_A(x)\big]\big[\bar\psi_B(0)\hamma^\mu\Xi_{1B}(0)\big]
+[\bar\psi_A(x)\hamma_\mu\Xi_{2A}(x)\big]\big[\Bxi_{1B}(0)\hamma^\mu\psi_B(0)\big]
\nonumber\\
&&\hspace{-1mm}
+~[\Bxi_{2A}(x)\hamma_\mu\psi_A(x)\big]\big[\Bxi_{1B}(0)\hamma^\mu\psi_B(0)\big]
+[\bar\psi_A(x)\hamma_\mu\Xi_{2A}(x)\big]\big[\bar\psi_B(0)\hamma^\mu\Xi_{1B}(0)\big]
~+~x\leftrightarrow 0.
\nonumber
\end{eqnarray}
As we prove in appendix \ref{secfivap}, the leading power correction comes from last two lines in eq. (\ref{7newlines}). We will consider them in turn.

\subsubsection{Last two lines  in eq. (\ref{7newlines}) \label{67lines}}

Using eq. (\ref{fildz0}) and separating color-singlet matrix elements, we 
rewrite the  sixth  line in eq. (\ref{7newlines}) as
\begin{eqnarray}
&&\hspace{-1mm}
[\Bxi_{2A}(x)\hamma_\mu\psi_A(x)\big]\big[\bar\psi_B(0)\hamma^\mu\Xi_{1B}(0)\big]
+[\bar\psi_A(x)\hamma_\mu\Xi_{2A}(x)\big]\big[\Bxi_{1B}(0)\hamma^\mu\psi_A(0)\big]
\\
&&\hspace{-1mm}
=~{g^2\over (N_c^2-1)s^2}\Big(\big[\big(\bsi_A{1\over\alpha}\big)(x)\gamma^j\slashed{p}_2\hamma_\mu A_k(0)\psi_A(x)\big]
\big[\bsi_B(0)\hamma^\mu\slashed{p}_1\gamma^k B_j(x){1\over\beta}\psi_B(0)\big]
\nonumber\\
&&\hspace{-1mm}
+~
\big[\bsi_A(x)\hamma^\mu\slashed{p}_2\gamma^j A_k(0){1\over\alpha}\psi_A(x)\big]
\big[\big(\bsi_B{1\over\beta}\big)(0)\gamma^k\slashed{p}_1\hamma_\mu B_j(x)\psi_B(0)\big]\Big)
~+~x\leftrightarrow 0
\nonumber\\
&&\hspace{-1mm}
=~{g^2(a^2-1)\over (N_c^2-1)s^2}
\Big(\big[\big(\bsi_A{1\over\alpha}\big)(x)\slashed{p}_2(A_i+i\gamma_5\tilde{A}_i)(0)\psi_A(x)\big]
\big[\bar\psi_B(0)\slashed{p}_1(B^i(x)-i\gamma_5\tilde{B}^i)(x){1\over\beta}\psi_B(0)\big]
\nonumber\\
&&\hspace{-1mm}
+\big[\bar\psi_A(x)\slashed{p}_2[A_i(0)-i\gamma_5\tilde{A}_i(0)]{1\over \alpha}\psi_A(x)\big]
\big[\big(\bar\psi_B{1\over\beta}\big)(0)\slashed{p}_1[B^i(x)+i\gamma_5\tilde{B}^i(x)]\psi_B(0)\big]\Big)
~+~x\leftrightarrow 0,
\nonumber
\end{eqnarray}
where we used eqs. (\ref{formula}) and (\ref{9.11}).
For the forward matrix elements
\begin{eqnarray}
&&\hspace{-1mm}
\int\! dx_\bu~e^{-i\alpha_q x_\bu}\langle A|\big(\hat{\bar\psi}{1\over\alpha}\big)(x_\bu,x_\perp)\slashed{p}_2\hatA_i(0)\hsi(x_\bu,x_\perp)|A\rangle
\nonumber\\
&&\hspace{22mm}
=~-{1\over\alpha_q}\!\int\! dx_\bu~e^{-i\alpha_q x_\bu}
\!\int_{-\infty}^{x_\bu}\! dx'_\bu\langle A|\hat{\bar\psi}(x'_\bu,x_\perp){2\slashed{p}_2\over s}\hatF_{\ast i}(0)\hsi(x_\bu,x_\perp)|A\rangle,
\nonumber\\
&&\hspace{-1mm}
\int\! dx_\bu~e^{-i\alpha_q x_\bu}\langle A|\hat{\bar\psi}(x_\bu,x_\perp)\slashed{p}_2\hatA_i(0){1\over\alpha}\hsi(x_\bu,x_\perp)|A\rangle
\nonumber\\
&&\hspace{22mm}
=~{1\over\alpha_q}\!\int\! dx_\bu~e^{-i\alpha_q x_\bu}
\!\int_{-\infty}^{x_\bu}\! dx'_\bu\langle A|\hat{\bar\psi}(x_\bu,x_\perp){2\slashed{p}_2\over s}\hatF_{\ast i}(0)\hsi(x'_\bu,x_\perp)|A\rangle,
\nonumber\\
&&\hspace{-1mm}
\int\! dx_\ast~e^{-i\beta_q x_\ast}\langle B|\hat{\bar\psi}(0)\slashed{p}_1\hatA_i(x_\ast,x_\perp){1\over\beta}\hsi(0)|B\rangle
\nonumber\\
&&\hspace{-1mm}
=~-{1\over\beta_q}\!\int\! dx_\ast~e^{-i\beta_q x_\ast}
\!\int_{-\infty}^0\! dx'_\ast\langle B|\hat{\bar\psi}(0){2\slashed{p}_1\over s}\hatF_{\bu i}(x_\ast,x_\perp)\hsi(x'_\ast,0_\perp)|B\rangle,
\nonumber\\
&&\hspace{-1mm}
\int\! dx_\ast~e^{-i\beta_q x_\ast}\langle B|\big(\hat{\bar\psi}{1\over\beta}\big)(0)\slashed{p}_1\hatA_i(x_\ast,x_\perp)\hsi(0)|B\rangle
\nonumber\\
&&\hspace{-1mm}
=~{1\over\beta_q}\!\int\! dx_\ast~e^{-i\beta_q x_\ast}
\!\int_{-\infty}^0\! dx'_\ast\langle B|\hat{\bar\psi}(x'_\ast,0_\perp){2\slashed{p}_1\over s}\hatF_{\bu i}(x_\ast,x_\perp)\hsi(0)|B\rangle.
\label{forwards}
\end{eqnarray}
The corresponding contribution to 
$W(\alpha_q,\beta_q,x_\perp)$ takes the form
\begin{eqnarray}
&&\hspace{-1mm}
{g^2e^2(a^2-1)\over 8(2\pi)^4s_W^2c_W^2(N_c^2-1)Q^2s^2}\!\int\! dx_\bu dx_\ast~e^{-i\alpha_qx_\bu-i\beta_q x_\ast}
\label{67term}\\
&&\hspace{-1mm}
\times~\Big\{\!\int_{-\infty}^{x_\bu}\! dx'_\bu\!\int_{-\infty}^0\! dx'_\ast\Big[
\langle A|\hat{\bar\psi}(x'_\bu,x_\perp){2\slashed{p}_2\over s}
\big[\hatF_{\ast i}(0)+i\gamma_5\hat\tilF_{\ast i}(0)\big]\hsi(x_\bu,x_\perp)|A\rangle
\nonumber\\
&&\hspace{22mm}
\times~\langle B|\hat{\bar\psi}(0){2\slashed{p}_1\over s}
\big[\hatF_{\bu}^{\ i}(x_\ast,x_\perp)-i\gamma_5\hat\tilF_{\bu}^{\ i}(x_\ast,x_\perp)\big]\hsi(x'_\ast,0_\perp)|B\rangle
\nonumber\\
&&\hspace{-1mm}
+~\langle A|\hat{\bar\psi}(x_\bu,x_\perp){2\slashed{p}_2\over s}
\big[\hatF_{\ast i}(0)-i\gamma_5\hat\tilF_{\ast i}(0)\big]\hsi(x'_\bu,x_\perp)|A\rangle
\nonumber\\
&&\hspace{-1mm}
\times~\langle B|
\hat{\bar\psi}(x'_\ast,0_\perp){2\slashed{p}_1\over s}
\big[\hatF_{\bu}^{\ i}(x_\ast,x_\perp)+i\gamma_5\hat\tilF_{\bu}^{\ i}(x_\ast,x_\perp)\big]\hsi(0)
|B\rangle\Big]
~+~x\leftrightarrow 0\Big\}
\Big(1~+~O\big({m_\perp^2\over s}\big)\Big).
\nonumber
\end{eqnarray}

Similarly, for the seventh line in eq. (\ref{7newlines}) using eqs. (\ref{fildz0}) and (\ref{9.12}) one obtains
\begin{eqnarray}
&&\hspace{-1mm}
[\Bxi_{2A}(x)\hamma_\mu\psi_A(x)\big]\big[\Bxi_{1B}(0)\hamma^\mu\psi_B(0)\big]
+[\bar\psi_A(x)\hamma_\mu\Xi_{2A}(x)\big]\big[\bar\psi_B(0)\hamma^\mu\Xi_{1B}(0)\big]
~+~x\leftrightarrow 0
\label{4.50}\\
&&\hspace{-1mm}
=~{g^2\over (N_c^2-1)s^2}\Big(\big[\big(\bsi_A{1\over\alpha}\big)(x)\gamma^j\slashed{p}_2\hamma_\mu A_k(0)\psi_A(x)\big]
\big[\big(\bsi_B{1\over\beta}\big)(0)\gamma^k\slashed{p}_1\hamma^\mu B_j(x)\psi_B(0)\big]
\nonumber\\
&&\hspace{-1mm}
+~
\big[\bsi_A(x)\hamma_\mu\slashed{p}_2\gamma^j A_k(0){1\over\alpha}\psi_A(x)\big]
\big[\bsi_B(0)\hamma^\mu\slashed{p}_1\gamma^k B_j(x){1\over\beta}\psi_B(0)\big]
\Big)
~+~x\leftrightarrow 0
\nonumber\\
&&\hspace{-1mm}
=~{g^2\over (N_c^2-1)s^2}\Big((1+a^2)
\big\{\big[\big(\bsi_A{1\over\alpha}\big)(x)\slashed{p}_2[A_i(0)+i\gamma_5\tilde{A}_i(0)]\psi_A(x)\big]
\nonumber\\
&&\hspace{44mm}
\times~\big[\big(\bar\psi_B{1\over\beta}\big)(0)\slashed{p}_1[B^i(x)+i\gamma_5\tilde{B}^i(x)]\psi_B(0)\big]
\nonumber\\
&&\hspace{-1mm}
+~\big[\bar\psi_A(x)\slashed{p}_2[A_i(0)-i\gamma_5\tilde{A}_i(0)]\big({1\over \alpha}\psi_A\big)(x)\big]
\big[\bar\psi_B(0)\slashed{p}_1[B^i(x)-i\gamma_5\tilde{B}^i(x)]\big({1\over\beta}\psi_B\big)(0)\big]\big\}
\nonumber\\
&&\hspace{-1mm}
-~2a
\big\{\big[\big(\bsi_A{1\over\alpha}\big)(x)\slashed{p}_2[A_i(0)+i\gamma_5\tilde{A}_i(0)]\psi_A(x)\big]
\big[\big(\bar\psi_B{1\over\beta}\big)(0)\slashed{p}_1[\gamma_5B^i(x)+i\tilde{B}^i(x)]\psi_B(0)\big]
\nonumber\\
&&\hspace{-1mm}
+~\big[\bar\psi_A(x)\slashed{p}_2[A_i(0)-i\gamma_5\tilde{A}_i(0)]\big({1\over \alpha}\psi_A\big)(x)\big]
\big[\bar\psi_B(0)\slashed{p}_1[\gamma_5B^i(x)-i\tilde{B}^i(x)]\big({1\over\beta}\psi_B\big)(0)\big]\big\}\Big)
+~x\leftrightarrow 0.
\nonumber
\end{eqnarray}

Using eq. (\ref{forwards}) one obtains the contribution to $W(\alpha_q,\beta_q,x_\perp)$ in the form
\begin{eqnarray}
&&\hspace{-1mm}
-{g^2e^2(a^2+1)\over 8(2\pi)^4s_W^2c_W^2(N_c^2-1)Q^2s^2}\!\int\! dx_\bu dx_\ast~e^{-i\alpha_qx_\bu-i\beta_q x_\ast}
\label{7term}\\
&&\hspace{-1mm}
\times~\Big\{\!\int_{-\infty}^{x_\bu}\! dx'_\bu\!\int_{-\infty}^0\! dx'_\ast
\Big[
\langle A|\hat{\bar\psi}(x'_\bu,x_\perp){2\slashed{p}_2\over s}
\big[\hatF_{\ast i}(0)+i\gamma_5\hat\tilF_{\ast i}(0)\big]\hsi(x_\bu,x_\perp)|A\rangle
\nonumber\\
&&\hspace{22mm}
\times~\langle B|
\hat{\bar\psi}(x'_\ast,0_\perp){2\slashed{p}_1\over s}
\big[\hatF_{\bu}^{\ i}(x_\ast,x_\perp)+i\gamma_5\hat\tilF_{\bu}^{\ i}(x_\ast,x_\perp)\big]\hsi(0)|B\rangle
\nonumber\\
&&\hspace{-1mm}
+~\langle A|\hat{\bar\psi}(x_\bu,x_\perp){2\slashed{p}_2\over s}
\big[\hatF_{\ast i}(0)-i\gamma_5\hat\tilF_{\ast i}(0)\big]\hsi(x'_\bu,x_\perp)|A\rangle
\nonumber\\
&&\hspace{22mm}
\times~
\langle B|\hat{\bar\psi}(0){2\slashed{p}_1\over s}
\big[\hatF_{\bu}^{\ i}(x_\ast,x_\perp)-i\gamma_5\hat\tilF_{\bu}^{\ i}(x_\ast,x_\perp)\big]\hsi(x'_\ast,0_\perp)|B\rangle\Big]
~+~x\leftrightarrow 0\Big\}.
\nonumber
\end{eqnarray}
Here we used the fact that the last term in eq. (\ref{4.50}) $\big(\sim{2a\over s^2}\big)$ 
\begin{eqnarray}
&&\hspace{-1mm}
-~2a
\Big[
\langle A|\hat{\bar\psi}(x'_\bu,x_\perp){2\slashed{p}_2\over s}
\big[\hatF_{\ast i}(0)+i\gamma_5\hat\tilF_{\ast i}(0)\big]\hsi(x_\bu,x_\perp)|A\rangle
\nonumber\\
&&\hspace{22mm}
\times~\langle B|
\hat{\bar\psi}(x'_\ast,0_\perp){2\slashed{p}_1\over s}
\big[\gamma_5\hatF_{\bu}^{\ i}(x_\ast,x_\perp)+i\hat\tilF_{\bu}^{\ i}(x_\ast,x_\perp)\big]\hsi(0)|B\rangle
\nonumber\\
&&\hspace{22mm}
+~\langle A|\hat{\bar\psi}(x_\bu,x_\perp){2\slashed{p}_2\over s}
\big[\hatF_{\ast i}(0)-i\gamma_5\hat\tilF_{\ast i}(0)\big]\hsi(x'_\bu,x_\perp)|A\rangle
\nonumber\\
&&\hspace{22mm}
\times~
\langle B|\hat{\bar\psi}(0){2\slashed{p}_1\over s}
\big[\gamma_5\hatF_{\bu}^{\ i}(x_\ast,x_\perp)-i\hat\tilF_{\bu}^{\ i}(x_\ast,x_\perp)\big]\hsi(x'_\ast,0_\perp)|B\rangle\Big]
\end{eqnarray}
 gives no contribution since 
\begin{eqnarray}
&&\hspace{-1mm}
\langle A|\hat{\bar\psi}(x'_\bu,x_\perp){2\slashed{p}_2\over s}
\big[\hatF_{\ast i}(0)\pm i\gamma_5\hat\tilF_{\ast i}(0)\big]\hsi(x_\bu,x_\perp)|A\rangle
~\sim~x_i,
\nonumber\\
&&\hspace{-1mm}
\langle B|
\hat{\bar\psi}(x'_\ast,0_\perp){2\slashed{p}_1\over s}
\big[\gamma_5\hatF_{\bu i}(x_\ast,x_\perp)\pm i\hat\tilF_{\bu i}(x_\ast,x_\perp)\big]\hsi(0)|B\rangle~\sim~\epsilon_{ij}x^j
\end{eqnarray}
same as in eq. (\ref{4.29}).

Next, using parametrizations (\ref{lastparam})  from the next section we obtain the contribution of the 6th and 7th lines in eq. (\ref{7newlines}) 
in the form
\begin{eqnarray}
&&\hspace{-1mm}
W^{\rm 6+7th}(\alpha_q,\beta_q,q_\perp)
~=~{e^2\over 8s_W^2c_W^2(N_c^2-1)Q^2}\!\int\! d^2k_\perp ~(k,q-k)_\perp
\Big[\Big\{2(1+a_u^2)
\nonumber\\
&&\hspace{-1mm}
\times~
\big[j_{1u}^{\rm tw3}(\alpha_q,k_\perp)j_{2u}^{\rm tw3}(\beta_q,q_\perp-k_\perp) 
-~
\tilj_{1u}^{\rm tw3}(\alpha_q,k_\perp)\tilj_{2u}^{\rm tw3}(\beta_q,q_\perp-k_\perp)\big]
\nonumber\\
&&\hspace{-1mm}
+~
(1-a_u^2)
\big[j_{1u}^{\rm tw3}(\alpha_q,k_\perp)j_{1u}^{\rm tw3}(\beta_q,q_\perp-k_\perp) 
+\tilj_{1u}^{\rm tw3}(\alpha_q,k_\perp)\tilj_{1u}^{\rm tw3}(\beta_q,q_\perp-k_\perp)~
\nonumber\\
&&\hspace{-1mm}
+~j_{2u}^{\rm tw3}(\alpha_q,k_\perp)j_{2u}^{\rm tw3}(\beta_q,q_\perp-k_\perp) 
+\tilj_{2u}^{\rm tw3}(\alpha_q,k_\perp)\tilj_{2u}^{\rm tw3}(\beta_q,q_\perp-k_\perp)\big]~+~\alpha_q\leftrightarrow\beta_q\Big\}~
\nonumber\\
&&\hspace{-1mm}
+~\Big\{u\leftrightarrow c\Big\}+\Big\{u\leftrightarrow d\Big\}+\Big\{u\leftrightarrow s\Big\}\Big]\Big(1~+~O\big({m_\perp^2\over s}\big)\Big),
\label{thirdPC}
\end{eqnarray}
where  $\alpha_q\leftrightarrow\beta_q$ contribution comes  as usually from the $(x\leftrightarrow 0)$ term in eq. (\ref{67term}).

\subsubsection{Parametrization of TMDs from section \ref{67lines} \label{67param}}
We  parametrize TMDs from section \ref{67lines} as follows
\begin{eqnarray}
&&\hspace{-1mm}
{g\over 8\pi^3s}\!\int\! d^2x_\perp dx_\bu~e^{-i\alpha x_\bu+i(k,x)_\perp}
\!\int_{-\infty}^{x_\bu}\! dx'_\bu\langle A|\hat{\bsi}(x'_\bu,x_\perp){2\slashed{p}_2\over s}
\big[\hatF_{\ast i}(0)+i\gamma_5\hat\tilF_{\ast i}(0)\big]\hsi(x_\bu,x_\perp)|A\rangle
\nonumber\\
&&\hspace{-1mm}
=~k_i\big[j_1^{\rm tw3}(\alpha,k_\perp^2)+i\tilj_1^{\rm tw3}(\alpha,k_\perp^2)\big],
\nonumber\\
&&\hspace{-1mm}
{g\over 8\pi^3s}\!\int\! d^2x_\perp dx_\bu~e^{-i\alpha x_\bu+i(k,x)_\perp}
\!\int_{-\infty}^{x_\bu}\! dx'_\bu\langle A|\hat{\bsi}(x_\bu,x_\perp){2\slashed{p}_2\over s}
\big[\hatF_{\ast i}(0)-i\gamma_5\hat\tilF_{\ast i}(0)\big]\hsi(x'_\bu,x_\perp)|A\rangle
\nonumber\\
&&\hspace{-1mm}
=~k_i\big[j_2^{\rm tw3}(\alpha,k_\perp^2)-i\tilj_2^{\rm tw3}(\alpha,k_\perp^2)\big].
\end{eqnarray}
By complex conjugation we get
\begin{eqnarray}
&&\hspace{-1mm}
{g\over 8\pi^3s}\!\int\! d^2x_\perp dx_\bu~e^{-i\alpha x_\bu+i(k,x)_\perp}
\!\int_{-\infty}^0\! dx'_\bu\langle A|\bar\hsi(0){2\slashed{p}_2\over s}[\hatF_{\ast i}(x)-i\gamma_5\hat\tilF_{\ast i}(x)]\hsi(x'_\bu,0_\perp)|A\rangle
\nonumber\\
&&\hspace{-1mm}
=~k_i\big[j_1^{\rm tw3}(\alpha,k_\perp^2)-i\tilj_1^{\rm tw3}(\alpha,k_\perp^2)\big],
\nonumber\\
&&\hspace{-1mm}
{g\over 8\pi^3s}\!\int\! d^2x_\perp dx_\bu~e^{-i\alpha x_\bu+i(k,x)_\perp}
\!\int_{-\infty}^0\! dx'_\bu\langle A|\bar\hsi(x'_\bu,0_\perp){2\slashed{p}_2\over s}
\big[\hatF_{\ast i}(x)+i\gamma_5\hat\tilF_{\ast i}(x)\big]\hsi(0)|A\rangle
\nonumber\\
&&\hspace{-1mm}
=~k_i\big[j_2^{\rm tw3}(\alpha,k_\perp^2)+i\tilj_2^{\rm tw3}(\alpha,k_\perp^2)\big].
\label{lastparam}
\end{eqnarray}
Target matrix elements are obtained by usual substitutions $\alpha\leftrightarrow\beta$, $\slashed{p}_2\leftrightarrow\slashed{p}_1$, $x_\bu\leftrightarrow x_\ast$, and $\hatF_{\ast i}\leftrightarrow \hatF_{\bu i}$.

For completeness let us present the explicit form of the gauge links in an arbitrary gauge:
\begin{eqnarray}
&&\hspace{-1mm}
\hat{\bsi}(x'_\bu,x_\perp)
\hatF_{\ast i}(0)\hsi(x_\bu,x_\perp)
~\rightarrow~
\\
&&\hspace{-1mm}
\hat{\bsi}(x'_\bu,x_\perp)[x'_\bu,-\infty_\bu]_x[x_\perp,0_\perp]_{-\infty_\bu}[-\infty_\bu,0]_{0_\perp}\hatF_{\ast i}(0)
[0,-\infty_\bu]_{0_\perp}[0_\perp,x_\perp]_{-\infty_\bu}[-\infty_\bu,x_\bu]_x\hsi(x_\bu,x_\perp).
\nonumber
\end{eqnarray}

\section{Results and estimates \label{sec:results}}
Combining eqs. (\ref{WLT}), (\ref{firstPC}), (\ref{2ndPC}), and (\ref{thirdPC}) we get the leading term and
first power corrections to $W(q)$ in the kinematic region $s\gg Q^2\gg q_\perp^2$ in the form
\begin{eqnarray}
&&\hspace{-1mm}
W(p_A,p_B,q)
~=~-{e^2\over 8s_W^2c_W^2N_c}\!\int\! d^2k_\perp\bigg[\Big\{(1+a_u^2)
\Big[1-2{(k,q-k)_\perp\over Q^2}\Big]
\nonumber\\
&&\hspace{-1mm}
\times~
f_{1u}(\alpha_z,k_\perp)\barf_{1u}(\beta_z,q_\perp-k_\perp) 
+~2(a_u^2-1){k_\perp^2(q-k)_\perp^2\over m^2_NQ^2}
h^\perp_{1u}(\alpha_z,k_\perp)\barh^\perp_{1u}(\beta_z,q_\perp-k_\perp) 
\nonumber\\
&&\hspace{-1mm}
+~{2k_\perp^2(q-k)_\perp^2\over (N_c^2-1)Q^2m^2_N}(a_u^2-1)[h_{u}^{\rm tw3}(\alpha_z,k_\perp)
\barh_{u}^{\rm tw3}(\beta_z,q_\perp-k_\perp) 
+\tilh_{u}^{\rm tw3}(\alpha_z,k_\perp)\tilde\barh_{u}^{\rm tw3}(\beta_z,q_\perp-k_\perp)]
\nonumber\\
&&\hspace{-1mm}
-~{N_c\over N_c^2-1}{(k,q-k)_\perp\over Q^2}
\nonumber\\
&&\hspace{-1mm}
\times~
\Big(2(1+a_u^2)
\big[j_{1u}^{\rm tw3}(\alpha_z,k_\perp)j_{2u}^{\rm tw3}(\beta_z,q_\perp-k_\perp) 
-~
\tilj_{1u}^{\rm tw3}(\alpha_z,k_\perp)\tilj_{2u}^{\rm tw3}(\beta_z,q_\perp-k_\perp)\big]
\nonumber\\
&&\hspace{-1mm}
+~(1-a_u^2)\big[j_{1u}^{\rm tw3}(\alpha_z,k_\perp)j_{1u}^{\rm tw3}(\beta_z,q_\perp-k_\perp) 
+j_{2u}^{\rm tw3}(\alpha_z,k_\perp)j_{2u}^{\rm tw3}(\beta_z,q_\perp-k_\perp) 
\nonumber\\
&&\hspace{-1mm}
+~\tilj_{1u}^{\rm tw3}(\alpha_z,k_\perp)\tilj_{1u}^{\rm tw3}(\beta_z,q_\perp-k_\perp)
+\tilj_{2u}^{\rm tw3}(\alpha_z,k_\perp)\tilj_{2u}^{\rm tw3}(\beta_z,q_\perp-k_\perp)\big]\Big)
\nonumber\\
&&\hspace{-1mm}
+~(\alpha_z\leftrightarrow\beta_z)\Big\}
~+~\Big\{u\leftrightarrow c\Big\}+\Big\{u\leftrightarrow d\Big\}+\Big\{u\leftrightarrow s\Big\}\bigg]\Big(1
~+~O\big({m_\perp^2\over s}\big)\Big),
\label{resultZ}
\end{eqnarray}
where the momentum of the produced Z-boson is $q = \alpha_z p_1 + \beta_z p_2 + q_\perp$.

For completeness, let us present our final result in the transverse coordinate space
\begin{eqnarray}
&&\hspace{-1mm}
W(p_A,p_B,q)
~=~-{e^2\over 8s_W^2c_W^2N_c}\!\int\! {d^2b_\perp\over 4\pi^2}~e^{i(q,b)_\perp}
\bigg[\Big\{(1+a_u^2)\Big[f_{1u}(\alpha_z,b_\perp)\barf_{1u}(\beta_z,b_\perp) 
\nonumber\\
&&\hspace{7mm}
+~{2\over Q^2}\partial^\perp_if_{1u}(\alpha_z,b_\perp)\partial^\perp_i\barf_{1u}(\beta_z,b_\perp) \Big]
+{2(a_u^2-1)\over m^2_NQ^2}
\partial_\perp^2h^\perp_{1u}(\alpha_z,b_\perp)\partial_\perp^2\barh^\perp_{1u}(\beta_z,b_\perp) 
\nonumber\\
&&\hspace{-1mm}
+~{2(a_u^2-1)\over (N_c^2-1)Q^2m^2_N}[\partial_\perp^2h_{u}^{\rm tw3}(\alpha_z,b_\perp)
\partial_\perp^2\barh_{u}^{\rm tw3}(\beta_z,b_\perp) 
+\partial_\perp^2\tilh_{u}^{\rm tw3}(\alpha_z,b_\perp)\partial_\perp^2\tilde\barh_{u}^{\rm tw3}(\beta_z,b_\perp)]
\nonumber\\
&&\hspace{-1mm}
+~{N_c\over (N_c^2-1)Q^2}
\nonumber\\
&&\hspace{-1mm}
\times~
\Big(2(1+a_u^2)
\big[\partial^\perp_ij_{1u}^{\rm tw3}(\alpha_z,b_\perp)\partial^\perp_ij_{2u}^{\rm tw3}(\beta_z,b_\perp) 
-~
\partial^\perp_i\tilj_{1u}^{\rm tw3}(\alpha_z,b_\perp)\partial^\perp_i\tilj_{2u}^{\rm tw3}(\beta_z,b_\perp)\big]
\nonumber\\
&&\hspace{-1mm}
+~(1-a_u^2)\big[\partial^\perp_ij_{1u}^{\rm tw3}(\alpha_z,b_\perp)\partial^\perp_ij_{1u}^{\rm tw3}(\beta_z,b_\perp) 
+\partial^\perp_ij_{2u}^{\rm tw3}(\alpha_z,b_\perp)\partial^\perp_ij_{2u}^{\rm tw3}(\beta_z,b_\perp) 
\nonumber\\
&&\hspace{-1mm}
+~\partial^\perp_i\tilj_{1u}^{\rm tw3}(\alpha_z,b_\perp)\partial^\perp_i\tilj_{1u}^{\rm tw3}(\beta_z,b_\perp)
+\partial^\perp_i\tilj_{2u}^{\rm tw3}(\alpha_z,b_\perp)\partial^\perp_i\tilj_{2u}^{\rm tw3}(\beta_z,b_\perp)\big]\Big)
\nonumber\\
&&\hspace{-1mm}
+~(\alpha_z\leftrightarrow\beta_z)\Big\}
~+~\Big\{u\leftrightarrow c\Big\}+\Big\{u\leftrightarrow d\Big\}+\Big\{u\leftrightarrow s\Big\}\bigg]\Big(1
~+~O\big({m_\perp^2\over s}\big)\Big),
\label{resultZcoord}
\end{eqnarray}
where $f_{1u}(\alpha_z,b_\perp)\equiv\int\! d^2k_\perp e^{-i(k,b)_\perp}f_{1u}(\alpha_z,k_\perp)$ etc.

Note that in the leading order in $N_c$ power corrections are expressed in terms
of leading power functions $f_1$ and $h_1^\perp$. To estimate the order of magnitude
of power corrections, one can assume that ${1\over N_c}$ is a good parameter and leave only first
term in the r.h.s. of eq. (\ref{resultZ}): 
\begin{eqnarray}
&&\hspace{-1mm}
W(p_A,p_B,q)
~=~-{e^2\over 8s_W^2c_W^2N_c}\!\int\! d^2k_\perp\bigg[\Big\{(1+a_u^2)
\Big[1-2{(k,q-k)_\perp\over Q^2}\Big]
\label{resultN}\\
&&\hspace{7mm}
\times~
f_{1u}(\alpha_z,k_\perp)\barf_{1u}(\beta_z,q_\perp-k_\perp) +~2(a_u^2-1){k_\perp^2(q-k)_\perp^2\over m^2_NQ^2}
h^\perp_{1u}(\alpha_z,k_\perp)\barh^\perp_{1u}(\beta_z,q_\perp-k_\perp) 
\nonumber\\
&&\hspace{14mm}
+~(\alpha_z\leftrightarrow\beta_z)\Big\}
~+~\Big\{u\leftrightarrow c\Big\}+\Big\{u\leftrightarrow d\Big\}+\Big\{u\leftrightarrow s\Big\}\bigg]\Big(1
~+~O\big({m_\perp^2\over s}\big)~+~O\big({1\over N_c}\big)\Big).
\nonumber
\end{eqnarray}
Next, eq. (\ref{resultN}) is a tree-level formula and for an estimate we
should specify the rapidity cutoffs for $f_1$'s and $h_1^\perp$'s. As we discussed in section \ref{sec:funt},
the rapidity cutoff for $f_1(\alpha_z,k_\perp^2)$ is $\sigma_a$ and for $f_1(\alpha_z,k_\perp^2)$ $\sigma_b$, 
where $\sigma_a$ and $\sigma_b$ are rapidity bounds for central fields. Since we calculated only tree diagrams made of $C$-fields 
we have $\sigma_a=\beta_z$ and $\sigma_b=\alpha_z$ in eq. (\ref{resultZ}). 
\footnote{In general, we should integrate over $C$-fields in the leading log approximation and match the logs to the 
double-log and/or single-log evolution of TMDs.}

Next, power corrections become sizable at $q_\perp^2\gg m^2_N$ where we probe the perturbative tails of TMD's $f_1\sim{1\over k_\perp^2}$ and 
$h_1^\perp\sim{1\over k_\perp^4}$ \cite{Zhou:2008fb}. 
So, as long as  $m^2_N \ll k_\perp^2 \ll Q^2$ we can approximate
\begin{equation}
f_1(\alpha_z,k_\perp^2)~\simeq~{f(\alpha_z)\over k_\perp^2},~~h_1^\perp(\alpha_z,k_\perp^2)~\simeq~{m^2_Nh(\alpha_z)\over k_\perp^4} ,
~~~~~~\barf_1\simeq{\barf(\alpha_z)\over k_\perp^2},~~\barh_1^\perp\simeq{m^2_N\barh(\alpha_z)\over k_\perp^4} 
\label{6.7}
\end{equation}
(up to logarithmic corrections). Similarly, for the target we can use the estimate
\begin{equation}
f_1(\beta_z,k_\perp^2)~\simeq~{f(\beta_z)\over k_\perp^2},~~h_1^\perp(\beta_z,k_\perp^2)~\simeq~{m^2_N h(\beta_z)\over k_\perp^4} ,
~~~~~~\barf_1\simeq{\barf(\beta_z)\over k_\perp^2},~~\barh_1^\perp\simeq{m^2_N\barh(\beta_z)\over k_\perp^4} 
\label{6.8}
\end{equation}
as long as $m^2_N \ll k_\perp^2 \ll Q^2$. 

Substituting this to eq. (\ref{resultZ}) we get
the following estimate of the strength of power corrections for Z-boson production
\begin{eqnarray}
&&\hspace{-1mm}
W(p_A,p_B,q)
~=~-{e^2\over 8s_W^2c_W^2N_c}
\!\int\! d^2k_\perp{1\over k_\perp^2(q-k)_\perp^2}
\label{6.4}\\
&&\hspace{-1mm}
\times~\Big[\Big\{(1+a_u^2)\Big[1-2{(k,q-k)_\perp\over Q^2}\Big][f_u(\alpha_z)\barf_u(\beta_z)+\barf_u(\alpha_z)f_u(\beta_z)]
\nonumber\\
&&\hspace{-1mm}
+2(a_u^2 - 1)[h_u(\alpha_z)\barh_u(\beta_z)+\barh_u(\alpha_z)h_u(\beta_z)]
{m^2_N\over Q^2}\Big\}
~+~\Big\{u\leftrightarrow c\Big\}+\Big\{u\leftrightarrow d\Big\}+\Big\{u\leftrightarrow s\Big\}\Big]
\nonumber\\
&&\hspace{-1mm}
\simeq~-{e^2\over 8s_W^2c_W^2N_c}
\!\int\! d^2k_\perp{1\over k_\perp^2(q-k)_\perp^2}\Big[1-2{(k,q-k)_\perp\over Q^2}\Big]
\nonumber\\
&&\hspace{-1mm}
\times~\Big[\Big\{(1+a_u^2)[f_u(\alpha_z)\barf_u(\beta_z)+\barf_u(\alpha_z)f_u(\beta_z)]
\Big\}
~+~\Big\{u\leftrightarrow c\Big\}+\Big\{u\leftrightarrow d\Big\}+\Big\{u\leftrightarrow s\Big\}\Big].
\nonumber
\end{eqnarray}
Here we used the fact that due to the ``positivity constraint'' $h_1^\perp(x,k_\perp^2)\leq {m_N\over |k_\perp|}f_1^\perp(x,k_\perp^2)$ \cite{Bacchetta:1999kz}, we can safely assume
that the numbers $f(x)$ and $h(x)$ in eqs. (\ref{6.7}) and (\ref{6.8}) are of the same order of magnitude so the last term in the third line 
in eq. (\ref{6.4}) $\sim {m^2_N \over Q^2}$ can be neglected.
Thus, the relative weight of the leading term and power correction is determined by the factor
$1-2{(k,q-k)_\perp\over Q^2}$. The integrals over $k_\perp$ are logarithmic and should be cut from below
by $m^2_N$ and from above by $Q^2$ so we get an estimate
\begin{eqnarray}
&&\hspace{-1mm}
W(p_A,p_B,q)
~=~-{\pi e^2\over 4s_W^2c_W^2N_c}\Big[{1\over q_\perp^2}\ln{q_\perp^2\over m^2_N}+{1\over Q^2}\ln{Q^2\over q_\perp^2}\Big]
\label{estimate}\\
&&\hspace{-1mm}
\times~\Big[\Big\{(1+a_u^2)[f_u(\alpha_z)\barf_u(\beta_z)+\barf_u(\alpha_z)f_u(\beta_z)]\Big\}
~+~\Big\{u\leftrightarrow c\Big\}+\Big\{u\leftrightarrow d\Big\}+\Big\{u\leftrightarrow s\Big\}\Big],
\nonumber
\end{eqnarray}
where we assumed that the first term is determined by the logarithmical region $q_\perp^2\gg k_\perp^2\gg m^2_N$ and the second by $Q^2\gg k_\perp^2\gg q_\perp^2$.
By this estimate, the power correction reaches the level of few percent at $q_\perp\geq$ 20 GeV. Of course, when $q_\perp^2$ increases,
the correction becomes bigger, but the validity of the approximation ${q_\perp^2\over Q^2}\ll 1$ worsens. Moreover, we have ignored all 
logarithmic (and double-log) evolutions which can significantly change the relative strength of power corrections.

\section{Power corrections for Drell-Yan process \label{sec:DY}}

In this section we consider $\gamma^\ast$ contribution to the cross section of the Drell-Yan process which is determined by the hadronic tensor
\begin{eqnarray}
&&\hspace{-1mm}
W_{\mu\nu}(p_A, p_B, q)~=~\int\!d^2x_\perp ~e^{i(q,x)_\perp} W_{\mu\nu}(\alpha_q,\beta_q,x_\perp),
\label{defWmunu}\\
&&\hspace{-1mm}
W_{\mu\nu}(\alpha_q,\beta_q,x_\perp)~\equiv~
{1\over 16\pi^4}{2\over s}\!\int\! dx_\bu dx_\ast ~e^{-i\alpha_qx_\bu-i\beta_q x_\ast}\langle p_A,p_B|J^{\rm  em}_\mu(x_\bu,x_\ast,x_\perp)J^{\rm  em}_\nu(0)|p_A,p_B\rangle,
\nonumber
\end{eqnarray}
 where $J^{\rm  em}_\mu~=~e_u\bsi_u\gamma_\mu \psi_u+e_d\bsi_d\gamma_\mu \psi_d+e_s\bsi_s\gamma_\mu \psi_s+e_c\bsi_c\gamma_\mu \psi_c$ 
 is the electromagnetic current for active flavors in our kinematical region.
 
From the results of the present paper it is easy to extract power corrections to $W^\mu_\mu$.  
\footnote{The problem of calculating power corrections for $W_{\mu\nu}$ with non-convoluted indices is a separate issue which we hope to address in a different publication.}
We replace constants $a_u$ in eq. (\ref{resultZ}) by $e_f^2$ and remove factors ``1'' from expressions like $a^2\pm 1$. One can formally set $a_u\rightarrow \infty$ in 
$\hamma_\mu\equiv\gamma_\mu(a_u-\gamma_5)$, divide the result (\ref{resultZ}) by $a_u^2$,  
and multiply by $e_u^2$. After that, we repeat the procedure for other flavors
and get
\begin{eqnarray}
&&\hspace{-1mm}
W^\mu_\mu(p_A,p_B,q)
~=~-{2\over N_c}\!\int\! d^2k_\perp\Big[\Big\{e_u^2
\big[1-2{(k,q-k)_\perp\over Q^2}\big]
f_{1u}(\alpha_q,k_\perp)\barf_{1u}(\beta_q,q_\perp-k_\perp) 
\nonumber\\
&&\hspace{-1mm}
+~2e_u^2{k_\perp^2(q-k)_\perp^2\over m^2_NQ^2}\big[
h^\perp_{1u}(\alpha_q,k_\perp)\barh^\perp_{1u}(\beta_q,q_\perp-k_\perp) 
\nonumber\\
&&\hspace{22mm}
+~\frac{1}{N^2_c-1}\big(h_{u}^{\rm tw3}(\alpha_q,k_\perp)
\barh_{u}^{\rm tw3}(\beta_q,q_\perp-k_\perp) 
+\tilh_{u}^{\rm tw3}(\alpha_q,k_\perp)\tilde\barh_{u}^{\rm tw3}(\beta_q,q_\perp-k_\perp)\big)\big]
\nonumber\\
&&\hspace{-1mm}
-~e_u^2{N_c\over N_c^2-1}{(k,q-k)_\perp\over Q^2}
\big[2j_{1u}^{\rm tw3}(\alpha_q,k_\perp)j_{2u}^{\rm tw3}(\beta_q,q_\perp-k_\perp) 
-2\tilj_{1u}^{\rm tw3}(\alpha_q,k_\perp)\tilj_{2u}^{\rm tw3}(\beta_q,q_\perp-k_\perp)
\nonumber\\
&&\hspace{-1mm}
-~j_{1u}^{\rm tw3}(\alpha_q,k_\perp)j_{1u}^{\rm tw3}(\beta_q,q_\perp-k_\perp) 
-j_{2u}^{\rm tw3}(\alpha_q,k_\perp)j_{2u}^{\rm tw3}(\beta_q,q_\perp-k_\perp) 
\nonumber\\
&&\hspace{-1mm}
-~\tilj_{1u}^{\rm tw3}(\alpha_q,k_\perp)\tilj_{1u}^{\rm tw3}(\beta_q,q_\perp-k_\perp)
-\tilj_{2u}^{\rm tw3}(\alpha_q,k_\perp)\tilj_{2u}^{\rm tw3}(\beta_q,q_\perp-k_\perp)\big]
+~(\alpha_q\leftrightarrow\beta_q)\Big\}
\nonumber\\
&&\hspace{-1mm}
~+~\Big\{u\leftrightarrow c\Big\}+\Big\{u\leftrightarrow d\Big\}+\Big\{u\leftrightarrow s\Big\}\Big].
\label{resultDY}
\end{eqnarray}
Let us present also the large-$N_c$ estimate similar to eq. (\ref{6.4})
\begin{eqnarray}
&&\hspace{-1mm}
W^\mu_\mu(p_A,p_B,q)
~=~-{2\over N_c}
\!\int\! d^2k_\perp{1\over k_\perp^2(q-k)_\perp^2}
\label{7.3}\\
&&\hspace{-1mm}
\times~\Big[\Big\{e_u^2\Big[1-2{(k,q-k)_\perp\over Q^2}\Big][f_u(\alpha_q)\barf_u(\beta_q)+\barf_u(\alpha_q)f_u(\beta_q)]
\nonumber\\
&&\hspace{-1mm}
+~2e_u^2[h_u(\alpha_q)\barh_u(\beta_q)+\barh_u(\alpha_q)h_u(\beta_q)]
{m^2_N\over Q^2}\Big\}
~+~\Big\{u\leftrightarrow c\Big\}+\Big\{u\leftrightarrow d\Big\}+\Big\{u\leftrightarrow s\Big\}\Big]
\nonumber\\
&&\hspace{-1mm}
\simeq~-{2\over N_c}
\!\int\! d^2k_\perp{1\over k_\perp^2(q-k)_\perp^2}\Big[1-2{(k,q-k)_\perp\over Q^2}\Big]
\nonumber\\
&&\hspace{-1mm}
\times~\big\{e_u^2[f_u(\alpha_q)\barf_u(\beta_q)+\barf_u(\alpha_q)f_u(\beta_q)]
~+~(u\leftrightarrow c)+(u\leftrightarrow d)+(u\leftrightarrow s)\big\}.
\nonumber
\end{eqnarray}
Obviously, the relative strength of leading-twist terms and power corrections is the same as for Z-boson production so 
from our na\"ive estimate (\ref{estimate}) one should expect power corrections of order of few percent starting from $q_\perp\sim {1\over 4}Q$.

\section{Conclusions and outlook}

In this paper we have calculated the higher-twist power correction to Z-boson production (and Drell-Yan process) 
in the kinematical region $s\gg Q^2\gg q_\perp^2$. Our back-of-the-envelope estimation of importance
of power corrections tells that they reach a few percent of the leading-twist result at $q_\perp\sim {1\over 4}Q$ 
which surprisingly agrees with the same estimate made in ref. \cite{Scimemi:2017etj} by comparing
leading-order fits to experimental data.

Of course, we made our estimate without taking into account the TMD evolution, notably the most essential
double-log (Sudakov) evolution. One should evolve projectile TMD  from
$\sigma_a=\beta_q$ to $\tigma_a={q_\perp^2\over\alpha_q s} =\beta_q{q_\perp^2\over Q^2}$,
 target TMDs from $\sigma_b=\alpha_q$ to
$\tigma_b={q_\perp^2\over\beta_q s}=\alpha_q{q_\perp^2\over Q^2}$, and match to the result of leading-log
calculation of integral over central fields in the rapidity interval between $\tigma_a$ and $\tigma_b$. 

To accurately match these evolutions, we hope to use logic  borrowed from the operator product expansion. We write down a general formula (\ref{W5})
\begin{equation}
\hspace{-1mm}  
W~=~\frac{1}{(2\pi)^4}\!\int \! d^4x  e^{-iqx}
\sum_{m,n}\! \int\! dz_m c_{m,n}(q,x)
\langle p_A|\hat\Phi_A(z_m)|p_A\rangle\!\int\! dz'_n\langle p_B| \hat\Phi_B(z'_n)|p_B\rangle,   
\label{W7}
\end{equation}
where the coefficient functions $c_{m,n}(q,x)$ are determined by integrals over $C$-fields and do not depend
on the form of projectile or target. To find these coefficients in the first-loop order, we integrate over $C$-fields in eq. (\ref{funtc}) 
with action $S_C=S_{\rm QCD}(C+ A+B, \psi_C + \psi_A + \psi_B)-S_{\rm QCD}(A, \psi_A)-S_{\rm QCD}(B, \psi_B)$ but {\it without} any rapidity restrictions on $C$-fields, 
and subtract matrix elements  of the operators $\hat\Phi_A(z_m)\hat\Phi_B(z'_n)$ in the background fields $A$, $\psi_A$ and $B$, $\psi_B$ multiplied by tree-level
coefficients. Both the  integrals over $C$-fields in eq. (\ref{funtc}) and matrix elements of $\hat\Phi_A(z_m)\hat\Phi_B(z'_n)$ 
will have rapidity divergencies which will be canceled in their sum so what remains are the logarithms (or double logs) of the ratio of kinematical
variables ($Q^2$ in our case) to the rapidity cutoffs $\sigma_a$ of the operators  $\hat\Phi_A(z_m)$  and $\sigma_b$ of $\hat\Phi_B(z'_n)$. 
Using the above logic we hope to avoid the problem of double-counting of fields which arises when integrals over longitudinal momenta of $C$-fields got
pinched at small momenta (see the discussion in the end of secttion \ref{sect:ApprSol}). The work is in progress.

It should be mentioned that,  as discussed in ref. \cite{Balitsky:2017flc}, our rapidity factorization is different 
from the standard factorization 
scheme for particle production in hadron-hadron 
scattering, namely splitting  the diagrams in collinear to projectile part, collinear to 
target part, hard factor, and soft factor \cite{Collins:2011zzd}. Here we factorize only in rapidity
and the $Q^2$ evolution arises from $k_\perp^2$ dependence of the rapidity evolution kernels, same
as in the BK (and NLO BK \cite{Balitsky:2008zza}) equations. Also, since matrix elements of TMD operators
with our rapidity cutoffs are UV-finite \cite{Balitsky:2015qba,Balitsky:2016dgz}, the only UV divergencies 
in our approach are usual
UV divergencies absorbed in the effective QCD coupling.

It is worth noting that recently the treatment of power corrections was performed in the framework of SCET theory
(see e.g. refs. \cite{Moult:2017rpl,Goerke:2017ioi,Beneke:2017ztn}). 
However, since our rapidity factorization is different from factorization  used by  SCET, the detailed comparison 
of power corrections to Z-boson (or Higgs) production would be possible when SCET  result for TMD
corrections in the form of ${1\over m_Z^2}$ times matrix elements of quark-antiquark-gluon operators will be available.

Let us note that  we obtained power corrections for Drell-Yan hadronic tensor  convoluted over Lorentz indices.
 It would be interesting (and we plan) to calculate the higher-twist correction to full DY hadronic tensor. 
Also, it is well known that for semi-inclusive deep inelastic scattering (SIDIS) and for DY process
the leading-order TMDs have different directions of Wilson lines: one to $+\infty$ and
another to $-\infty$ \cite{Brodsky:2002cx,Collins:2002kn}. We think that the same directions of Wilson lines will stay on  
in the case of power corrections and we plan to study this question in forthcoming publications.

The authors are grateful to  S. Dawson,  A. Prokudin, T. Rogers, 
 R. Venugopalan, and A. Vladimirov for valuable discussions. This material is based upon work 
 supported by the U.S. Department of Energy, Office of Science, Office of Nuclear Physics under contracts DE-AC02-98CH10886 and DE-AC05-06OR23177
 and by U.S. DOE grant DE-FG02-97ER41028.

\section{Appendix}
\subsection{Next-to-leading quark fields \label{sec:nloquarks}}

In this Section we present the explicit expressions for the next-to-leading quark fields $\Psi^{(1)}$. It is convenient to separate these
fields in ``left'' and ``right'' components:
\begin{equation}
\Psi^{(1)}~=~\Psi_{1}^{(1)}+\Psi_{2}^{(1)},~~~~~\Psi_{1}^{(1)}\equiv{\slashed{p}_1\slashed{p}_2\over s}\Psi^{(1)},~~~~
\Psi_{2}^{(1)}\equiv{\slashed{p}_2\slashed{p}_1\over s}\Psi^{(1)}.
\end{equation}
The next-to-leading term in the expansion of the fields (\ref{kvarkfildz})
has the form:
\begin{eqnarray}
&&
\Psi_{1A}^{(1)}~=~-{g\slashed{p}_1\over s\beta}\gamma^i B_i\psi_A
-{g\gamma^i\over s^2}\slashed{p}_1\slashed{p}_2{1\over \beta}\calp_i{1\over \alpha}
\gamma^j B_j\psi_A
-{2g\over s^2}\slashed{p}_1\slashed{p}_2{1\over \beta}(\mathbb{A}_\bu^{[1]})^{(0)}\psi_A,
\label{fildz1}\\
&&
\Psi_{2A}^{(1)}~=~-{2g\slashed{p}_2\slashed{p}_1\over s^2}B_\ast{1\over\alpha}\psi_A
-{g\gamma^i\over s^2}\slashed{p}_2\slashed{p}_1{1\over \alpha}\calp_i{1\over \beta}\gamma^j B_j\psi_A
-{2g\over s^2}\slashed{p}_2\slashed{p}_1{1\over \alpha}(\mathbb{A}_\ast^{[1]})^{(0)}\psi_A
\nonumber\\
&&\hspace{16mm}
-~ \frac{2g^2\gamma^j}{s^2\alpha^2}\slashed{p}_2 B_\ast
 B_j\psi_A + \frac{g\gamma^i\gamma^j}{s^2}\slashed{p}_2 \frac{1}{\alpha}\calp_i \frac{1}{\beta}\calp_j
\frac{1}{\alpha}\gamma^k B_k\psi_A,
\nonumber\\
&&
\Psi_{1B}^{(1)}~=~-{g\slashed{p}_2\over s\alpha}\gamma^i A_i\psi_B
-{g\gamma^i\over s^2}\slashed{p}_2\slashed{p}_1{1\over \alpha}\calp_i{1\over \beta}\gamma^j A_j\psi_B
-{2g\over s^2}\slashed{p}_2\slashed{p}_1{1\over \alpha}(\mathbb{A}_\ast^{[1]})^{(0)}\psi_B,
\nonumber\\
&&
\Psi_{2B}^{(1)}~=~-{2g\slashed{p}_1\slashed{p}_2\over s^2}A_\bu{1\over\beta}\psi_B
-{g\gamma^i\over s^2}\slashed{p}_1\slashed{p}_2{1\over \beta}\calp_i{1\over \alpha}\gamma^jA_j\psi_B
-{2g\over s^2}\slashed{p}_1\slashed{p}_2{1\over \beta}(\mathbb{A}_\bu^{[1]})^{(0)}\psi_B
\nonumber\\
&&\hspace{16mm}
-~ \frac{2g^2\gamma^j}{s^2\beta^2}\slashed{p}_1 A_\bullet A_j \psi_B 
+ \frac{g\gamma^i\gamma^j}{s^2}\slashed{p}_1\frac{1}{\beta}\calp_i \frac{1}{\alpha} \calp_j
\frac{1}{\beta} \gamma^k A_k \psi_B,
\nonumber\\
&&
\Bsi_{1A}^{(1)}~=~-\bar\psi_A\gamma^i B_i{g\slashed{p}_1\over s(\beta-i\epsilon)}
-\bar\psi_A\gamma^j B_j\slashed{p}_2\slashed{p}_1{1\over \alpha-i\epsilon}\calp_i{1\over \beta-i\epsilon}{g\gamma^i\over s^2}
-{2g\over s^2}\bar\psi_A (\mathbb{A}_\bu^{[1]})^{(0)}{1\over \beta-i\epsilon}\slashed{p}_2\slashed{p}_1,
\nonumber\\
&&
\Bsi_{2A}^{(1)}~=~-\bar\psi_A{1\over\alpha-i\epsilon}B_\ast{2g\slashed{p}_1\slashed{p}_2\over s^2}
-\bar\psi_A\gamma^j B_j\slashed{p}_1\slashed{p}_2{1\over \beta-i\epsilon}\calp_i{1\over \alpha-i\epsilon}{g\gamma^i\over s^2}
-{2g\over s^2}\bar\psi_A (\mathbb{A}_\ast^{[1]})^{(0)}{1\over \alpha-i\epsilon}\slashed{p}_1\slashed{p}_2
\nonumber\\
&&\hspace{16mm}
-~ \bar\psi_A B_j B_\ast\slashed{p}_2 \frac{2g^2\gamma^j}{s^2\alpha^2}
 + \bar\psi_A \gamma^k B_k \frac{1}{\alpha-i\epsilon}\calp_j  \frac{1}{\beta-i\epsilon}\calp_i \frac{1}{\alpha-i\epsilon}\slashed{p}_2 \frac{g\gamma^j \gamma^i}{s^2}
 ,
\nonumber\\
&&
\Bsi_{1B}^{(1)}~=~-\bar\psi_B\gamma^i A_i{g\slashed{p}_2\over s(\alpha-i\epsilon)}
-\bar\psi_B\gamma^j A_j\slashed{p}_1\slashed{p}_2{1\over \beta-i\epsilon}\calp_i{1\over \alpha-i\epsilon}{g\gamma^i\over s^2}
-{2g\over s^2}\bar\psi_B (\mathbb{A}_\ast^{[1]})^{(0)}{1\over \alpha-i\epsilon}\slashed{p}_1\slashed{p}_2,
\nonumber\\
&&
\Bsi_{2B}^{(1)}~=~-\bsi_B{1\over\beta-i\epsilon}A_\bu{2g\slashed{p}_2\slashed{p}_1\over s^2}
-\bar\psi_B\gamma^j A_j\slashed{p}_2\slashed{p}_1{1\over \alpha-i\epsilon}\calp_i{1\over \beta-i\epsilon}{g\gamma^i\over s^2}
-{2g\over s^2}\bar\psi_B (\mathbb{A}_\bu^{[1]})^{(0)}{1\over \beta-i\epsilon}\slashed{p}_2\slashed{p}_1
\nonumber\\
&&\hspace{16mm}
-~ \bar\psi_B A_j A_\bullet \slashed{p}_1\frac{2g^2\gamma^j}{s^2\beta^2}    
+ \bar\psi_B \gamma^k A_k  \frac{1}{\beta}\calp_j \frac{1}{\alpha}\calp_i \frac{1}{\beta} \slashed{p}_1\frac{g\gamma^j\gamma^i}{s^2},
\nonumber
\end{eqnarray}
where $\calp_i=i\partial_i+gA_i+gB_i$, see eq. (\ref{3.12}). 
The expressions for $\Bsi$ should be read from right to left, e.g.
\begin{eqnarray}
\bar\psi_A\gamma^j B_j\slashed{p}_2\slashed{p}_1{1\over \alpha-i\epsilon}\calp_i{1\over \beta-i\epsilon}(x){\gamma^i\over s^2}
~\equiv~\int\! dz~ \bar\psi_A(z)\gamma^j B_j(z)\slashed{p}_2\slashed{p}_1
(z|{1\over \alpha-i\epsilon}\calp_i{1\over \beta-i\epsilon}|x){\gamma^i\over s^2}
\nonumber\\
\end{eqnarray}
(and ${1\over\alpha}\equiv{1\over\alpha+i\epsilon}$ ${1\over\beta}\equiv{1\over\beta+i\epsilon}$as usual).
It is easy to see that the power counting of quark fields has the form (cf eq. (\ref{psi0})):
\begin{eqnarray}
&&\Psi_{1A}^{(1)}\sim\Psi_{1B}^{(1)}\sim\Psi_{2A}^{(1)}\sim\Psi_{2B}^{(1)}~\sim~{m_\perp^{7/2}\over s}.
\label{psi1}
\end{eqnarray}
The gluon fields $\mathbb{A}_\bu^{(0)}$ and $\mathbb{A}_\ast^{(0)}$ were  calculated in ref. \cite{Balitsky:2017flc}:
\begin{eqnarray}
&&\hspace{-1mm}
\mathbb{A}^{(0)}_\bu~=~A_\bu+(\mathbb{A}^{[1]}_\bu)^{(0)}, ~~~~~~(\mathbb{A}^{[1]a}_\bu)^{(0)}~=~{g\over 2\alpha}A_j^{ab}B^{jb},~~~~~
\nonumber\\
&&\hspace{-1mm}
\mathbb{A}^{(0)}_\ast~=~B_\ast+(\mathbb{A}^{[1]}_\ast)^{(0)}, ~~~~~~(\mathbb{A}^{[1]a}_\ast)^{(0)}~=~-{g\over 2\beta}A_j^{ab}B^{jb}
\label{glufildz}
\end{eqnarray}
and their power counting reads
\begin{equation}
gA_\bu~\sim~gB_\ast~\sim m_\perp^2,~~~~~~~gA_i~\sim~gB_i~\sim m_\perp.
\end{equation}

\subsection{Formulas with Dirac matrices \label{diracs}}
In the gauge $A_\bu=0$ the field $A_i$ can be represented as 
\begin{equation}
A_i(x_\bu,x_\perp)~=~{2\over s}\!\int_{-\infty}^{x_\bu}\!dx'_\bu~A_{\ast i}(x'_\bu,x_\perp)
\end{equation}
(see eq. (\ref{AfromF})). It is convenient to define a ``dual'' field
\begin{equation}
\tilde{A}_i(x_\bu,x_\perp)~=~{2\over s}\!\int_{-\infty}^{x_\bu}\!dx'_\bu~\tilde{A}_{\ast i}(x'_\bu,x_\perp),
~~~\tilde{B}_i(x_\ast,x_\perp)~=~{2\over s}\!\int_{-\infty}^{x_\ast}\!dx'_\ast~\tilde{B}_{\bu i}(x'_\ast,x_\perp),
\end{equation}
 where $\tilF_{\mu\nu}=\half\epsilon_{\mu\nu\lambda\rho}F^{\lambda\rho}$ as usual. 
\footnote{We use conventions from {\it Bjorken \& Drell} where $\epsilon^{0123}=-1$ and
$
\gamma^\mu\gamma^\nu\gamma^\lambda=g^{\mu\nu}\gamma^\lambda +g^{\nu\lambda}\gamma^\mu-g^{\mu\lambda}\gamma^\nu
-i\epsilon^{\mu\nu\lambda\rho}\gamma_\rho\gamma_5
$.
Also, with this convention $\tigma_{\mu\nu}\equiv\half \epsilon_{\mu\nu\lambda\rho}\sigma^{\lambda\rho}=i\sigma_{\mu\nu}\gamma_5$.
}
With this definition, we get 
\begin{equation}
{2\over s}\epsilon_{\bu\ast ij}A^j=\tilde{A}_i,~~~{2\over s}\epsilon_{\bu\ast ij}B^j=-\tilde{B}_i
~~\Rightarrow~~~\tilde{A}_i\otimes\tilde{B}^i=-A_i\otimes B^i,~~\tilde{A}_i\otimes B^i= A_i\otimes\tilde{B}^i
\label{9.9}
\end{equation}
and therefore
\begin{eqnarray}
&&\hspace{-1mm}
A_k\gamma_i\slashed{p}_2\gamma^j\otimes B_j\gamma^i\slashed{p}_1\gamma^k~=~
\slashed{p}_2(A_i-i\tilde{A}_i\gamma_5)\otimes \slashed{p}_1(B^i-i\tilde{B}^i\gamma_5),
\nonumber\\
&&\hspace{-1mm}
A_k\gamma^j\slashed{p}_2\gamma_i\otimes B_j\gamma^k\slashed{p}_1\gamma^i~=~
\slashed{p}_2(A_i+i\tilde{A}_i\gamma_5)\otimes \slashed{p}_1(B^i+i\tilde{B}^i\gamma_5),
\nonumber\\
&&\hspace{-1mm}
A_k\gamma_i\slashed{p}_2\gamma_j\otimes B^j\gamma^k\slashed{p}_1\gamma^i
~=~\slashed{p}_2(A_i-i\tilde{A}_i\gamma_5)\otimes\slashed{p}_1(B^i+i\tilde{B}^i\gamma_5),
\nonumber\\
&&\hspace{-1mm}
A_k\gamma_j\slashed{p}_2\gamma_i\otimes B^j\gamma^i\slashed{p}_1\gamma^k
~=~\slashed{p}_2(A_i+i\tilde{A}_i\gamma_5)\otimes\slashed{p}_1(B^i-i\tilde{B}^i\gamma_5).
\label{formula}
\end{eqnarray}
We will also need
\begin{eqnarray}
&&\hspace{-1mm}
A_k\gamma_i\slashed{p}_2\gamma^j\gamma_5\otimes B_j\gamma^i\slashed{p}_1\gamma^k\gamma_5~=~A_k\gamma_i\slashed{p}_2\gamma^j\otimes B_j\gamma^i\slashed{p}_1\gamma^k,
\nonumber\\
&&\hspace{-1mm}
A_k\gamma^j\slashed{p}_2\gamma_i\gamma_5\otimes B_j\gamma^k\slashed{p}_1\gamma^i\gamma_5~=~
A_k\gamma^j\slashed{p}_2\gamma_i\otimes B_j\gamma^k\slashed{p}_1\gamma^i,
\nonumber\\
&&\hspace{-1mm}
A_k\gamma_i\slashed{p}_2\gamma_j\gamma_5\otimes B^j\gamma^k\slashed{p}_1\gamma^i\gamma_5
~=~-A_k\gamma_i\slashed{p}_2\gamma_j\otimes B^j\gamma^k\slashed{p}_1\gamma^i,
\nonumber\\
&&\hspace{-1mm}
A_k\gamma_j\slashed{p}_2\gamma_i\gamma_5\otimes B^j\gamma^i\slashed{p}_1\gamma^k\gamma_5
~=~-A_k\gamma_j\slashed{p}_2\gamma_i\otimes B^j\gamma^i\slashed{p}_1\gamma^k
\label{9.11}
\end{eqnarray}
and hence
\begin{eqnarray}
&&\hspace{-1mm}
A_k\gamma_i\slashed{p}_2\gamma^j(a-\gamma_5)\otimes B_j\gamma^i\slashed{p}_1\gamma^k(a-\gamma_5)
\label{9.12}\\
&&\hspace{-1mm}
=~(a^2+1)\slashed{p}_2(A_i-i\tilde{A}_i\gamma_5)\otimes \slashed{p}_1(B^i-i\tilde{B}^i\gamma_5)
-2a\slashed{p}_2(A_i-i\tilde{A}_i\gamma_5)\otimes \slashed{p}_1(\gamma_5B^i-i\tilde{B}^i),
\nonumber\\
&&\hspace{-1mm}
A_k\gamma^j\slashed{p}_2\gamma_i(a-\gamma_5)\otimes B_j\gamma^k\slashed{p}_1\gamma^i(a-\gamma_5)
\nonumber\\
&&\hspace{-1mm}
=~(a^2+1)\slashed{p}_2(A_i+i\tilde{A}_i\gamma_5)\otimes \slashed{p}_1(B^i+i\tilde{B}^i\gamma_5)
-2a\slashed{p}_2(A_i+i\tilde{A}_i\gamma_5)\otimes \slashed{p}_1(\gamma_5B^i+i\tilde{B}^i),
\nonumber\\
&&\hspace{-1mm}
A_k\gamma_i\slashed{p}_2\gamma_j(a-\gamma_5)\otimes B^j\gamma^k\slashed{p}_1\gamma^i(a-\gamma_5)~
=~(a^2-1)\slashed{p}_2(A_i-i\tilde{A}_i\gamma_5)\otimes \slashed{p}_1(B^i+i\tilde{B}^i\gamma_5),
\nonumber\\
&&\hspace{-1mm}
A_k\gamma_j\slashed{p}_2\gamma_i(a-\gamma_5)\otimes B^j\gamma^i\slashed{p}_1\gamma^k(a-\gamma_5)
~
=~(a^2-1)\slashed{p}_2(A_i+i\tilde{A}_i\gamma_5)\otimes \slashed{p}_1(B^i-i\tilde{B}^i\gamma_5).
\nonumber
\end{eqnarray}
Next, using formula
\begin{eqnarray}
&&\hspace{-1mm}
\tigma_{\mu\nu}\otimes\tigma_{\alpha\beta}~
=~-\half(g_{\mu\alpha}g_{\nu\beta}-g_{\nu\alpha}g_{\mu\beta})\sigma_{\xi\eta}\otimes\sigma^{\xi\eta}
\nonumber\\
&&\hspace{-1mm}
+~g_{\mu\alpha}\sigma_{\beta\xi}\otimes\sigma_\nu^{~\xi}-g_{\nu\alpha}\sigma_{\beta\xi}\otimes\sigma_\mu^{~\xi}-g_{\mu\beta}\sigma_{\alpha\xi}\otimes\sigma_\nu^{~\xi}+g_{\nu\beta}\sigma_{\alpha\xi}\otimes\sigma_\mu^{~\xi}
-\sigma_{\alpha\beta}\otimes\sigma_{\mu\nu}
\label{tigmi}
\end{eqnarray}
we get
\begin{eqnarray}
&&\hspace{-1mm}
A^k \slashed{p}_2 \gamma_j \otimes B^j \slashed{p}_1 \gamma_k - A^k \slashed{p}_2 \gamma_j \gamma_5 \otimes B^j \slashed{p}_1 \gamma_k \gamma_5
\label{9.13}\\
&&\hspace{-1mm}
=~ A^k \slashed{p}_2 \gamma_j \otimes B^j \slashed{p}_1 \gamma_k   + A^k \slashed{p}_2 \gamma_k \otimes B^j \slashed{p}_1 \gamma_j  - \frac{s}{4} A^j \sigma_{\xi\eta}\otimes B_j \sigma^{\xi\eta} - \frac{s}{2} A^k \sigma_{ki}\otimes B^j \sigma_j^{\ i}
\nonumber\\
&&\hspace{-1mm}
-~ A^j \slashed{p}_1 \gamma_k\otimes B_j \slashed{p}_2\gamma^k   + \frac{s}{2} A^i \sigma_{jk}\otimes B_i \sigma^{j k} - \frac{2}{s}A^k \sigma_{\ast\bullet}\otimes B_k \sigma_{\ast \bullet} - 2 A^k \slashed{p}_2\gamma_j \otimes B_k \slashed{p}_1\gamma^j.
\nonumber
\end{eqnarray}
For appendix \ref{sec:sublead} we also need
\begin{eqnarray}
&&\hspace{-3mm}
{2\over s}(\hatp_2\gamma^i\hatp_1\otimes \hatp_1 B_i+\hatp_2\gamma^i\hatp_1\gamma_5\otimes \hatp_1\gamma_5 B_i)
= -\gamma^i\otimes \hatp_1(B_i+i\tilde{B}_i\gamma_5)-\gamma^i\gamma_5\otimes \hatp_1\gamma_5(B_i+i\tilde{B}_i\gamma_5),
\nonumber\\
&&\hspace{-3mm}
\gamma_k\gamma^i\hatp_2\otimes B_i\gamma^k
+ \gamma_k\gamma^i\hatp_2\gamma_5\otimes B_i\gamma^k\gamma_5
~=~\hatp_2\otimes (B_i +i\tilde{B}_i\gamma_5)\gamma^i+\hatp_2\gamma_5\otimes (B_i+i\tilde{B}_i\gamma_5)\gamma^i\gamma_5,
\nonumber\\
&&\hspace{-3mm}
 \slashed{p}_2 \gamma^i\gamma^j\otimes\slashed{p}_1 B_j
~+~ \slashed{p}_2\gamma_5 \gamma^i\gamma^j\otimes\slashed{p}_1\gamma_5 B_j
=\slashed{p}_2\otimes (B^i+i\tilde{B}^i\gamma_5)\slashed{p}_1 + \gamma_5\slashed{p}_2\otimes \gamma_5(B^i+i\tilde{B}^i\gamma_5)\slashed{p}_1,
\nonumber\\
&&\hspace{-3mm}
 \slashed{p}_2 \gamma^i\gamma^j\otimes\slashed{p}_1 B_j
~+~ \slashed{p}_2\gamma_5 \gamma^i\gamma^j\otimes\slashed{p}_1\gamma_5 B_j
=\slashed{p}_2\otimes \slashed{p}_1(B^i-i\tilde{B}^i\gamma_5) + \slashed{p}_2\gamma_5\otimes \slashed{p}_1\gamma_5(B^i-i\tilde{B}^i\gamma_5).
\nonumber\\
\label{8.13}
\end{eqnarray}

\subsection{Subleading power corrections\label{sec:sublead}}

\subsubsection{Second, third, and fourth lines in eq. (\ref{6lines}) \label{234lines}}

In this appendix we show that second, third, and fourth lines in eq. (\ref{6lines}) yield subleading power corrections and can be neglected in our approximation.

Let us consider for example the last term in the third line of eq. (\ref{6lines}). The Fierz transformation (\ref{fierz}) yields
\begin{eqnarray}
&&\hspace{-1mm}
[\bar\psi_A(x)\hamma_\mu\psi_B(x)\big]\big[\bar\psi_B(0)\hamma^\mu\Xi_{2A}(0)\big]
\nonumber\\
&&\hspace{-1mm}
=~{1+a^2\over 2}\big\{\big[\bar\psi_A^{m}(x)\gamma_\alpha\Xi^{n}_{2A}(0)\big]\big[\bar\psi_B^{n}(0)\gamma^\alpha\psi^{m}_{B}(x)\big]
~+~(\gamma_\alpha\otimes\gamma^\alpha\leftrightarrow\gamma_\alpha\gamma_5\otimes\gamma^\alpha\gamma_5)\big\}
\nonumber\\
&&\hspace{-1mm}
-~a\big\{\big[\bar\psi_A^{m}(x)\gamma_\alpha\Xi^{n}_{2A}(0)\big]\big[\bar\psi_B^{n}(0)\gamma^\alpha\gamma_5\psi^{m}_{B}(x)\big]
+(\gamma_\alpha\otimes\gamma^\alpha\gamma_5\leftrightarrow\gamma_\alpha\gamma_5\otimes\gamma^\alpha)\big\}
\nonumber\\
&&\hspace{-1mm}
+~(1-a^2)\big\{\big[\bar\psi_A^{m}(x)\Xi^{n}_{2A}(0)\big]\big[\bar\psi_B^{n}(0)\psi^{m}_{B}(x)\big]
-(1\otimes1\leftrightarrow\gamma_5\otimes\gamma_5)\big\}
\nonumber\\
&&\hspace{-1mm}
=~{1+a^2\over 2}\big\{\big[\bar\psi_A^{m}(x)\gamma_i\Xi^{n}_{2A}(0)\big]\big[\bar\psi_B^{n}(0)\gamma^i\psi^{m}_{B}(x)\big]
~+~(\gamma_i\otimes\gamma^i\leftrightarrow\gamma_i\gamma_5\otimes\gamma^i\gamma_5)\big\}
\nonumber\\
&&\hspace{-1mm}
-~a\big\{\big[\bar\psi_A^{m}(x)\gamma_i\Xi^{n}_{2A}(0)\big]\big[\bar\psi_B^{n}(0)\gamma^i\gamma_5\psi^{m}_{B}(x)\big]
+(\gamma_i\otimes\gamma^i\gamma_5\leftrightarrow\gamma_i\gamma_5\otimes\gamma^i)\big\}
\nonumber\\
&&\hspace{-1mm}
+~(1-a^2)\big\{\big[\bar\psi_A^{m}(x)\Xi^{n}_{2A}(0)\big]\big[\bar\psi_B^{n}(0)\psi^{m}_{B}(x)\big]
-(1\otimes1\leftrightarrow\gamma_5\otimes\gamma_5)\big\}~+~O\big({m_\perp^8\over s}\big).
\label{4.14}
\end{eqnarray}
Sorting out the color-singlet contributions 
\footnote
{Recall that after the promotion of background fields to operators we can still move those operators freely since all of them commute, 
see the footnotes on pp. 6 and 14.}
we get
\begin{eqnarray}
&&\hspace{-1mm}
\langle A,B|(\bsi_A^m (B_j)^{nk}\psi_A^k)(\bsi_B^n \psi_B^m)|A,B\rangle
~
=~\langle A,B|(\bsi_A^m\psi_A^k)(\bsi_B^n (B_j)^{nk} \psi_B^m)|A,B\rangle
\nonumber\\
&&\hspace{-1mm}
=~{1\over N_c}\langle A|(\bsi_A^l\psi_A^l)|A\rangle\langle B|(\bsi_B^n B_j^{nk}\psi_B^k)|B\rangle
\label{4.15}
\end{eqnarray}
and therefore
\begin{eqnarray}
&&\hspace{-1mm}
-N_c\big[\bar\psi_A(x)\hamma_\mu\psi_{B}(x)\big]\big[\bar\psi_B(0)\hamma^\mu\Xi_{2A}(0)\big]
\label{4.16}\\
&&\hspace{-1mm}
=~{1+a^2\over 2s}
\big\{\big[\bar\psi_A(x)\gamma_i\slashed{p}_2\gamma^j{1\over\alpha}\psi_A(0)\big]
\big[\bar\psi_B(0)gB_j(0)\gamma^i\psi_B(x)\big]
~+~~(\gamma_i\otimes\gamma^i\leftrightarrow\gamma_i\gamma_5\otimes\gamma^i\gamma_5)\big\}
\nonumber\\
&&\hspace{-1mm}
+~{1-a^2\over s}\big\{\big[\bar\psi_A(x)\slashed{p}_2\gamma^j{1\over\alpha}\psi_A(0)\big]
\big[\bar\psi_B(0)gB_j(0)\psi_B(x)\big]
~-~~(1\otimes1\leftrightarrow\gamma_5\otimes\gamma_5)\big\}
\nonumber\\
&&\hspace{-1mm}
-~{a\over s}\big\{\big[\bar\psi_A(x)\gamma_i\gamma_5\slashed{p}_2\gamma^j{1\over\alpha}\psi_A(0)\big]
\big[\bar\psi_B(0)gB_j(0)\gamma^i\psi_B(x)\big]
~+~~(\gamma_i\gamma_5\otimes\gamma^i\leftrightarrow\gamma_i\otimes\gamma^i\gamma_5)\big\},
\nonumber
\end{eqnarray}
where ${1\over \alpha}\equiv{1\over\alpha+i\epsilon}$, see eq. (\ref{3.25}).

For the forward matrix elements we get
\begin{eqnarray}
&&\hspace{-1mm}
{1\over s}\!\int\! dx_\perp dx_\bu~e^{-i\alpha_qx_\bu +i(k,x)_\perp}
\langle A|\hat{\bar\psi}(x_\bu,x_\perp)\slashed{p}_2\Gamma{1\over\alpha}\hat\psi(0)|A\rangle
\nonumber\\
&&\hspace{-1mm}
=~{1\over\alpha_qs }\!\int\! dx_\perp dx_\bu~e^{-i\alpha_q x_\bu +i(k,x)_\perp}
\langle A|\hat{\bar\psi}(x_\bu,x_\perp)\slashed{p}_2\Gamma\hat\psi(0)|A\rangle~\sim~{1\over\alpha_q}f_\Gamma(\alpha_q,k_\perp^2),
\label{4.17}
\end{eqnarray}
where $\Gamma$ is any of  $\gamma$-matrices with transverse indices.
Next, consider 
\begin{eqnarray}
&&\hspace{-3mm}
\int\! dx_\perp dx_\ast~e^{-i\beta_q x_\ast+i(k,x)_\perp}
\langle B|\hat{\bsi}(0)g\hatB_i(0)\hsi(x_\ast,x_\perp)|B\rangle
\label{4.18}\\
&&\hspace{-3mm}
=~{2\over s }\int\! dx_\perp dx_\ast~e^{-i\beta_q x_\ast+i(k,x)_\perp}\!\int_{-\infty}^0\! \!\!dx'_\ast
\langle B|\hat{\bsi}(0)g\hatF_{\bu i}(x'_\ast,0_\perp)\hsi(x_\ast,x_\perp)|B\rangle
~\sim~{k_i\over m}f^{\rm tw 3}(\beta_q,k_\perp^2),
\nonumber
\end{eqnarray}
where $f^{\rm tw 3}(\beta_q,k_\perp^2)$ is some function of order one (by power counting (\ref{fildz}) this matrix element (\ref{4.18}) is $\sim 1$). 
Also, this function may have only logarithmical singularities 
in $\beta_q$ as $\beta_q\rightarrow 0$ but {\it not} the power behavior ${1\over \beta_q}$. 
\footnote{
Large $x_\ast$ correspond to low-$x$ domain where  matrix elements can be calculated in
a shock-wave background of the target particle. The typical propagator in the shock-wave external field has 
a factor $e^{-i{p_\perp^2\over\alpha s}(x-z)_\ast}$ where $z_\ast$ is a position of a shock wave and $p_\perp$ 
is of order of characteristic transverse momentum \cite{Balitsky:1995ub,Balitsky:2001gj}. The integration over large $x_\ast$ gives then 
$\big(\beta_q+{p_\perp^2\over\alpha s}\big)^{-1}$  and since the integration over $\alpha$ is restricted from 
above by $\sigma_a$,  such terms cannot give ${1\over\beta_q}$ (cf. refs \cite{Balitsky:2015qba,Balitsky:2016dgz}). 
}
The corresponding contribution to $W(q)$ of eq. (\ref{defW}) is proportional  to
\begin{eqnarray}
&&\hspace{-1mm}
{1\over\alpha_q s}\int\! \dhd^2k_\perp
f_{\Gamma_i}(\alpha_q,k_\perp)k_if^{\rm tw 3}(\beta_q,k_\perp^2)~\sim {q_\perp^2\over\alpha_qs}W^{\rm lt}~\ll~
{q_\perp^2\over Q^2}W^{\rm lt}
\label{4.19}
\end{eqnarray}
so it can be neglected in comparison to the contributions $\sim{q_\perp^2\over Q^2}W^{\rm lt}$
(recall that we assume that Z-boson is produced in a central range of rapidity so
${q_\perp^2\over\alpha_qs}\simeq {q_\perp^2\over Q\sqrt s}\ll {q_\perp^2\over Q^2}$). In a similar way one can show
that the remaining three terms in the second and third lines of eq. (\ref{6lines}) give small contributions to $W(q)$.

Next, it is easy to see that the matrix element of the fourth line of eq. (\ref{6lines}) vanishes. Indeed, let us consider the first term in the fourth line 
and perform Fierz transformation (\ref{fierz}):
\begin{eqnarray}
&&\hspace{-1mm}
[\Bxi_{2A}(x)\hamma_\mu\psi_B(x)\big]\big[\bar\psi_B(0)\hamma^\mu\Xi_{2A}(0)\big]
\nonumber\\
&&\hspace{-1mm}
=~{1+a^2\over 2}\big\{\big[\Bxi_{2A}^{m}(x)\gamma_\alpha\Xi^{n}_{2A}(0)\big]\big[\bar\psi_B^{n}(0)\gamma^\alpha\psi^{m}_{B}(x)\big]
~+~(\gamma_\alpha\otimes\gamma^\alpha\leftrightarrow\gamma_\alpha\gamma_5\otimes\gamma^\alpha\gamma_5)\big\}
\nonumber\\
&&\hspace{-1mm}
-~a\big(\big[\Bxi_{2A}^{m}(x)\gamma_\alpha\Xi^{n}_{2A}(0)\big]\big[\bar\psi_B^{n}(0)\gamma^\alpha\gamma_5\psi^{m}_{B}(x)\big]
+(\gamma_\alpha\otimes\gamma^\alpha\gamma_5\leftrightarrow\gamma_\alpha\gamma_5\otimes\gamma^\alpha)\big\}
\nonumber\\
&&\hspace{-1mm}
+~(1-a^2)\big(\big[\Bxi_{2A}^{m}(x)\Xi^{n}_{2A}(0)\big]\big[\bar\psi_B^{n}(0)\psi^{m}_{B}(x)\big]
-(1\otimes1\leftrightarrow\gamma_5\otimes\gamma_5)\big\}.
\end{eqnarray}
From the explicit form of $\Xi_{2A}$ and $\Bxi_{2A}$ in eq. (\ref{fildz0}) we see that the last term in the r.h.s. vanishes while the first two are small. Indeed, 
\begin{eqnarray}
&&\hspace{-1mm}
\langle A,B|{2\over s}\big[\Bxi_{2A}^{m}(x)\slashed{p}_1\Xi^{n}_{2A}(0)\big]\big[\bar\psi_B^{n}(0)\slashed{p}_2\psi^{m}_{B}(x)\big]
|A,B\rangle
\\
&&\hspace{-1mm}
=~
\langle A,B|{2\over s}\big[\big(\bar\psi_A^k{1\over\alpha}\big)(x)\gamma^i
{\slashed{p}_2\over s}\gamma^j{1\over \alpha}\psi_A^l(0)\big]
\big[\bar\psi_B^{n}(0)g^2\hat{B}^{km}_i(x)\hat{B}_j^{nl}(0)\slashed{p}_2\psi^{m}_{B}(x)\big]|A,B\rangle
\nonumber\\
&&\hspace{-1mm}
=~
{2\over sN_c}\langle A|\big(\hat\bsi{1\over\alpha}\big)(x)\gamma^i
{\slashed{p}_2\over s}\gamma^j{1\over \alpha}\hsi(0)|A\rangle
\langle B|\hat\bsi(0)g^2\hatA_j(0)\hatA_i(x)\slashed{p}_2\hsi(x)|B\rangle~\sim~O\big({m_\perp^8\over s}\big),
\nonumber
\end{eqnarray}
so the contribution to $W$ is of order of ${m_\perp^4\over s^2}W^{\rm lt}$.

\subsubsection{Second to fifth lines in eq. (\ref{7newlines})\label{secfivap}}
Here we show that second to fifth lines in eq. (\ref{7newlines}) either vanish or can be neglected.

Obviously, matrix element of the operator in the second line vanishes. Formally,
\begin{eqnarray}
&&\hspace{-1mm}
\int\! dx_\bu~e^{-i\alpha_q x_\bu}\langle A|\hat{\bar\psi}(x_\bu,x_\perp)\hamma_\mu\hsi(x_\bu,x_\perp)|A\rangle
~=~\delta(\alpha_q)\langle A|\hat{\bar\psi}(0)\hamma_\mu\hsi(0)|A\rangle,
\nonumber\\
&&\hspace{-1mm}
\int\! dx_\ast~e^{-i\beta_q x_\ast}\langle B|\hat{\bar\psi}(0)\hamma_\mu\hsi(0)|B\rangle
~=~\delta(\beta_q)\langle B|\hat{\bar\psi}(0)\hamma_\mu\hsi(0)|B\rangle
\label{xxvanish}
\end{eqnarray}
and, non-formally, one hadron cannot produce Z-boson on his own. For a similar reason,
matrix elements of the operators in the third and fourth lines in eq. (\ref{7newlines}) vanish - either projectile or target
matrix element will be of eq. (\ref{xxvanish}) type. In addition, from the explicit form $\Xi$'s in eq. (\ref{fildz0}) it is easy to see that
 the fifth line in eq. (\ref{7newlines})  can be rewritten as follows:
\begin{eqnarray}
&&\hspace{-1mm}
[\Bxi_{2A}(x)\hamma_\mu\Xi_{2A}(x)\big]\big[\bar\psi_B(0)\hamma^\mu\psi_B(0)\big]
+[\bar\psi_A(x)\hamma_\mu\psi_{A}(x)\big]\big[\Bxi_{1B}(0)\hamma^\mu\Xi_{1B}(0)\big]
~+~x\leftrightarrow 0
\nonumber\\
&&\hspace{-1mm}
=~\big[\big(\bar\psi_A{1\over\alpha}\big)(x)\gamma^igB_i(x){\slashed{p}_2\over s}\hamma_\mu
{\slashed{p}_2\over s}\gamma^kgB_k(x){1\over \alpha}\psi_A(x)\big]\big[\bar\psi_B(0)\hamma^\mu\psi_B(0)\big]
\nonumber\\
&&\hspace{-1mm}
+~[\bar\psi_A(x)\hamma_\mu\psi_{A}(x)\big]\big[\big(\bar\psi_B{1\over \beta}\big)(0)\gamma^igA_i(0){\slashed{p}_1\over s}
\hamma^\mu{\slashed{p}_1\over s}\gamma^kgA_k(0){1\over\beta}\psi_B(0)\big]
~+~x\leftrightarrow 0
\nonumber\\
&&\hspace{-1mm}
=~2\big[\big(\bar\psi_A{1\over\alpha}\big)(x)\gamma^igB_i(x){\slashed{p}_2\over s^2}(a-\gamma_5)
\gamma^kgB_k(x){1\over \alpha}\psi_A(x)\big]\big[\bar\psi_B(0)\slashed{p}_2(a-\gamma_5)\psi_B(0)\big]
\nonumber\\
&&\hspace{-1mm}
+~2[\bar\psi_A(x)\slashed{p}_1(a-\gamma_5)\psi_{A}(x)\big]\big[\big(\bar\psi_B{1\over \beta}\big)(0)\gamma^igA_i(0){\slashed{p}_1\over s^2}
(a-\gamma_5)\gamma^kgA_k(0){1\over\beta}\psi_B(0)\big]
~+~x\leftrightarrow 0.
\nonumber\\
\end{eqnarray}
From the power counting (\ref{fildz}) we see that this term is $\sim {m_\perp^8\over s}$, so we are left with the 
contribution of  the last two lines in eq. (\ref{7newlines}).

\subsubsection{Gluon power corrections from $\calj_{A}^\mu(x)\calj_{A\mu}(0)$ terms \label{gluterms}}
There is one more contribution which should be discussed and neglected: 
\begin{eqnarray}
&&\hspace{-1mm}
\calj_{A}^\mu(x)\calj_{A\mu}(0)~=~{e^2\over 16s_W^2c_W^2}
\nonumber\\
&&\hspace{-1mm}
\times~
\big[\big(\bar\psi_A (x)+\Bxi_{2A}(x)\big)\hamma_\mu\big(\psi_A(x)+\Xi_{2A}(x)\big)\big]
\big[\big(\bar\psi_A(0)+\Bxi_{2A}(0)\big)\hamma^\mu\big(\psi_A(0)+\Xi_{2A}(0)\big)\big]
\nonumber\\
&&\hspace{-1mm}
=~{e^2\over 16s_W^2c_W^2}\Big(\big[\Bxi_{2A}(x)\hamma_\mu\psi_A(x)\big]\big[\bar\psi_A(0)\hamma^\mu\Xi_{2A}(0)\big]
+\big[\bar\psi_A(x)\hamma_\mu\Xi_{2A}(x)\big]\big[\Bxi_{2A}(0)\hamma^\mu\psi_A(0)\big]
\nonumber\\
&&\hspace{-1mm}
+~\big[\Bxi_{2A}(x)\hamma_\mu\psi_A(x)\big]\big[\Bxi_{2A}(0)\hamma^\mu\psi_A(0)\big]
+\big[\bar\psi_A(x)\hamma_\mu\Xi_{2A}(x)\big]\big[\bar\psi_A(0)\hamma^\mu\Xi_{2A}(0)\big]\Big),
\label{5.15}
\end{eqnarray}
where we neglected terms which cannot contribute to $W$ due to the reason discussed after eq. (\ref{xxvanish}), i.e.
that one hadron (``A'' or ``B'') cannot produce Z-boson on its own. 

Let us  consider the first term in the r.h.s of this equation
\begin{eqnarray}
&&\hspace{-1mm}
\big[\Bxi_{2A}(x)\hamma_\mu\psi_A(x)\big]\big[\bar\psi_A(0)\hamma^\mu\Xi_{2A}(0)\big]
\label{5.16}\\
&&\hspace{-1mm}
\stackrel{\rm f. int.}{\rightarrow}~-{g^2\over 2N_c(N_c^2-1)}\langle A|\big(\hat{\bar\psi}{1\over\alpha}\big)(x)\gamma_i{\slashed{p}_2\over s}\hamma_\mu\hsi(x)
\hat{\bar\psi}(0)\hamma^\mu{\slashed{p}_2\over s}\gamma_j{1\over\alpha}\hsi(0)|A\rangle
\langle B|\hatA^{ai}(x)\hatA^{aj}(0)|B\rangle
\nonumber\\
&&\hspace{-1mm}
=~-{g^2\over 2N_c(N_c^2-1)}\langle A|\big(\hat{\bar\psi}{1\over\alpha}\big)(x)\gamma_i{\slashed{p}_2\over s}\hamma_k\hsi(x)
\hat{\bar\psi}(0)\hamma^k{\slashed{p}_2\over s}\gamma_j{1\over\alpha}\hsi(0)|A\rangle
\nonumber\\
&&\hspace{77mm}
\times\langle B|\hatA^{ai}(x)\hatA^{aj}(0)|B\rangle~+~O\big({m_\perp^8\over s}\big),
\nonumber
\end{eqnarray}
where $\stackrel{\rm f. int.}{\rightarrow}$ denotes functional integration over $A$ and $B$ fields in eq. (\ref{W3}).

The matrix element $\langle B|\hatA^{ai}(x)\hatA^{aj}(0)|B\rangle$ for unpolarized hadrons can be proportional either
to $2x^ix^j+x_\perp^2g^{ij}$ or to $g^{ij}$. Since the former structure does not contribute 
due to 
\begin{equation}
(2x^ix^j+x_\perp^2g^{ij})\gamma_i\slashed{p}_2\gamma^k\otimes \gamma_k\slashed{p}_2\gamma_j~=~0
\end{equation}
we get
\begin{eqnarray}
&&\hspace{-1mm}
\langle A,B|\big[\Bxi_{2A}(x)\hamma_\mu\psi_A(x)\big]\big[\bar\psi_A(0)\hamma^\mu\Xi_{2A}(0)\big]|A,B\rangle
\label{gluonterm}\\
&&\hspace{11mm}
=~-{g^2\over 2N_c(N_c^2-1)s^2}\langle A|\hat{\bar\psi}{1\over\alpha}(x)\slashed{p}_2(a-\gamma_5)\hsi(x)
\hat{\bar\psi}(0)\slashed{p}_2(a-\gamma_5){1\over\alpha}\hsi(0)|A\rangle
\nonumber\\
&&\hspace{77mm}
\times~\langle B|\hatA^a_i(x)\hatA^{ai}(0)|B\rangle~+~O\big({m_\perp^8\over s}\big).
\nonumber
\end{eqnarray}
For the forward target matrix element one obtains
\begin{eqnarray}
&&\hspace{-1mm}
\!\int\! dx_\ast ~e^{-i\beta_qx_\ast}~\langle B|\hatA^a_i(x)\hatA^{ai}(0)|B\rangle
\label{gluonvklad}\\
&&\hspace{11mm}
=~
{4\over s^2}\int\! dx_\ast ~e^{-i\beta_qx_\ast}\!\int_{-\infty}^{x_\ast}\! dx'_\ast \!\int_{-\infty}^{0}\! dx''_\ast~
\langle B|\hat{F}^a_{\bu i}(x'_\ast,x_\perp)\hat{F}_\bu^{ai}(x''_\ast,0_\perp)|B\rangle
\nonumber\\
&&\hspace{11mm}
=~{4\over\beta_q^2s^2}\!\int\! dx_\ast ~e^{-i\beta_qx_\ast}\langle B|\hat{F}^a_{\bu i}(x_\ast,x_\perp)\hat{F}_\bu^{ai}(0)|B\rangle
~=~-{1\over\beta_q}8\pi^2\alpha_s\cald_g(\beta_q,x_\perp),
\nonumber
\end{eqnarray}
where we used parametrization (4.6) from ref. \cite{Balitsky:2017flc}. Since the gluon TMD $\cald_g(x_B,x_\perp)$ behaves
only logarithmically as $x_B\rightarrow 0$ \cite{Balitsky:2016dgz}, the contribution of eq. (\ref{gluonterm}) to $W(q)$ is of order of 
${m_\perp^2\over\beta_q s}\ll{m_\perp^2\over Q^2}$ (note that the projectile TMD in the r.h.s. of eq. (\ref{5.16}) does not have ${1\over\alpha_q}$ 
terms for the same reason as in eq. (\ref{5.7})). Similarly, all other terms in eq. (\ref{5.15}) are either 
${m_\perp^2\over\beta_q s}$ or ${m_\perp^2\over\alpha_q s}$ so they can be neglected.
\footnote{
It is worth mentioning that if Z-boson is produced in the region of rapidity close to the projectile,
the contribution (\ref{gluonvklad}) may be the most important since gluon parton densities  at small $x_B$ 
are larger than quark ones.}

\subsubsection{Power corrections from $\Psi^{(1)}$ fields\label{psi1sup}}

First, let us notice that  terms like
\begin{equation}
\bsi_A(x)\hamma_\mu\psi_A(x)\bsi_A(0)\hamma^\mu\Psi_{A}^{(1)}(0),~~~
\bsi_A(x)\hamma_\mu\psi_A(x)\bsi_B(0)\hamma^\mu\Psi_{B}^{(1)}(0)
\end{equation}
give zero  contribution since $\bsi_A(x)\hamma_\mu\psi_A(x)$ does not depends on $x_\ast$ 
so \\
$\int\! dx_\ast~e^{-i\beta_qx_\ast}~=~2\pi\delta(\beta_q)$.

Let us consider now the last two lines in the power expansion (\ref{4.6}) of 
$\calj_{AB}^\mu(x)\calj_{BA\mu}(0)$:
\begin{eqnarray}
&&\hspace{-1mm}
\Bsi_A^{(1)}(x)\hamma^\mu\Psi_B^{(0)}(x)\Bsi_B^{(0)}(0)\hamma_\mu\Psi_A^{(0)}(0)
+\Bsi_A^{(0)}(x)\hamma^\mu\Psi_B^{(1)}(x)\Bsi_B^{(0)}(0)\hamma_\mu\Psi_A^{(0)}(0)
\label{5.2}\\
&&\hspace{-1mm}
+~\Bsi_A^{(0)}(x)\hamma^\mu\Psi_B^{(0)}(x)\Bsi_B^{(1)}(0)\hamma_\mu\Psi_A^{(0)}(0)
+\Bsi_A^{(0)}(x)\hamma^\mu\Psi_B^{(0)}(x)\Bsi_B^{(0)}(0)\hamma_\mu\Psi_A^{(1)}(0)~+~...
\nonumber
\end{eqnarray}
After Fierz transformation (\ref{fierz}) the first term in the above equation turns to 
\begin{eqnarray}
&&\hspace{-1mm}
\Bsi_A^{(0)}(x)\hamma^\mu\Psi_B^{(0)}(x)\Bsi_B^{(0)}(0)\hamma_\mu\Psi_A^{(1)}(0)
\label{5.3}\\
&&\hspace{-1mm}
=~{1+a^2\over 2}\big\{\big[\Bsi_A^{m(0)}(x)\gamma_\alpha\Psi_A^{n(1)}(0)\big]\big[\Bsi_B^{n(0)}(0)\gamma^\alpha\Psi_B^{m(0)}(x)\big]
~+~(\gamma_\alpha\otimes\gamma^\alpha\leftrightarrow\gamma_\alpha\gamma_5\otimes\gamma^\alpha\gamma_5)\big\}
\nonumber\\
&&\hspace{-1mm}
-~a\big\{\big[\Bsi_A^{m(0)}(x)\gamma_\alpha\Psi_A^{n(1)}(0)\big]\big[\Bsi_B^{n(0)}(0)\gamma^\alpha\gamma_5\Psi_B^{m(0)}(x)\big]
+(\gamma_\alpha\otimes\gamma^\alpha\gamma_5\leftrightarrow\gamma_\alpha\gamma_5\otimes\gamma^\alpha)\big\}
\nonumber\\
&&\hspace{-1mm}
+~(1-a^2)\big\{\big[\Bsi_A^{m(0)}(x)\Psi_A^{n(1)}(0)\big]\big[\Bsi_B^{n(0)}(0)\Psi_B^{m(0)}(x)\big]
-(1\otimes1\leftrightarrow\gamma_5\otimes\gamma_5)\big\}
\nonumber\\
&&\hspace{-1mm}
=~{1+a^2\over 2}\big\{ {2\over s}\big[\bar\psi_A^{m}(x)\slashed{p}_2\Psi_{1A}^{n(1)}(0)\big]
\big[\bar\psi_B^{n}(0)\slashed{p}_1\psi^{m}_{B}(x)\big]
~+~(\slashed{p}_2\otimes\slashed{p}_1\leftrightarrow\slashed{p}_2\gamma_5\otimes\slashed{p}_1\gamma_5)\big\}
\nonumber\\
&&\hspace{-1mm}
-~a\big\{ {2\over s}\big[\bar\psi_A^{m}(x)\slashed{p}_2\Psi_{1A}^{n(1)}(0)\big]\big[\bar\psi_B^{n}(0)\slashed{p}_1\gamma_5\psi^{m}_{B}(x)\big]
+~(\slashed{p}_2\otimes\slashed{p}_1\gamma_5\leftrightarrow\slashed{p}_2\gamma_5\otimes\slashed{p}_1)\big\}
~+~O\big({m_\perp^8\over s}\big).
\nonumber
\end{eqnarray}
Since
\begin{eqnarray}
&&\hspace{-1mm}
\slashed{p}_2\Psi_{1A}^{(1)}~=~
-{g\slashed{p}_2\slashed{p}_1\over s\beta}\gamma^iB_i\Psi_A
+{g\gamma^i\over s}\slashed{p}_2{1\over \beta}\calp_i{1\over \alpha}\gamma^jB_j\Psi_A
-{2g\over s}\slashed{p}_2{1\over \beta}(\mathbb{A}_\bu^{[1]})^{(0)}\Psi_A
\label{5.4}
\end{eqnarray}
we get
\begin{eqnarray}
&&\hspace{-5mm}
\Bsi_A^{(0)}(x)\hamma^\mu\Psi_B^{(0)}(x)\Bsi_B^{(0)}(0)\hamma_\mu\Psi_A^{(1)}(0)
\nonumber\\
&&\hspace{-1mm}
=~-{1+a^2\over s^2}g\big\{\big[\bar\psi_A^{m}(x)
\Big(\slashed{p}_2\slashed{p}_1\gamma^i\big({1\over\beta}B_i\big)
+\slashed{p}_2\gamma^i\gamma^j{1\over \beta}\calp_iB_j{1\over \alpha}
+2\slashed{p}_2{1\over \beta}(\mathbb{A}_\bu^{[1]})^{(0)}\Big)^{nk}\psi_A^k(0)\big]
\nonumber\\
&&\hspace{44mm}
\times~\big[\bar\psi_B^{n}(0)\slashed{p}_1\psi^{m}_{B}(x)\big]
~+~(\slashed{p}_2\otimes\slashed{p}_1\leftrightarrow\slashed{p}_2\gamma_5\otimes\slashed{p}_1\gamma_5)\big\}
\nonumber\\
&&\hspace{5mm}
+~{2a\over s^2}g\big\{\big[\bar\psi_A^{m}(x)\Big(\slashed{p}_2\slashed{p}_1\gamma^i\big({1\over\beta}B_i\big)
+\slashed{p}_2\gamma^i\gamma^j{1\over \beta}\calp_i B_j{1\over \alpha}
+2\slashed{p}_2{1\over \beta}(\mathbb{A}_\bu^{[1]})^{(0)}\Big)^{nk}\psi_A^k(0)\big]
\nonumber\\
&&\hspace{44mm}
\times~
\big[\bar\psi_B^{n}(0)\slashed{p}_1\gamma_5\psi^{m}_{B}(x)\big]
~+~(\slashed{p}_2\otimes\slashed{p}_1\gamma_5\leftrightarrow\slashed{p}_2\gamma_5\otimes\slashed{p}_1)\big\}.
\label{5.5}
\end{eqnarray}

Let us start with the first term in parentheses in the second line of eq. (\ref{5.5}). Using eq. (\ref{8.13}) the corresponding matrix element  can be rewritten as
\begin{eqnarray}
&&\hspace{-1mm}
-{1+a^2\over 2sN_c}g\langle A|\hbsi(x)\gamma^i\hsi(0)|A\rangle
\langle B|\hbsi(0)\slashed{p}_1\big({1\over\beta}(\hatA_i+i\gamma_5\hat\tilA_i)\big)(0)\hsi(x)|B\rangle
\nonumber\\
&&\hspace{-1mm}
-~{1+a^2\over 2sN_c}g\langle A|\hbsi(x)\gamma^i\gamma_5\hsi(0)|A\rangle
\langle B|\hbsi(0)\slashed{p}_1\big({1\over\beta}(\gamma_5\hatA_i+i\hat\tilA_i)\big)(0)\hsi(x)|B\rangle.
\label{5.6}
\end{eqnarray}
Let us consider 
\begin{eqnarray}
&&\hspace{-1mm}
\int\! dx_\ast~e^{-i\beta_qx_\ast}\langle B|\hbsi(0)\slashed{p}_1\big({1\over\beta}(\hatA_i+i\gamma_5\hat\tilA_i)\big)(0)\hsi(x_\ast,x_\perp)|B\rangle
\label{5.7}\\
&&\hspace{-1mm}
=~{2i\over s}\int\! dx_\ast~e^{-i\beta_qx_\ast}\!\int_{-\infty}^0\!dx'_\ast~x'_\ast 
\langle B|\hbsi(0)\slashed{p}_1\big[\hatF_{\bu i}(x'_\ast,0_\perp)+i\gamma_5\hat\tilF_{\bu i}(x'_\ast,0_\perp)\big]\hsi(x_\ast,x_\perp)|B\rangle,
\nonumber
\end{eqnarray}
where we used
$$
{1\over\beta+i\epsilon}\hatA_k(z_\ast,z_\perp)~=~-i\int_{-\infty}^{z_\ast}\!dz'_\ast~\hatA_k(z'_\ast, z_\perp)
~=~-{2i\over s}\int_{-\infty}^{z_\ast}\! dz'_\ast~(z-z')_\ast \hatF_{\bu k}(z'_\ast, z_\perp).
$$
Let us compare this matrix element to that of eq. (\ref{4.37}):
\begin{eqnarray}
&&\hspace{-1mm}
\int\! dx_\ast~e^{-i\beta_qx_\ast}\langle B|\big(\hat{\bar\psi}{1\over\beta-i\epsilon}\big)(0)\hatA_i(0)\Gamma\hsi(x_\ast,x_\perp)|B\rangle
\label{5.8}\\
&&\hspace{-1mm}
=~-{1\over\beta_q}\!\int\! dx_\ast~e^{-i\beta_qx_\ast}\langle B|
\Big[\hat{\bar\psi}(0)\hatA_i(0)+\!\int_{-\infty}^0\!\!dx'_\ast
~\hat{\bar\psi}(x'_\ast,0_\perp){2\over s}\hat{F}_{\bu i}(0)\Big]\Gamma\hsi(x_\ast,x_\perp)|B\rangle.
\nonumber
\end{eqnarray}
We see that ${1\over \beta _z}$ in eq. (\ref{5.8}) is traded for an extra $x'_\ast\sim1$ in eq. (\ref{5.7})  (recall that $x'_\ast$ in the target matrix elements 
is inversely proportional to characteristic $\beta$'s in the target which are of order 1). Consequently, power correction due to matrix element  (\ref{5.7})
can be neglected in our kinematic region since the matrix element (\ref{5.6}) is $\sim {m_\perp^2\over s}$ rather than $\sim {m_\perp^2\over \alpha_q\beta_qs}$.
Similarly, the second term in eq. (\ref{5.6}) does not contribute in our kinematical region.

This is the same reason why we neglected power corrections (\ref{4.19}). In general, as we discussed in ref. \cite{Balitsky:2017flc},
the way to figure out  integrations that give ${1\over\beta_q}$ 
is very simple: take $\beta_q\rightarrow 0$ and check if there is an infinite
integration of the type $\int_{-\infty}^{x_\ast}dx'_\ast$ without any integrand. Similarly, the factor ${1\over\alpha_q}$ 
can be figured out from (possible) unrestricted integrals over $x'_\bu$ after one sets $\alpha_q=0$ 
in the relevant matrix element. Note that to get the terms $\sim{1\over Q^2}={1\over\alpha_q\beta_qs}$ we need to find contributions which satisfy both of the above conditions.

Next, consider the second term  in parentheses in the second line of eq. (\ref{5.5}). Using eq. (\ref{4.35}) the corresponding matrix element  can be rewritten as
\begin{eqnarray}
&&\hspace{-1mm}
-{1+a^2\over s^2}g\big[\bar\psi_A^{m}(x)\slashed{p}_2\gamma^i\gamma^j\big({1\over \beta}\calp_i B_j\big)^{nk}{1\over \alpha}\psi_A^k(0)\big]\big[\bar\psi_B^{n}(0)\slashed{p}_1\psi^{m}_{B}(x)\big]
\nonumber\\
&&\hspace{-1mm}
=~-{1+a^2\over s^2}g\big[
\bar\psi_A^{m}(x)\slashed{p}_2\gamma^i\gamma^j\big({1\over \beta}B_j\big)^{nk}{1\over \alpha}i\partial_i\psi_A^k(0)
+g\bar\psi_A^{m}(x)\slashed{p}_2\gamma^i\gamma^j\big(A_i{1\over \beta}B_j\big)^{nk}{1\over \alpha}\psi_A^k(0)
\nonumber\\
&&\hspace{-1mm}
+~\bar\psi_A^{m}(x)\slashed{p}_2\gamma^i\gamma^j{1\over \beta}\big(i\partial_iB_j+gB_i B_j\big)^{nk}{1\over \alpha}\psi_A^k(0)\big]\big[\bar\psi_B^{n}(0)\slashed{p}_1\psi^{m}_{B}(x)\big]
\nonumber\\
&&\hspace{-1mm}
\stackrel{\rm f. int.}{\rightarrow}~-{1+a^2\over N_cs^2}g\langle A|\hbsi(x)\slashed{p}_2\gamma^i\gamma^j{1\over \alpha}i\partial_i\hsi(0)|A\rangle
\langle B|\hbsi(0)\slashed{p}_1\big({1\over \beta}\hatA_j\big)(0)\hsi(x)|B\rangle
\nonumber\\
&&\hspace{-1mm}
+~{1+a^2\over N_c(N_c^2-1)s^2}g^2\langle A|\hbsi(x)\slashed{p}_2\gamma^i\gamma^j\hatA_i{1\over \alpha}\hsi(0)|A\rangle
\langle B|\hbsi(0)\slashed{p}_1\big[{1\over \beta}\hatA_j(0)\big]\hsi(x)|B\rangle
\nonumber\\
&&\hspace{-1mm}
-~{1+a^2\over N_cs^2}g\langle A|\hbsi(x)\slashed{p}_2\gamma^i\gamma^j{1\over \alpha}\psi_A(0)|A\rangle
\langle B|\hbsi(0)\slashed{p}_1\big[{1\over \beta}(i\partial_i\hatA_j+g\hatA_i\hatA_j)(0)\big]\hsi(x)|B\rangle.
\end{eqnarray}
Using eq. (\ref{8.13}) we get
\begin{eqnarray}
&&\hspace{-1mm}
-{1+a^2\over s^2}g\big[\bar\psi_a^{m}(x)\slashed{p}_2\gamma^i\gamma^j\big({1\over \beta}\calp_i B_j\big)^{nk}{1\over \alpha}\psi_a^k(0)\big]\big[\bar\psi_b^{n}(0)\slashed{p}_1\psi^{m}_b(x)\big]
+(\slashed{p}_2\otimes\slashed{p}_1\leftrightarrow\slashed{p}_2\gamma_5\otimes\slashed{p}_1\gamma_5)
\nonumber\\
&&\hspace{-1mm}
\stackrel{\rm f. int.}{\rightarrow}~-{1+a^2\over N_cs^2}g\langle A|\hbsi(x)\slashed{p}_2{1\over \alpha}i\partial_i\hsi(0)|A\rangle
\langle B|\hbsi(0)\slashed{p}_1\big({1\over \beta}(\hatA^i-i\hat\tilA^i\gamma_5)(0)\big)\hsi(x)|B\rangle
\nonumber\\
&&\hspace{-1mm}
+~{1+a^2\over N_c(N_c^2-1)s^2}g^2\langle A|\hbsi(x)\slashed{p}_2\hatA_i{1\over \alpha}\hsi(0)|A\rangle
\langle B|\hbsi(0)\slashed{p}_1\big({1\over \beta}(\hatA^i-i\hat\tilA^i\gamma_5)(0)\big)\hsi(x)|B\rangle
\nonumber\\
&&\hspace{-1mm}
-~{1+a^2\over N_cs^2}g\langle A|\hbsi(x)\slashed{p}_2{1\over \alpha}\psi_A(0)|A\rangle
\langle B|\hbsi(0)\slashed{p}_1\big({1\over \beta}(i\partial_i\hatA^i+g\hatA_i\hatA^i)(0)\big)\hsi(x)|B\rangle
\nonumber\\
&&\hspace{-1mm}
-~{2(1+a^2)\over N_cs^3}g\langle A|\hbsi(x)\slashed{p}_2{1\over \alpha}\psi_A(0)|A\rangle
\langle B|\hbsi(0)\slashed{p}_1\gamma_5\big({1\over \beta}\hat\tilF_{\ast\bu}(0)\big)\hsi(x)|B\rangle 
\nonumber\\
&&\hspace{77mm}
+~(\slashed{p}_2\otimes\slashed{p}_1\leftrightarrow\slashed{p}_2\gamma_5\otimes\slashed{p}_1\gamma_5).
\label{5.10}
\end{eqnarray}
It is easy to see that projectile matrix elements lead to terms $\sim{1\over\alpha_q}$ after integration over $x_\bu$, for example
\begin{eqnarray}
&&\hspace{-1mm}
\int\! dx_\bu ~e^{-i\alpha_q x_\bu}\langle A|\hbsi(x)\slashed{p}_2\hatA_i(0){1\over \alpha}\hsi(0)|A\rangle={2\over\alpha_q s}\!\int\!dx_\bu~e^{i\alpha_q x_\bu}\!\int_{-\infty}^{x_\bu}\! dx'_\bu
\\
&&\hspace{-1mm}
\times~\langle A|\hbsi(0)\slashed{p}_2[\hatF_{\ast i}(x_\bu,-x_\perp)\hsi(x'_\bu,-x_\perp)+\hatF_{\ast i}(x'_\bu,-x_\perp)\hsi(x_\bu,-x_\perp)|A\rangle.
\nonumber
\end{eqnarray}
On the other hand, the target matrix elements cannot give a ${1\over\beta_q}$ factor. For the first two lines 
in the r.h.s of eq. (\ref{5.10}) we proved this in eq. (\ref{5.7}) above. As to the last lines in eq. (\ref{5.10}), the target
matrix element can be rewritten as
\begin{eqnarray}
&&\hspace{-2mm}
\int\! dx_\ast ~e^{-i\beta_q x_\ast}
\langle B|\hbsi(0)\slashed{p}_1\big({1\over \beta}(i\partial_i\hatA^i+g\hatA_i\hatA^i)(0)\big)\hsi(x_\ast,x_\perp)|B\rangle
=-{2i\over s}\int\! dx_\ast ~e^{-i\beta_q x_\ast}\!\int_{-\infty}^0\!dx'_\ast
\nonumber\\
&&\hspace{-2mm}
\times~\langle B|\hbsi(0)\slashed{p}_1\Big(\!\int_{-\infty}^{x'_\ast}\! dx''_\ast
\Big[i\hat{D}^i\hat{F}_{\bu i}(x''_\ast,0_\perp)+\frac{2g}{s}\hat{F}_\bu^{~i}(x''_\ast,0_\perp)\!\int_{-\infty}^{x''_\ast}\! dx'''_\ast 
\hat{F}_{\bu i}(x'''_\ast,0_\perp)
\Big]\Big)\hsi(x_\ast,x_\perp)
|B\rangle
\nonumber
\end{eqnarray}
and
\begin{eqnarray}
&&\hspace{-1mm}
\int\! dx_\ast ~e^{-i\beta_q x_\ast}
\langle B|\hbsi(0)\slashed{p}_1\gamma_5\big({1\over \beta}\hat\tilF_{\ast\bu}(0)\big)\hsi(x_\ast,x_\perp)|B\rangle
\\
&&\hspace{8mm}
=~-i\!\int\! dx_\ast ~e^{-i\beta_q x_\ast}\!\int_{-\infty}^0\!dx'_\ast
~\langle B|\hbsi(0)\slashed{p}_1\gamma_5\hat\tilF_{\ast\bu}(x'_\ast,0_\perp)\hsi(x_\ast,x_\perp)
|B\rangle.
\nonumber
\end{eqnarray}
We see that at $\beta_q=0$ there are no unrestricted integration over any longitudinal variable so 
the r.h.s. of these equations cannot give ${1\over\beta_q}$ factor and therefore the contribution to $W(q)$ 
is $\sim{m_\perp^2\over\alpha_qs}\ll {m_\perp^2\over Q^2}$.

Finally, we should consider the third term in eq. (\ref{5.5}) 
\begin{eqnarray}
&&\hspace{-1mm}
-2{1+a^2\over s^2}g\big[\bar\psi_A^{m}(x)
\slashed{p}_2{1\over \beta}(\mathbb{A}_\bu^{[1]nl})^{(0)}\psi_A^l(0)\big]
\big[\bar\psi_B^{n}(0)\slashed{p}_1\psi^{m}_B(x)\big]
\nonumber\\
&&\hspace{-1mm}
=~-{1+a^2\over s^2}g^2\big[\bar\psi_A^{m}(x)\slashed{p}_2\big({1\over\alpha}A_j(0)\big)^{nk}\psi_A^l(0)\big]
\big[\bar\psi_B^{n}(0)\slashed{p}_1\big({1\over\beta}B^j(0)\big)^{kl}\psi^{m}_B(x)\big]
\nonumber\\
&&\hspace{-1mm}
+~{1+a^2\over s^2}g^2\big[\bar\psi_A^{m}(x)\slashed{p}_2\big({1\over\alpha}A_j(0)\big)^{kl}\psi_A^l(0)\big]
\big[\bar\psi_B^{n}(0)\slashed{p}_1\big({1\over\beta}B^j(0)\big)^{nk}\psi^{m}_B(x)\big],
\end{eqnarray}
where we used eq. (\ref{glufildz}) for  $(\mathbb{A}^{[1]}_\bu)^{(0)}$. 
Taking projectile and target matrix elements and separating color-singlet contributions using Eq. (\ref{4.35}), we 
obtain
\begin{eqnarray}
&&\hspace{-1mm}
-{(1+a^2)N_c\over s^2(N_c^2-1)}g^2\langle A|\hbsi(x)\slashed{p}_2\big({1\over \alpha}\hatA_j(0)\big)\hsi(0)|A\rangle
\langle B|\hbsi(0)\slashed{p}_1({1\over\beta}\hatA^j(0)\big)\hsi(x)|B\rangle.
\end{eqnarray}
It has been demonstrated in Eq. (\ref{5.7}) that such matrix elements cannot give 
${1\over \alpha_q}$ and ${1\over \beta_q}$ so their contribution to $W(q)$ is small in our kinematical region. 
Moreover, since the above arguments do not depend on presence (or absence) of $\gamma_5$, we proved that
all terms in Eq. (\ref{5.5}) give small contributions at $\alpha_q,\beta_q\ll 1$. In a similar way, one can demonstrate
that the other three terms in Eq. (\ref{5.2}) can be neglected.

\bibliography{fact1}
\bibliographystyle{JHEP}

\end{document}